\documentclass[journal]{IEEEtran}
\usepackage{cite}
\usepackage{amsmath,amssymb,placeins}
\usepackage{algorithmic}
\usepackage{graphicx}
\usepackage{textcomp}
\usepackage[mathscr]{euscript}
\usepackage{lipsum}
\usepackage{cuted}
\usepackage{bm}
\usepackage{multicol}
\usepackage{array}
\usepackage{makecell}
\usepackage{diagbox}
\usepackage[colorlinks=true,linkcolor=gray,citecolor=black]{hyperref}
\usepackage[symbol]{footmisc}
\usepackage[utf8]{inputenc}
\usepackage{caption}
\usepackage{tabularx}

\usepackage[mathscr]{euscript}
\usepackage[usenames]{color}
\usepackage[table]{xcolor}
\usepackage{amsfonts}
\usepackage{latexsym}
\usepackage{subfigure}
\usepackage{tikz}
\usepackage{caption}
\usepackage{floatrow}
\usepackage{multirow}
\usepackage{epstopdf}
\usepackage[english]{babel}

\usepackage{amsthm}
\newtheorem{theorem}{Theorem}
\newtheorem{lemma}{Lemma}
\newtheorem{corollary}{Corollary}

\newtheorem{proposition}{Proposition}
\newtheorem{remark}{Remark}

\newtheorem{example}{Example}

\newcommand{\ccn}{\color{black}}

\newtheorem{case}{\textit{Case}}
\def\BibTeX{{\rm B\kern-.05em{\sc i\kern-.025em b}\kern-.08em
		T\kern-.1667em\lower.7ex\hbox{E}\kern-.125emX}}
		
		\def\@fnsymbol#1{\ensuremath{\ifcase#1\or *\or \dagger\or \ddagger\or
   \mathsection\or \mathparagraph\or \|\or **\or \dagger\dagger
   \or \ddagger\ddagger \else\@ctrerr\fi}}
\newcommand{\Z}{\mathbb{Z}}
{\renewcommand{\baselinestretch}{1.2}

\begin{document}
\title{New Correlation Bound and Construction of Quasi-Complementary Sequence Sets} 
\author{Palash~Sarkar,~Chunlei~Li,~Sudhan~Majhi,~and  %\IEEEmembership{}
       Zilong~Liu~%\IEEEmembership{Senior Member,~IEEE,}
%      \thanks{The work of Palash Sarkar and Chunlei Li was supported by the Research Council of Norway (No. 311646/O70). Sudhan Majhi's  research was supported by SERB GoI, Core Research Grant (CRG) with grant No. CRG/2022/000529 and Empowerment And Equity Opportunities For Excellence In Science (EEQ) with grant No. EEQ/2022/001018. The work of Z. Liu was supported in part by the UK Engineering and Physical Sciences Research Council under Grants EP/X035352/1 and EP/Y000986/1, by the Royal Society under Grant IEC$\backslash$R3$\backslash$223079, and by the Research Council of Norway under Grant 311646/070. } 
\thanks{Palash Sarkar and Chunlei Li are with the Department of Informatics, Selmer Center, University of Bergen, Norway, e-mail: {\tt palash.sarkar@uib.no; chunlei.li@uib.no}.}% <-this % stops a space
\thanks{Sudhan Majhi is with the Department of Electrical Communication Engineering, Indian Institute of Science, Bangalore, India, e-mail:{\tt smajhi@iisc.ac.in}.}
\thanks{Zilong Liu is with the School of Computer Science and Electronic Engineering, University of Essex, UK, e-mail:{\tt zilong.liu@essex.ac.uk}.}}
\IEEEpeerreviewmaketitle
\maketitle
\begin{abstract}
Quasi-complementary sequence sets (QCSSs) have attracted sustained research interests for simultaneously supporting more active users in multi-carrier code-division multiple-access (MC-CDMA) systems compared to  complete complementary codes (CCCs).
In this paper, we investigate a novel class of QCSSs composed of multiple CCCs.  
We derive a new aperiodic correlation lower bound for this type of QCSSs, which is tighter than the existing bounds for QCSSs.
We then present a systematic construction of such QCSSs with a small alphabet size and low maximum correlation magnitude, and also 	 
show that the constructed aperiodic QCSSs can meet the newly derived bound asymptotically.
\end{abstract}
\section{Introduction}
\label{sec:intro}
As a generalization of the Golay complementary pair \cite{golay1961}, the complementary sequence set introduced by Tseng and Liu  \cite{chinchong} 
consists of $M\geq 2$ constituent sequences of length $L$ having zero aperiodic auto-correlation sum for all nonzero time shifts.  
A complementary sequence set is usually arranged as an $M\times L$ matrix (known as a complementary matrix or complementary code). 
A set of $K$ complementary codes  with the same order $(M,L)$ is called a mutually orthogonal complementary sequence set (MOCSS) if any two distinct complementary codes  have zero aperiodic cross-correlation sums for all time shifts \cite{rati}.
A MOCSS has its size $K\leq M$ and it is known as
a complete complementary code (CCC) when the equality is reached.
Due to the ideal auto- and cross-correlation properties, CCCs have a salient  feature for supporting interference-free 
multi-carrier code-division multiple-access (MC-CDMA) communication where users are assigned with different complementary codes from a  CCC \cite{chen2007next,frdlu,hator}.

To support more users in MC-CDMA systems, the notion of low-correlation zone complementary sequence set (CSS),  which refers to a set of complementary codes  or codes  having low maximum correlation magnitudes within a time-shift zone around the origin, was proposed \cite{lcz_lb}; in particular, when the maximum correlation magnitude within the zone is zero, it reduces to a zero-correlation zone CSS \cite{pskaccess,psktcom,pa_pbf}. 
By extending the low correlation zone to all the non-trivial time-shifts, quasi-complementary sequence sets (QCSSs) with uniformly low  maximum correlation magnitude have been investigated in \cite{liuqcss}.
A QCSS-based MC-CDMA system is expected to accommodate larger amount of asynchronous time-offsets, whilst supporting more users \cite{samad_majhi,hhchen_mccdma}.
%%%%%%%%%%%%%%%%%%%%%%%%%%%%%%%%%%%%%%%%%%%%%%%%%%%%%%%Correlaton Bounds%%%%%%%%%%%%%%%%%%%%%%%%%%%%%%%%%%%%%%%%%%%%
\subsection{Existing Works on the Constructions and the Correlation Bounds of QCSSs}
In this subsection, we recall some basics and known results on QCSSs. 
Let $q$ be a positive integer and $\mathcal{A}_q=\{\xi_q^i\, | 0\leq i<q \}$, where $\xi_q=\exp(2\pi\sqrt{-1}/q)$ is a $q$-th primitive root of unity. 
We denote by $\mathcal{A}_q^{M\times L}$ the set of all $M\times L$ matrices over $\mathcal{A}_q$.
A subset of $\mathcal{A}_q^{M\times L}$ is termed a $(K,M,L,\theta)$-QCSS over $\mathcal{A}_q$ if it consists of $K$ matrices in $\mathcal{A}_q^{M\times L}$ and its maximum magnitude of aperiodic correlation sums equals a positive value $\theta$. 
The multipath interference and multiuser interference in QCSS-based MC-CDMA system are closely related to the maximum correlation sum magnitude $\theta$, which is 
desired to be small. 
In the literature, several researchers have studied the lower bound on $\theta$.
Welch in \cite{crlbw} first gave the following lower bound:
\begin{equation}\label{lb1}
\theta\geq ML\sqrt{\frac{\frac{K}{M}-1}{K(2L-1)-1}}.
\end{equation}
In 2014, Liu, Guan and Mow \cite{crlbzl} extended the idea of Levenshtein bound \cite{leven} for $M\geq 2$ and provided a tighter correlation lower bound for the case of $K \geq 3M$ and $L \geq 2$:
\begin{equation}\label{intlczbnd}
\theta\geq \sqrt{ML\left(1-2\sqrt{\frac{M}{3K}}\right)}.
\end{equation}
With respect to a lower bound, the optimality of a QCSS can be evaluated in terms of  the optimality factor $\rho$, which is defined as the ratio of its maximum correlation magnitude $\theta$ and the lower bound \cite{liuqcss}.
%$\rho=\frac{\theta}{\theta'}$, where $\theta'$ is the lower bound on $\theta$.
A $(K,M,L,\theta)$-QCSS is said to be optimal if $\rho=1$, near-optimal if $1<\rho\leq 2$, and asymptotically optimal if $\rho$ tends to $1$ for sufficiently large $L$, with respect to a lower bound, which is usually taken as the best known one.

For periodic QCSSs, the first known optimal and near-optimal QCSSs were proposed in \cite{liuqcss} with the aid of Singer difference sets. Several other constructions on periodic QCSSs using various algebraic tools, such as difference sets and characters over finite fields, can be found in \cite{li_qcss_cl,liqcss2,liqcss1,luo_qccs}.
Aperiodic QCSSs with asymptotically optimal correlation properties have been developed with the help of various tools, such as permutation functions and Florentine rectangles \cite{liano,zhu_qccs,avikr_qccs}.
The QCSSs in \cite{zhu_qccs} and \cite{avikr_qccs} appear as a collection of CCCs with low inter-set cross-correlation properties.
In practice\ccn, this type of QCSSs can be useful in a multi-cell (or multi-cluster) mobile network where each cell is assigned with a distinctive CCC for interference-free MC-CDMA communication; at the same time, each cell also receives multiuser interference from other neighbouring cells \cite{fhlu,tamaw}. In this setting, the low inter-set cross-correlation property may permit minimum inter-cell interference, whilst achieving zero intra-cell interference due to the ideal correlation properties of CCCs.
Besides low correlation, it is also desirable to design QCSSs over an alphabet of small size  for the ease of practical implementations \cite{luo_qccs}. 
\subsection{Motivations and Contributions}
Motivated by the promising applications of QCSSs in MC-CDMA systems\ccn, in this paper we are interested in investigating aperiodic QCSSs that are composed of multiple CCCs. Fundamentally, we aim to understand the theoretical trade-offs between different parameters of this type of QCSSs. Furthermore, we target at developing systematic constructions with both desirable correlation properties and flexible parameters. 

Our first contribution in this paper is  the derivation of a new lower bound on the maximum correlation magnitude of the new type of QCSSs. The new bound is obtained by a revisit to the generalized Levenshtein bound for QCSSs in \cite{crlbzl} with extra consideration on a special feature of such QCSSs. 
 Several forms of this new lower bound are derived by setting proper weighting vectors in the bounding function. As listed in Table \ref{nupur1}, they 
are shown to be tighter than the lower bound in \cite{crlbzl}. Here it is worth noting that the bound in \cite{crlbzl} was proposed for generic QCSSs and that 
new bounds in Table \ref{nupur1} should be used to evaluate 
aperiodic QCSSs composed of multiple CCCs. It is to be noted that the QCSSs reported in \cite{zhu_qccs} and \cite{avikr_qccs} satisfy our proposed aperiodic correlation lower bound.   
\begin{table}[]
	\caption{Aperiodic correlation lower bounds for $(K,M,L,\theta)$-QCSSs composed of  $(M,L)$-CCCs}\label{nupur1}
	\resizebox{\columnwidth}{!}{
		\begin{tabular}{|c|l|l|l|}
			\hline
			\multicolumn{1}{|c|}{$N=K/M$} & Derived correlation lower bound                                                                      & Derivation                     & Constraints                 \\ \hline
			\multicolumn{1}{|c|}{$N=2$}                    & $\theta^2 \geq {\frac{ML^2}{2L-1}}$                                                                & Corollary 1                    & $L,M\geq 2$                 \\ \hline
			\multirow{3}{*}{$N=3$}                        
			& $\theta^2\geq \frac{ML^2}{2L-1}$                                                 & Corollary 1                    & $3\leq L\leq 25$, $M\geq 2$ \\ 
			\cline{2-4}
			& \multicolumn{1}{c|}{$\theta^2\geq {ML\left(1-\frac{L^2(2\pi^2+4N-16)-N\pi^2}{16L^2(N-1)}\right)}$} & Corollary 2                    & $L> 25$, $M\geq 2$       \\ \cline{2-4}
			& $\theta^2\geq \frac{ML}{3}$                                                                           & \cite{crlbzl} & $L,M\geq 2$                 \\ \hline
			\multirow{3}{*}{$N=4$}                         & $\theta^2\geq  {ML\left(1-\frac{L^2(2\pi^2+4N-16)-N\pi^2}{16L^2(N-1)}\right)}$                           & Corollary 2                    & $L\geq 5$, $M\geq 2$   
			\\ \cline{2-4}
			& \multicolumn{1}{l|}{$\theta^2\geq ML(1-\frac{L-1.2}{2L-1})$} & Corollary 2                    & $L=4, M\geq 2$    
			   \\ \cline{2-4}
			& $\theta^2\geq ML\left(1-\frac{1}{\sqrt{3}}\right)$                                                   & \cite{crlbzl} & $L,M\geq 2$                 \\ \hline
			\multirow{2}{*}{$N>4$}                         & $\theta^2\geq {ML\left(1-\frac{\pi\sqrt{N(2L^2-N)}-4L}{4(N-1)L}\right)}$                           & Corollary 2                    & $L\geq 5$, $M\geq 2$        \\ \cline{2-4}
			& $\theta^2\geq ML\left(1-\frac{2}{\sqrt{3N}}\right)$                                                   & \cite{crlbzl} & $L,M\geq 2$                 \\ \hline
	\end{tabular}} 
\end{table}

In the construction of QCSSs,  multivariate functions have turned to be an effective tool to generate sequences with flexible parameters. 
Multivariate functions were studied in \cite{palash_ccc} to design CCCs with flexible parameters and then soon followed by \cite{rajen_zcp} to construct Z-complementary sequences and by \cite{tao} to construct Golay complementary array set.
Our second contribution in this paper is a systematic construction framework of aperiodic QCSSs using multivariate functions from a graphical perspective.
We consider the multivariate functions from  $\mathbb{Z}_p^m$ to $\mathbb{Z}_q$, where $p$ is an arbitrary prime divisor of $q$. This type of multivariate functions are referred to as $q$-ary functions in this paper for ease of presentation. \ccn With a graphical approach, we utilize $q$-ary functions in $m$ variables to construct a $(p^{n+1}(p-1),p^{n+1},p^m,p^m)$-QCSS over $\mathcal{A}_q$, where $1\leq n<m$. The key requirement on the employed $q$-ary function $f$ is that the graph of each restriction of $f$ on certain $n$ variables
yields a Hamiltonian path with edges having identical weights of $q/p$. 
We show that  such $q$-ary functions give rise to $p-1$  distinct $(p^{n+1}, p^m)$-CCCs, and a QCSS composed of these CCCs has maximum correlation magnitude $p^m$. Notice that
 the alphabet size $q$ of proposed QCSSs can be as small as $p$, 
which is different from the known QCSSs as listed in Table \ref{comtab}, for which the alphabet size is required to be at least the sequence length.  
To the best of our knowledge, it is the first time in the literature that a construction of aperiodic QCSSs can maintain a small alphabet size
irrespective of the sequence length and set size.
Finally, it is shown that the proposed QCSSs can asymptotically meet the newly derived aperiodic correlation lower bound. 

The structure of this paper is outlined as follows: In Section II, we introduce the essential mathematical tools utilized in this work. 
Section III derives the proposed new tighter correlation lower bounds for QCSSs comprised of multiple CCCs.
In Section IV, we present our contributions related to the construction of QCSSs that can meet the correlation lower bound introduced in Section III. Lastly, Section V concludes this work.
\begin{table*}[t]\small
	\centering
	\caption{The parameters of the exisiting and proposed \ccn aperiodic QCSSs}\label{comtab}
	\resizebox{\textwidth}{!}{
	\begin{tabular}{ |c|c|c|c|c|c|c| }
			\hline
			Ref.& $K$ & $M$ & $L$ & $\theta$ & Alphabet &Constraints \\\hline\hline
			Th. 1 \cite{liano}& $u(u+1)$ & $u$ & $u$ & $u$ & $\mathbb{Z}_u$ & $u$ is power of prime  \\\hline
			Th. 2 \cite{liano} & $u^2$ & $u$ & $u-1$ & $u$ & $\mathbb{Z}_u$ & $u$ is power of prime, $u\geq 5$ \\ \hline
			\cite{zhu_qccs} &  $N(t_0-1)$ & $N$ & $N$ & $N$ & $\mathbb{Z}_N$ & \makecell{ $N$ ($\geq 5$) is odd positive integer with\\ $t_0$ as its smallest prime factor} \\ \hline
			\cite{avikr_qccs} & $N\times F(N)$ & $N$ & $N$ & $N$ & $\mathbb{Z}_N$ & \makecell{$N$ ($\geq 2$) is any integer,\\ $F(N)$ is the maximum number of rows for which\\ $F(N)\times N$ Florentine rectangle exists} \\ \hline
			Proposed & $p^{n+1}(p-1)$ & $p^{n+1}$ & $p^m$ & $p^m$ & $\mathbb{Z}_q$ & \makecell{ $p$ is a prime number,\\$m$ is any positive integer,\\ $n$ ($\leq m-1$) is any non-negetive integer,\\
				$q$ is a positive multiple of $p$ }	\\ \hline 	
	\end{tabular}}
\end{table*}
%%%%%%%%%%%%%%%%%%%%%%%%%%%%%%%%%%%%%%%%%%%%%%%%%%%%%%%%%%%%%%%%%%%%%%%%%%%%%%%%%%%%%%%%%%%%%%%%%%%%%%%%%%%%%%%
%%%%%%%%%%%%%%%%%%%%%%%%%%%%%%%%%%%%%%%%%%%%%%%%%%%%%%%%%%%%%%%%%%%%%%%%%%%%%%%%%%%%%%%%%%%%%%%%%%%%%%%%%%%%%%%
                                        
\section{Preliminaries}
We first define some notations which will be used throughout the paper: 
\begin{itemize}
  \item $\mathbb{Z}_t=\mathbb{Z}/t\mathbb{Z}$ is the set of all integers modulo $t$
 \item  $q$ is a positive integer and $p$ is an arbitrary prime divisor of $q$
 \item $\xi_q=\exp(2\pi\sqrt{-1}/q)$ is a primitive $q$-th root of unity
 \item $\mathcal{A}_q=\{\xi_q^i\,:\, 0\leq i< q\}$ and
 $\mathcal{A}_q^{M\times L}$ is the set of matrices over $\mathcal{A}_q$ 
 \item $\mathbf{0}_L$ denotes the zero vector of length $L$
 \item lower-case letters in bold, e.g., $\mathbf{a}$, $\mathbf{b}$, denote sequences of certain length
 \item upper-case letters in bold, e.g, $\mathbf{C}$, $\mathbf{X}, \mathbf{Y}$, denote matrices or codes over $\mathcal{A}_q$
 \item $T^u(\mathbf{a}) =(a_{L-u},\cdots,a_{L-1},a_0,\hdots,a_{L-u-1})$ for a sequence $\mathbf{a} = (a_0,a_1, \cdots, a_{L-1})$
 \item $|a|,\, a^*$ denote the magnitude and conjugate of a complex number $a$, respectively
 \item $\left<\mathbf{a},\mathbf{b}\right>=a_0b_0^*+a_1b_1^*+\hdots+a_{L-1}b_{L-1}^*$ denotes the inner product between two complex-valued sequences $\mathbf{a}=(a_0,a_1,\dots, a_{L-1})$ and $\mathbf{b}=(b_0,b_1,\dots, b_{L-1})$ \ccn
 \item $\mathbf{a}\cdot \mathbf{b}$ denotes the inner product for two real-valued sequences $\mathbf{a}$ and $\mathbf{b}$  
 \item $\lceil a\rfloor$ denotes the integer closest to a real number $a$
 \item $[a,b]$ denotes a closed interval consisting of real numbers $x$ satisfying $a\leq x \leq b$
 \item $[a:b] = [a,\hdots, b]$ for integers $a\leq b$ 
 \item $\emptyset$ denotes the empty set
 \item $|S|$ denotes the size of a set $S$
\end{itemize}
\subsection{Aperiodic Auto- and Cross-Correlation}
For any two complex-valued sequences $\mathbf{a}=(a_0,a_1,\hdots,a_{L-1})$ and $\mathbf{b}=(b_0,b_1,\hdots,b_{L-1})$ of length $L$, we define the aperiodic cross-correlation function (ACCF) at time-shift $\tau$, where  $0\leq |\tau|<L$ as
\begin{equation}\label{accf}
\Theta (\mathbf{a},\mathbf{b})(\tau)=
\begin{cases}
\sum_{\alpha=0}^{L-\tau-1}a_\alpha b^*_{\alpha+\tau}, & 0\leq \tau<L,\\
\sum_{\alpha=0}^{L+\tau-1}a_{\alpha-\tau} b^*_\alpha,&-L<\tau<0.
\end{cases}
\end{equation}
For $\mathbf{a}=\mathbf{b}$, the ACCF defined in (\ref{accf}) reduces to the aperiodic auto-correlation function (AACF) of 
$\mathbf{a}$, which will be denotd as $\Theta (\mathbf{a})(\tau)$ for short.

Let $\mathcal{C}=\{\mathbf{C}_1,\mathbf{C}_2,\hdots,\mathbf{C}_{K}\}$ be a collection of $K$ codes, each containing $M$ sequences of length $L$. By arranging each code as a two-dimensional matrix, we write  $\mathbf{C}_k$ as
$$\mathbf{C}_k=\begin{bmatrix}
\mathbf{c}_k^1\\
\mathbf{c}_k^2\\
\vdots\\
\mathbf{c}_k^{M}
\end{bmatrix}_{M\times L},$$
where $k=1,\hdots,K$. The ACCF (sum) between $\mathbf{C}_{k_1}$
and $\mathbf{C}_{k_2}$ for $1\leq k_1, k_2 \leq K$ is defined as 
\begin{equation}\label{accfc}
\Theta (\mathbf{C}_{k_1},\mathbf{C}_{k_2})(\tau)=\sum_{j=1}^{M}\Theta   (\mathbf{c}_{k_1}^j,\mathbf{c}_{k_2}^j)(\tau).  
\end{equation}
For $k_1=k_2=k$, the ACCF in (\ref{accfc}) reduces to the AACF of $\mathbf{C}_{k}$ and we denote it by $\Theta (\mathbf{C}_{k})(\tau)$. Define
\begin{equation}\nonumber
\begin{split}
\theta_A=&\max \{|\Theta (\mathbf{C}_k)(\tau)|:k=1,\hdots,K,~0<|\tau|<L\},\\ \theta_C=&\max\{|\Theta (\mathbf{C}_{k_1},\mathbf{C}_{k_2})(\tau)|:1\leq k_1\neq k_2 \leq K, \\&~~~~~~~0\leq |\tau|<L\}.
\end{split}
\end{equation}
The maximum correlation magnitude of $\mathcal{C}$ is given by $\theta=\max\{\theta_A,\theta_C\}$. This collection of  codes is called an aperiodic QCSS, denoted by $(K,M,L,\theta)$-QCSS. In particular, when $\theta=0$ and 
$K=M$, $\mathcal{C}$ is said to be a CCC, 
denoted by an $(M,L)$-CCC.

In order to investigate the lower bound on $\theta$ for QCSSs, we recall an interesting function from \cite[Eq. (17)]{crlbzl} below. 
For two $(K,M,L,\theta)$-QCSSs $\mathcal{X},\mathcal{Y}\subset \mathcal{A}_q^{M\times L}$, define the following function:
\begin{equation}\label{fcd}
\begin{split}
&F(\mathcal{X},\mathcal{Y})=\frac{1}{|\mathcal{X}||\mathcal{Y}|}\times\\&\sum_{\mathbf{X}\in\mathcal{X}}\sum_{\mathbf{Y}\in\mathcal{Y}}\sum_{u=0}^{2L-2}\sum_{v=0}^{2L-2}\!\!\big|\big<T^u(\mathbf{X},\mathbf{0}_{L-1}),T^v(\mathbf{Y},\mathbf{0}_{L-1})\big>\big|^2w_u w_v,
\end{split}
\end{equation}
with 
\begin{equation}
\begin{split}
&\Big<T^u(\mathbf{X},\mathbf{0}_{L-1}),T^v(\mathbf{Y},\mathbf{0}_{L-1})\Big>\\=&
\sum_{j=1}^{M}\Big<T^u(\mathbf{X}^j,\mathbf{0}_{L-1}),T^v(\mathbf{Y}^j,\mathbf{0}_{L-1})\Big>,
\end{split}
\end{equation}
where $T$ represents the right cyclic shift operator, 
$(\mathbf{X}^j,\mathbf{0}_{L-1}), (\mathbf{Y}^j,\mathbf{0}_{L-1})$ denote the concatenation of the $j$th row of $\mathbf{X}$, $\mathbf{Y}$ and $\mathbf{0}_{L-1}$, respectively, and 
	$\mathbf{w}=(w_0,w_1,\hdots,w_{2L-2})$ is a weight vector, satisfying 
	\begin{equation}\label{eq:weightvector}\nonumber
	\sum_{j=0}^{2L-2}w_j=1  \text{ and }  w_j \geq 0 \text{ for } 0\leq j\leq 2L-2.
	\end{equation}
In the case of $\mathcal{X}=\mathcal{Y}=\mathcal{C}$, a lower bound on $F(\mathcal{C},\mathcal{C})$ was derived in \cite{crlbzl}. 	
We represent the lower bound in the following lemma which will be used in Section III to obtain new correlation lower bound for $\theta$: 
\begin{lemma}[\cite{crlbzl}]\label{lma1}
	Let $\mathcal{C}$ be a $(K,M,L,\theta)$-QCSS. Then
	\begin{equation}
	F(\mathcal{C},\mathcal{C})\geq \sum_{u,v=0}^{2L-2}M(L-\tau_{u,v,L})w_u w_v,
	\end{equation}
	where $$0\leq \tau_{u,v,L}=\min\{|v-u|,2L-1-|v-u|\}\leq L-1.$$	
\end{lemma}
\smallskip
In the following, we present some basics on $q$-ary functions and it's relation with graphs and sequences. We also discuss some basic and necessary properties of sequences. 
\subsection{Sequences associated with $q$-ary Functions}
Let $q$ be a positive integer and $p$ is a prime divisor of $q$. For example, let $q=12=2^2 3$, in this case $p$ can be either $2$ or $3$. 
For a $q$-ary function $f:\Z_p^m\rightarrow\mathbb{Z}_q$, it defines a $\mathbb{Z}_q$-valued sequence 
$\mathbf{f}=\left(f_0,f_1,\hdots,f_{p^m-1}\right),$
where  the coordinate $f_i=f(i_0,i_1,\hdots,i_{m-1})$ for $0\leq i<p^m$ with $i=\sum_{j=0}^{m-1}i_{j}p^{m-j-1}$ and the arithmetic operations on variables in the function $f$ are taken modulo $q$. 
In the sequel we shall 
identify an integer $i$ with $0\leq i<p^m$ as its $p$-ary vector representation $(i_0, i_1, \dots, i_{m-1})$ when there is no ambiguity. \ccn
We define the complex-valued sequence associated with $f$, denoted by $\psi_q(f)$, as $$\psi_q(f)=\left(\xi_q^{f_0},
\xi_q^{f_1},\hdots,\xi_q^{f_{p^m-1}}\right).$$
When there is no ambiguity in the context, we will write $\psi_q(f)$ as $\psi(f)$ for simplicity. 
For $\mathbf{x}=\left(x_0,x_1,\hdots,x_{m-1}\right)\in \Z_p^m$ and a subset 
$J=\{j_0,j_1,\hdots,j_{n-1}\}\subset \Z_m$, we define
$\mathbf{x}_J=\left(x_{j_0},x_{j_1},\hdots,x_{j_{n-1}}\right)$ as the restriction of the vector $\mathbf{x}$ on $J$. 
For $\mathbf{c}=(c_0,c_1,\hdots,c_{n-1})\in\Z_p^n$ and $\mathbf{x}_J=\mathbf{c}$, we define the
complex-valued sequence corresponding to the restricted $q$-ary function $f\arrowvert_{\mathbf{x}_J=\mathbf{c}}$ as follows:
\begin{equation}\label{vect}
\begin{split}
\psi(f\arrowvert_{\mathbf{x}_J=\mathbf{c}})&=(a_0,a_1,\hdots,a_{p^m-1}) \text{ with }\\
a_i&=\begin{cases}
\xi_q^{f_i}, & i_J = \mathbf{c},\\
0,& \textnormal{otherwise}.
\end{cases}
\end{split}
\end{equation}
%%%%%%%%%%%%%%%%%%%%%%%%%%%%%%%%%%%Revision 2 Updation%%%%%%%%%%%%%%%%%%%%%%%%%%%%%%%%
%%%%%%%%%%%%%%%%%%%%%%%%%%%%%%%%%%%%%%%%%%%%%%%%%%%%%%%%%%%%%%%%%%%%%%%%%%%%%%%%%%%%%%
For any two functions $f,g:\mathbb{Z}_p^m\rightarrow \mathbb{Z}_q$, below we define a set of ordered pairs $(\gamma,\delta)$ to calculate the ACCF between two $q$-ary restricted functions, $\psi(f\arrowvert_{\mathbf{x}_J=\mathbf{c}_1})$ and  $\psi(g\arrowvert_{\mathbf{x}_{J}=\mathbf{c}_2})$, where $\mathbf{c}_i \in\Z_p^n$ for $i=1,2$, at a time-shift  $0\leq \tau<p^m$  as follows:
\begin{equation}\label{cross_corr_restrict_c1_c2}
\begin{split}
\mathbf{B}_\tau(\mathbf{c}_1,\mathbf{c}_2)=&\{(\gamma,\delta):\delta=\gamma+\tau, 0\leq \gamma \leq p^m-\tau-1,\\& \gamma_{J}=\mathbf{c}_1, \delta_{J}=\mathbf{c}_2\}
\end{split}
\end{equation} 
where $\gamma_J, \, \delta_J$ correspond to the restrictions of the vector representations of $\gamma, \, \delta$, respectively.
Following (\ref{vect}), the complex-valued sequences $\psi(f\arrowvert_{\mathbf{x}_J=\mathbf{c}_1})$ and $\psi(g\arrowvert_{\mathbf{x}_J=\mathbf{c}_2})$ can be expressed as follows:
\begin{equation}\label{vect786786}
\begin{split}
 \psi(f\arrowvert_{\mathbf{x}_J=\mathbf{c}_i})&=(a_0,a_1,\hdots,a_{p^m-1}) \text{ with }\\
a_\gamma &=\begin{cases}
\xi_q^{f_\gamma}, & \gamma_J = \mathbf{c}_1,\\
0,& \textnormal{otherwise}, 
\end{cases}\\
\psi(g\arrowvert_{\mathbf{x}_J =\mathbf{c}_2})&=(b_0,b_1,\hdots,b_{p^m-1}) \text{ with }\\
b_\delta &=\begin{cases}
\xi_q^{g_\delta}, & \delta_J = \mathbf{c}_2,\\
0,& \textnormal{otherwise}.
\end{cases}
\end{split}
\end{equation}
From (\ref{accf}), (\ref{cross_corr_restrict_c1_c2}) and (\ref{vect786786}), the ACCF can be expressed as 
\begin{equation}\label{ccrf_rest1}
\begin{split}
\Theta(\psi(f\arrowvert_{\mathbf{x}_J=\mathbf{c}_1}),\psi(g\arrowvert_{\mathbf{x}_{J}=\mathbf{c}_2}))(\tau)
&=\sum_{\gamma = 0}^{p^m-\tau-1} a_{\gamma} b^*_{\delta} \\
&=\sum_{(\gamma,\delta)\in \mathbf{B}_\tau(\mathbf{c}_1,\mathbf{c}_2)} \xi_q^{{f_\gamma} - g_{\delta}}.
\end{split}
\end{equation}
When $\mathbf{c}_1=\mathbf{c}_2=\mathbf{c}$,  
we denote the notation $\mathbf{B}_\tau(\mathbf{c}_1,\mathbf{c}_2)$ in (\ref{cross_corr_restrict_c1_c2}) by $\mathbf{A}_\tau(\mathbf{c})$ which can be expressed as follows: 
\begin{equation}\label{corr_restric_c}
\begin{split}
 \mathbf{A}_\tau(\mathbf{c})=&\{(\gamma,\delta):0\leq \gamma\leq p^m-\tau-1,\delta=\gamma+\tau,\\&\gamma_{J}=\delta_{J}=\mathbf{c}\}.
 \end{split} 
\end{equation}
From (\ref{accf}), (\ref{vect786786}) and (\ref{corr_restric_c}), the ACCF between $\psi(f\arrowvert_{\mathbf{x}_J=\mathbf{c}})$ and $\psi(g\arrowvert_{\mathbf{x}_J=\mathbf{c}})$ can be expressed as
\begin{equation}\label{vect10987}
\Theta(\psi(f\arrowvert_{\mathbf{x}_J=\mathbf{c}}),\psi(g\arrowvert_{\mathbf{x}_{J}=\mathbf{c}}))(\tau)=\sum_{(\gamma,\delta)\in \mathbf{A}_\tau(\mathbf{c})}\xi_q^{f_\gamma - g_\delta}.
\end{equation}
When $f=g$, the ACCF in (\ref{vect10987}) reduces to the AACF of $\psi(f\arrowvert_{\mathbf{x}_J=\mathbf{c}})$.
%%%%%%%%%%%%%%%%%%%%%%%%%%%%%%%%%%%%%%%%%%%%%%%%%%%%%%%%%%%%%%%%%%%%%%%%%%%%%%%%%%%%%%%
%%%%%%%%%%%%%%%%%%%%%%%%%%%%%%%%%%%%%%%%%%%%%%%%%%%%%%%%%%%%%%%%%%%%%%%%%%%%%%%%%%%%%%%

The following example illustrates the sequences associated with a $q$-ary function and the calculation of ACCF between sequences associated with restricted $q$-ary functions.
	\begin{example}\label{Ex1}
		Assume $p=3$, $m=3$, and $q=3$. Let us consider  
		$f:\Z_3^3\rightarrow \mathbb{Z}_3$ as follows:
		\begin{equation}\nonumber
		f(x_0,x_1,x_2)=x_0x_2+2x_2x_1+2x_1^2+x_2+1.
		\end{equation}
		According to the above definitions, the associated sequences $\textbf{f}$, $\psi(f)$ and restricted sequences $\psi(f\arrowvert_{\mathbf{x}_J=\textbf{c}})$ w.r.t $J= \{0,2\}$ and $\textbf{c} \in \{(0,2), (1,2), (2,2)\}$ can be given in the Table \ref{3rdanki}, where the blanks for the last three rows $\psi(f\arrowvert_{\mathbf{x}_J=\textbf{c} })$ indicate that the corresponding coordinates take values of $0$.
		\begin{table*}
			\caption{Sequences corresponding to the ternary function $x_0x_2+2x_2x_1+2x_1^2+x_2+1$ \ccn }\label{3rdanki}
			\vspace*{-0.2cm}
		\begin{equation}\nonumber
		{
		\arraycolsep=.7pt\def\arraystretch{1}
		\begin{array}{|c|c|c|c|c|c|c|c|c|c|c|c|c|c|c|c|c|c|c|c|c|c|c|c|c|c|c|c|}
		\hline
		i&0&1&\cellcolor{gray!20}2&3&4&\cellcolor{gray!20}5&6&7&\cellcolor{gray!20}8&9&10&\cellcolor{gray!20}11&12&13&\cellcolor{gray!20}14&15&16&\cellcolor{gray!20}17&18&19&\cellcolor{gray!20}20&21&22&\cellcolor{gray!20}23&24&25&\cellcolor{gray!20}26 \\ \hline
		i_0i_1i_2 &  \mbox{\footnotesize 000} & \mbox{\footnotesize 001} & \mbox{\footnotesize 002} & \mbox{\footnotesize 010} & \mbox{\footnotesize 011} & \mbox{\footnotesize 012} & \mbox{\footnotesize 020} & \mbox{\footnotesize 021} & \mbox{\footnotesize 022}  &
		\mbox{\footnotesize 100} & \mbox{\footnotesize 101} & \mbox{\footnotesize 102} & \mbox{\footnotesize 110} & \mbox{\footnotesize 111} & \mbox{\footnotesize 112} & \mbox{\footnotesize 120} & \mbox{\footnotesize 121} & \mbox{\footnotesize 122}   & 
		\mbox{\footnotesize 200} & \mbox{\footnotesize 201} & \mbox{\footnotesize 202} & \mbox{\footnotesize 210} & \mbox{\footnotesize 211} & \mbox{\footnotesize 212} & \mbox{\footnotesize 220} & \mbox{\footnotesize 221} & \mbox{\footnotesize 222}   
		\\ \hline 
		\textbf{f} & 1&     2&     0&     0&     0&     0&     0&     2&     1&     1&     0&     2&     0&     1&     2&     0&     0&     0&     1&     1&  1&     0&     2&     1&     0&     1&     2
		\\ \hline 
		\psi(f) & \xi_3^1&     \xi_3^2&     \xi_3^0&     \xi_3^0&     \xi_3^0&     \xi_3^0&     \xi_3^0&     \xi_3^2&     \xi_3^1&     \xi_3^1&     \xi_3^0&     \xi_3^2&     \xi_3^0&   \xi_3^1&     \xi_3^2&     \xi_3^0&     \xi_3^0&     \xi_3^0&     \xi_3^1&     \xi_3^1&     \xi_3^1&     \xi_3^0&     \xi_3^2  &   \xi_3^1&     \xi_3^0&     \xi_3^1&     \xi_3^2
		\\ \hline 
		\hline 
		\psi(f\arrowvert_{\mathbf{x}_J=(0,2)}) & & & \xi_3^0 & & & \xi_3^0 & & & \xi_3^1& & & & & & & & & & & & & & & & & & 
		\\ \hline 
		\psi(f\arrowvert_{\mathbf{x}_J=(1,2)}) & & & & & & & & & & & & \xi_3^2 & & & \xi_3^2 & & & \xi_3^0 & & & & & & & & & 
		\\ \hline 
		\psi(f\arrowvert_{\mathbf{x}_J=(2,2)}) & & & & & & & & & & & & & & & & & & & & & \xi_3^1 & & & \xi_3^1 & & & \xi_3^2 
		\\ \hline 		
		\end{array}
	}
		\end{equation}
		\end{table*}
For $\mathbf{c}_1=(0,2)$, $\mathbf{c}_2=(1,2)$, and $\mathbf{c}=(0,2)$, from (\ref{cross_corr_restrict_c1_c2}) and (\ref{corr_restric_c}), $$\mathbf{B}_\tau(\mathbf{c}_1,\mathbf{c}_2)=\emptyset,~\forall~\tau\neq  3,6,9,12,15,$$ \text{and} $$\mathbf{A}_\tau(\mathbf{c})=\emptyset,~ \forall~\tau\neq 0,3,6.$$
Therefore, $$\Theta(\psi(f\arrowvert_{\mathbf{x}_J=\mathbf{c}_1}),\psi(f\arrowvert_{\mathbf{x}_J=\mathbf{c}_2}))(\tau)=0,\forall\tau\neq  3,6,9,12,15,$$ and  $$\Theta(\psi(f\arrowvert_{\mathbf{x}_J=\mathbf{c}}))(\tau)=0,~\forall~\tau\neq 0,3,6.$$ 
When $\tau \in \{3,6,9,12,15\}$, for example $\tau = 3$, from the row $\psi(f\arrowvert_{\mathbf{x}_J=(0,2)})$ in Table \ref{3rdanki} we see $\gamma_J = \mathbf{c}_1=(0,2)$ can hold only for $\gamma \in \{2, 5, 8\}$; furthermore, 
since $\tau =3$ we see that  $\delta = \gamma + \tau$ with $\delta_{J}=\mathbf{c}_2=(1,2)$ can only hold for $\delta = 11$, as indicated by the  row $\psi(f\arrowvert_{\mathbf{x}_J=(1,2)})$.
To summarize, we list {$\mathbf{A}_\tau(\mathbf{c})$ for $\tau=0,3,6$}, and {$\mathbf{B}_\tau(\mathbf{c}_1,\mathbf{c}_2)$ for $\tau=  3,6,9,12,15$ in Table \ref{3rdanki2}, where $\mathbf{A}_{\tau}, \mathbf{B}_\tau$ are used for simplicity.
\begin{table}[!b]
	\caption{$\mathbf{A}_\tau(\mathbf{c})$ for $\mathbf{c}=(0,2)$, and $\mathbf{B}_\tau(\mathbf{c}_1,\mathbf{c}_2)$ for $\mathbf{c}_2=(1,2)$ \ccn }\label{3rdanki2}
\begin{center} 
\resizebox{\columnwidth}{!}{	\begin{tabular}{|l|l|l|l|}
		\hline
		$\mathbf{B}_3$    & $\{(8,11)\}$               & \multirow{2}{*}{$\mathbf{A}_0$} & \multirow{2}{*}{$\{(2,2),(5,5),(8,8)\}$} \\ \cline{1-2}
		$\mathbf{B}_6$    & $\{(5,11),(8,14)\}$        &                                  &                                          \\ \hline
		$\mathbf{B}_9$    & $\{(2,11),(5,14),(8,17)\}$ & \multirow{2}{*}{$\mathbf{A}_3$} & \multirow{2}{*}{$\{(2,5),(5,8)\}$}       \\ \cline{1-2}
		$\mathbf{B}_{12}$ & $\{(2,14),(5,17)\}$        &                                  &                                          \\ \hline
		$\mathbf{B}_{15}$ & $\{(2,17)\}$               & $\mathbf{A}_6$                  & $\{(2,8)\}$                              \\ \hline
	\end{tabular}}
\end{center}
\end{table}
}
From the Table \ref{3rdanki2}, we  can express $\Theta(\psi(f\arrowvert_{\mathbf{x}_J=\mathbf{c}_1}),\psi(f\arrowvert_{\mathbf{x}_J=\mathbf{c}_2}))(\tau)$ and $\Theta(\psi(f\arrowvert_{\mathbf{x}_J=\mathbf{c}}))(\tau)$ for $\tau=3$ as follows:
$\Theta(\psi(f\arrowvert_{\mathbf{x}_J=\mathbf{c}_1}),\psi(f\arrowvert_{\mathbf{x}_J=\mathbf{c}_2}))(3)=\sum_{(\gamma,\delta)\in \mathbf{B}_3(\mathbf{c}_{1},\mathbf{c}_2) }\xi_3^{f_\gamma-f_{\delta}}=\xi_3^{f_8-f_{11}}=\xi_3^{1-2}=\xi_3^{-1}$, and 
$\Theta(\psi(f\arrowvert_{\mathbf{x}_J=\mathbf{c}}))(3)=\sum_{(\gamma,\delta)\in \mathbf{A}_3(\mathbf{c})}\xi_3^{f_\gamma-f_{\delta}}=\xi_3^{f_2-f_5}+\xi_3^{f_5-f_8}=\xi_3^{0-0}+\xi_3^{0-1}=1+\xi_3^{-1}$.
For other values of $\tau$, we can calculate the ACCFs similarly by following (\ref{ccrf_rest1}), (\ref{vect10987}) and the Table \ref{3rdanki2}. 
\end{example}
We can observe that there are $p^{m-n}$ nonzero components in $\psi(f\arrowvert_{\mathbf{x}_J=\mathbf{c}})$ for a choice of $\mathbf{c}$ in $\Z_p^n$. From (\ref{vect}) (and as illustrated in the Table \ref{3rdanki}), it is clear that the nonzero positions in $\psi(f\arrowvert_{\mathbf{x}_J=\mathbf{c}_1})$ and
$\psi(f\arrowvert_{\mathbf{x}_J=\mathbf{c}_2})$ for two distinct $\mathbf{c}_1$ and $\mathbf{c}_2$ in $\Z_p^n$ are always distinct. Therefore,
$\psi(f)$ can be expressed as 
\begin{equation}\label{Eq:seqrelation}
\psi(f)=\sum_{\mathbf{c}\in\Z_p^n} \psi(f\arrowvert_{\mathbf{x}_J=\mathbf{c}}).  
\end{equation}
With this relaltion, one can express the ACCF between two sequences $\psi(f)$ and $\psi(g)$ in terms of their correspoinding restricted sequences.
\begin{lemma}
	Let $f$ and $g$ be two $q$-ary functions in $m$ variables. The ACCF between $\psi(f)$ and $\psi(g)$ can be expressed as 
	\begin{equation}\nonumber
	\Theta (\psi(f),\psi(g))(\tau)=\!\!\!\displaystyle\sum_{\mathbf{c}_1,\mathbf{c}_2\in\Z_p^n}\!\!\!\Theta (\psi(f\arrowvert_{\mathbf{x}_J=\mathbf{c}_1}),\psi(g\arrowvert_{\mathbf{x}_J=\mathbf{c}_2}))(\tau).
	\end{equation}
	\begin{IEEEproof}
		Following the relation in \eqref{Eq:seqrelation}, we have 
		\begin{equation}\nonumber
		\begin{split}
		&\Theta (\psi(f),\psi(g))(\tau)\\=&~
		\Theta \left(\sum_{\mathbf{c}_1\in\Z_p^n}\psi( f\arrowvert_{\mathbf{x}_J=\mathbf{c}_1}),\sum_{\mathbf{c}_2\in\Z_p^n}\psi( g\arrowvert_{\mathbf{x}_J=\mathbf{c}_2})\right)(\tau)\\
		=&\sum_{\mathbf{c}_1\in\Z_p^n}\Theta\left(\psi( f\arrowvert_{\mathbf{x}_J=\mathbf{c}_1}),\sum_{\mathbf{c}_2\in\Z_p^n}\psi( g\arrowvert_{\mathbf{x}_J=\mathbf{c}_2})\right)(\tau)\\
		=&\sum_{\mathbf{c}_1, \mathbf{c}_2\in\Z_p^n}\Theta\left(\psi( f\arrowvert_{\mathbf{x}_J=\mathbf{c}_1}),\psi( g\arrowvert_{\mathbf{x}_J=\mathbf{c}_2})\right)(\tau).
		\end{split}
		\end{equation}
	\end{IEEEproof}
\end{lemma}
\subsection{Quadratic Functions and Graphs}
A quadratic $q$-ary function from $\Z_p^m$ to $\mathbb{Z}_q$ can be expressed as
\begin{equation}\label{somvf}\nonumber
f(x_0,x_1,\dots,x_{m-1})=\sum_{0\leq i,j<m}q_{i,j}x_ix_j+\sum_{0\leq j<m}c_j x_j+c,
\end{equation}
where 
$q_{i,j},c_j,c\in \mathbb{Z}_q$. For a quadratic 
$q$-ary function $f$, we define its graph $G(f)$ as a graph, in which 
there are $m$ vertices labeled as $x_i$, where there is an edge between vertices $x_i$ and $x_j$ if $q_{i,j}\neq 0$. 
%{Note that 
%non-cycle edges in $G(f)$ for a quadratic function $f$ depend only on those quadratic terms for which $i\neq j$ and $q_{i,j}\neq 0$.} 
A Hamiltonian path  in a graph is the path that visits each vertex exactly once. A graph contains only one vertex with no edges is also known as a Hamiltonain path \cite{pater2000}. For instance, Figure 1 represents the graph of $f(x_0,x_1,x_2)=x_0x_2+2x_2x_1+2x_1^2+x_2+1$ in Example \ref{Ex1}.
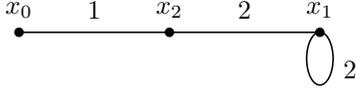
\begin{figure}
	\centering
	\begin{tikzpicture}[line width=.6pt]
	\filldraw (0, 0) circle [radius=1.5pt]
	(2, 0) circle [radius=1.5pt]
	(4, 0) circle [radius=1.5pt];
	\draw    (0, 0.3) node {$x_0$} 
	(1, 0.3) node {$1$} 
	(2, 0.3) node {$x_2$}
	(3, 0.3) node {$2$}
	(4, 0.3) node {$x_1$}; 
	\draw    (4, -0.35) ellipse [x radius=5 pt, y radius=10pt]
	(4.4, -0.5) node {$2$};  
	\draw(0,0) -- (2,0) -- (4,0);
	\end{tikzpicture}
	\caption{Graph of the function $x_0x_2+2x_2x_1+2x_1^2+x_2+1$}
	\label{fig_somvf}
\end{figure}

\section{Tighter Lower Bounds on the Maximum Correlation Magnitude of QCSSs}
In this section, we will further investigate the lower bound on the maximum correlation magnitude for  $(K,M,L,\theta)$-QCSSs that are composed of multiple CCCs. 

For the weight vector $\mathbf{w}$ in \eqref{fcd}, we define a quadratic form 
	\begin{equation}\label{quadratic_form_Q}
	Q(\mathbf{w},a)=a\sum_{u=0}^{2L-2} w_u^2+\sum_{u,v=0}^{2L-2}\tau_{u,v,L}w_uw_v,
	\end{equation}
	where $a$ is a real number. Below we present the first main theorem of this paper.
	\begin{theorem}\label{thm1}
		Let $N\geq 2$ and $\mathcal{C}$ be a collection of $N$ different $(M,L)$-CCCs.
		Then the maximum correlation magnitude $\theta$ of $\mathcal{C}$ satisfies
		\begin{equation}\nonumber
		\theta^2\geq \frac{M\left(L-Q\left(\mathbf{w},\frac{ML^2}{K}\right)\right)}{1-\frac{M}{K}},
		\end{equation}
		where $K=NM$.
	\end{theorem}
\begin{IEEEproof} Assume $\mathcal{C}=\cup_{i=1}^N \mathcal{C}_i$, where $\mathcal{C}_i$ is an $(M,L)$-CCC. Substituting $\mathcal{D}=\mathcal{C}$ in (\ref{fcd}), we have
	\begin{equation}\label{fccderiv}
	\begin{split}
	&|\mathcal{C}|^2	F(\mathcal{C},\mathcal{C})\\=&\!\!\sum_{\mathbf{X}\in\mathcal{C}}\!\sum_{\mathbf{Y}\in\mathcal{C}}\!\sum_{u=0}^{2L-2}\!\sum_{v=0}^{2L-2}\!\big|\big<T^u(\mathbf{X},\mathbf{0}_{L-1}),T^v(\mathbf{Y},\mathbf{0}_{L-1})\big>\big|^2w_u w_v\\
	=&\sum_{i=1}^{N}\!\sum_{\mathbf{X},\mathbf{Y}\in\mathcal{C}_i}\!\sum_{u,v=0}^{2L-2}\!\big|\big<T^u(\mathbf{X},\mathbf{0}_{L-1}),T^v(\mathbf{Y},\mathbf{0}_{L-1})\big>\big|^2w_u w_v\\
	+\!\!\!\!&\sum_{\substack{i,j=1\\i\neq j}}^{N}\!\sum_{\mathbf{X}\in\mathcal{C}_i}\!\sum_{\mathbf{Y}\in\mathcal{C}_j}\!\sum_{u,v=0}^{2L-2}\!\big|\big<T^u(\mathbf{X},\mathbf{0}_{L-1}),T^v(\mathbf{Y},\mathbf{0}_{L-1})\big>\big|^2\!\!w_u w_v\\
	=&S_1+S_2,	
	\end{split}
	\end{equation}
	where \begin{equation}\nonumber
	\begin{split}
	S_1=\sum_{i=1}^{N}\!\sum_{\mathbf{X},\mathbf{Y}\in\mathcal{C}_i}\!\sum_{u,v=0}^{2L-2}&\big|\big<T^u(\mathbf{X},\mathbf{0}_{L-1}),\\&~~~~~~T^v(\mathbf{Y},\mathbf{0}_{L-1})\big>\big|^2w_u w_v,
	\end{split}
	\end{equation} and
	
	\begin{equation}\nonumber
	\begin{split}
	S_2=\sum_{\substack{i,j=1\\i\neq j}}^{N}\sum_{\mathbf{X}\in\mathcal{C}_i}\sum_{\mathbf{Y}\in\mathcal{C}_j}\sum_{u,v=0}^{2L-2}&\big|\big<T^u(\mathbf{X},\mathbf{0}_{L-1}),\\&T^v(\mathbf{Y},\mathbf{0}_{L-1})\big>\big|^2w_u w_v.
	\end{split}
	\end{equation}
	As $\mathcal{C}_i$ is an $(M,L)$-CCC, for $\mathbf{X},\mathbf{Y}\in \mathcal{C}_i$ and $v\leq u$\ccn, 
	\begin{equation}\label{tuv1}
	\begin{split}
	&\big<T^u(\mathbf{X},\mathbf{0}_{L-1}),T^v(\mathbf{Y},\mathbf{0}_{L-1})\big>\\=&\Theta(\mathbf{X},\mathbf{Y})(\tau_{u,v,L})\\=&\begin{cases}
	ML,& \mathbf{X}=\mathbf{Y}, \tau_{u,v,L}=0,\\
	0,&\mathbf{X}=\mathbf{Y}, 1\leq \tau_{u,v,L}<L,\\
	0,&\mathbf{X}\neq\mathbf{Y}, 0\leq \tau_{u,v,L}<L,
	\end{cases}
	\end{split}
	\end{equation}
	and for $v>u$ \ccn,
	\begin{equation}\label{tuv2}
	\begin{split}
	&\big<T^u(\mathbf{X},\mathbf{0}_{L-1}),T^v(\mathbf{Y},\mathbf{0}_{L-1})\big>\\=&\Theta(\mathbf{X},\mathbf{Y})(-\tau_{u,v,L})\\=&\begin{cases}
	0,&\mathbf{X}=\mathbf{Y}, 1\leq \tau_{u,v,L}<L,\\
	0,&\mathbf{X}\neq\mathbf{Y}, 0< \tau_{u,v,L}<L.
	\end{cases}
	\end{split}
	\end{equation}
	Using (\ref{tuv1}) and (\ref{tuv2}) in $S_1$, we have
	\begin{equation}\label{sderiv1}
	\begin{split}
	S_1&=\sum_{i=1}^{N}\sum_{\mathbf{X},\mathbf{Y}\in\mathcal{C}_i}\sum_{u,v=0}^{2L-2}\big|\big<T^u(\mathbf{X},\mathbf{0}_{L-1}),\\&
	~~~~~~~~~~~~~~~~~~~~~~~T^v(\mathbf{Y},\mathbf{0}_{L-1})\big>\big|^2w_u w_v\\
	&=\sum_{i=1}^{N}\sum_{\substack{\mathbf{X},\mathbf{Y}\in\mathcal{C}_i\\\mathbf{X}=\mathbf{Y}}}\sum_{\substack{u,v=0\\u=v}}^{2L-2}\big|\big<T^u(\mathbf{X},\mathbf{0}_{L-1}),\\&~~~~~~~~~~~~~~~~~~~~~~~T^v(\mathbf{Y},\mathbf{0}_{L-1})\big>\big|^2w_u w_v\\
	&+\sum_{i=1}^{N}\sum_{\substack{\mathbf{X},\mathbf{Y}\in\mathcal{C}_i\\\mathbf{X}=\mathbf{Y}}}\sum_{\substack{u,v=0\\u\neq v}}^{2L-2}\big|\big<T^u(\mathbf{X},\mathbf{0}_{L-1}),\\&~~~~~~~~~~~~~~~~~~~~~~~T^v(\mathbf{Y},\mathbf{0}_{L-1})\big>\big|^2w_u w_v\\
	&+\sum_{i=1}^{N}\sum_{\substack{\mathbf{X},\mathbf{Y}\in\mathcal{C}_i\\\mathbf{X}\neq\mathbf{Y}}}\sum_{\substack{u,v=0}}^{2L-2}\big|\big<T^u(\mathbf{X},\mathbf{0}_{L-1}),\\&~~~~~~~~~~~~~~~~~~~~~~~T^v(\mathbf{Y},\mathbf{0}_{L-1})\big>\big|^2w_u w_v\\
	&=\sum_{i=1}^{N}\sum_{\mathbf{X}\in\mathcal{C}_i}\sum_{u=0}^{2L-2}\Theta^2(\mathbf{X})(0) w_u^2+0+0
	\\
	&
	=K M^2 L^2\sum_{u=0}^{2L-2} w_u^2,
	\end{split}
	\end{equation}
	where $K=MN$.
	Now,
	\begin{equation}\label{sderiv2}
	\begin{split}
	S_2&=\sum_{\substack{i,j=1\\i\neq j}}^{N}\sum_{\mathbf{X}\in\mathcal{C}_i}\sum_{\mathbf{Y}\in\mathcal{C}_j}\sum_{u,v=0}^{2L-2}\big|\big<T^u(\mathbf{X},\mathbf{0}_{L-1}),\\&~~~~~~~~~~~~~~~~~~~T^v(\mathbf{Y},\mathbf{0}_{L-1})\big>\big|^2w_u w_v\\
	&\leq \theta^2\sum_{\substack{i,j=1\\i\neq j}}^{N}\sum_{\mathbf{X}\in\mathcal{C}_i}\sum_{\mathbf{Y}\in\mathcal{C}_j}\sum_{u,v=0}^{2L-2}w_u w_v\\
	&=\theta^2K(K-M).
	\end{split}
	\end{equation}
	Combining (\ref{fccderiv}), (\ref{sderiv1}), and (\ref{sderiv2}) gives
	\begin{equation}\label{fccleq}
	F(\mathcal{C},\mathcal{C})=\frac{S_1+S_2}{K^2}\leq \frac{M^2 L^2 }{K}\sum_{u=0}^{2L-2} w_u^2+\theta^2\left(1-\frac{M}{K}\right).
	\end{equation}
	From Lemma \ref{lma1} and (\ref{fccleq}) it follows that
	\begin{equation}\nonumber
	\begin{split}
	&\frac{M^2 L^2 }{K}\sum_{u=0}^{2L-2} w_u^2+\theta^2\left(1-\frac{M}{K}\right)\\\geq& \sum_{u,v=0}^{2L-2}M(L-\tau_{u,v,L})w_u w_v\\=&ML-M\sum_{u,v=0}^{2L-2}\tau_{u,v,L}w_uw_v.
	\end{split}
	\end{equation}
	Therefore, we have
	\begin{equation}\label{lbcccderiv}
	\theta^2\geq \frac{ML-\frac{M^2L^2}{K}\sum_{u=0}^{2L-2} w_u^2-M\sum_{u,v=0}^{2L-2}\tau_{u,v,L}w_uw_v}{1-\frac{M}{K}}.
	\end{equation}
	The desired conclusion directly follows from the definition of $Q(\textbf{w}, a)$ in \eqref{quadratic_form_Q}.
\end{IEEEproof}	
\smallskip 
Theorem \ref{thm1} shows that the maximum correlation magnitude of $\mathcal{C}=\cup_{i=1}^N \mathcal{C}_i$ heavily depends on the weight vector $\textbf{w}$.
In order to obtain tighter correlation lower bound for $\theta$, our task now is to choose suitable weight vectors in (\ref{lbcccderiv}). We 
start with the weight vector $\textbf{w}$ from step functions.
\begin{corollary}\label{crol1}
	Suppose the weight vector $\mathbf{w}=(w_0, w_1, \dots, w_{2L-2})$ is given by 
	\begin{equation}\label{wv1}
	\begin{split}
	w_j=\begin{cases}
	\frac{1}{t},& j=0,1,\hdots,t-1,\\
	0,& j=t,t+1,\hdots,2L-2,
	\end{cases}
	\end{split}
	\end{equation}
	where $0<t\leq 2L-1$. Assume that $L\geq N $.
%	In this paper, we call the above weight vector as step function weight vectior.
	Then the lower bounds for $\theta$ are given as follows:
	\begin{itemize}
		\item when $N=2, 3$,  
		% the maximum value of the lower bound is achieved at $t=2L-1$, namely,
		%the aperiodic correlation lower bound appears as
		\begin{equation}\label{EqBound-N2komp}
		\theta^2 \geq ML \,\frac{L}{2L-1},
		\end{equation}
%		\item when $N=3$, $L\leq 8$,
%		\begin{itemize}
		%the maximum value of the aperiodic correlation lower bound  %achieved at $t=L$, namely, 
		%appears as
%	\item for $L\leq 8$,	
%	\begin{equation}\label{EqBound-N3komp}
%		\theta^2 \geq ML\left(\frac{1}{2}+\frac{1}{2L^2}\right),
%  	\end{equation}
%		\item for $L\geq 9$,
%		\begin{equation}\label{snig11}
%		\theta^2 \geq \frac{ML^2}{2L-1}.
%		\end{equation}
%		\end{itemize}
		\item when $N\geq 4$,		
		%the maximum value of the lower bound is achieved at 
		%$t=\left\lceil \sqrt{\frac{3L^2}{N}-1}\right\rceil$ or 
		%$t=\left\lfloor \sqrt{\frac{3L^2}{N}-1}\right\rfloor$, which is approximately %given by
		\begin{equation}\label{Eqtt}
		\theta^2   \geq ML \left( 1 - \frac{2\sqrt{{3L^2N}-N^2}-3L }{3L(N-1)}\right).
		\end{equation}
	\end{itemize}
	\begin{IEEEproof} 
As the full proof is lengthy, it is placed in  Appendix \ref{Appendix:A}. Here
we only provide the sketch of the proof. 
With weight vector $\textbf{w}$ given by \eqref{wv1}, in the calculation of \eqref{lbcccderiv} we need to consider two cases: $0<t\leq L$ and $L+1\leq t \leq 2L-1$.

\textit{Case 1:} $0<t\leq L$.	In this case, 
	\begin{equation}\nonumber
\begin{array}{c}
\sum\limits_{u,v=0}^{t-1}\tau_{u,v,L}=\sum\limits_{u=0}^{t-1}\frac{u(u+1)}{2}+\sum\limits_{u=0}^{t-1}\frac{(t-u-1)(t-u)}{2}
=\frac{t(t^2-1)}{3}.
\end{array}
\end{equation}
	Substituting the above equality into (\ref{lbcccderiv}), we obtain 
\begin{equation}\label{skthbndm<L}
\theta^2\geq  \frac{ML}{1-\frac{M}{K}} \left(1 - \frac{1}{3L}\left(t + \frac{3ML^2-K}{K t}\right)\right).
\end{equation}
We then need to find the value of $t$ that gives the maximum lower bound. 
For different choices of $K/M=N$, we obtain the following results:
\begin{itemize}
	\item For $N=2,3$, the maximum value of the lower bounds in (\ref{skthbndm<L}) is attained at $t=L$,  implying 
		\begin{equation}\label{skthEq_bound_N=2}
	\theta^2 \geq  \frac{ML}{1-M/K}\left(\frac{2}{3}-\frac{M}{K}+\frac{1}{3L^2}\right).
	\end{equation} 
%	\item  When $N=2,$, $t=L$ gives the maximum value of the lower bounds in (\ref{skthbndm<L}) as follows
%	\begin{equation}\label{skthEqBound-N3}
%	\theta^2 \geq ML\left(\frac{1}{2}+\frac{3}{2L^2}\right).
%	\end{equation}
	\item For $N>3$, the maximum value of the lower bounds is achieved at $
	t=\left \lceil \sqrt{\frac{3L^2}{N}-1} \right \rfloor$ $\in [1, L]$, and we have the following simplified lower bound:
	\begin{equation}\label{skthbndmaxm}
	\theta^2\geq ML\left(1- \frac{2\sqrt{N(3L^2-N)}-3L}{3L(N-1)}\right).
	\end{equation}
\end{itemize} 

\textit{Case 2:} $L<t\leq 2L-1.$ In this case, we obtain
\begin{equation}\label{skthtau_L+1_2L-1}
\begin{split}
\sum\limits_{u,v=0}^{t-1}\tau_{u,v,L} &=(t+1)(t-L)(L-1)\\&+(3Lt^2-t^3-3L^2t+t+2L^3-2L)/3.
\end{split}
\end{equation}	
From (\ref{lbcccderiv}) and	(\ref{skthtau_L+1_2L-1}), we have
\begin{equation}\label{skthEq4}
\theta^2 \geq \frac{M}{3(1-M/K)}\left(
t + \frac{a}{t} - \frac{b}{t^2} - 3(L-1)
\right),
\end{equation}
where 
\begin{equation}\label{skthkey}
\begin{split}
a&= (6L^2-6L+2)-3ML^2/K,  \text{ and }\\ b&=L(L-1)(2L-1).
\end{split}
\end{equation}
%Now we have to find that value of $t$ for which we obtain the maximum value of the lower bounds in (\ref{skthEq4}). 
To find the maximum lower bound, we 
analyze the function $f(x)=t + \frac{a}{t} - \frac{b}{t^2}$ over the interval $[L+1, 2L-1]$, where we consider both the 1st-order and 2nd-order derivatives.
For the different choices of $N$, we have the following results:
\begin{itemize}
	\item when $N=2, 3$, the maximum lower bound is achieved at $t=2L-1$, implying
	\begin{equation}\nonumber
		\theta^2\geq \frac{ML^2}{2L-1},
	\end{equation}
%	where in the case of $N=3$, $L$ is required to satisfy $L \geq %3N$.
%	\item  when $N=3$, the maximum value of the lower bound is achieved at $t=2L-1$. The lower bound is given by
%	\begin{equation}
%	\theta^2\geq \frac{ML^2}{2L-1},
%	\end{equation}
%	where $L\geq 9$.
	\item when $N>3$, the maximum lower bound is achieved at $t=L+1$, implying
	\begin{equation}\label{skthEq_BoundN3}
	\begin{split}
	\theta^2&  \geq 
%	\dfrac{M}{(1-M/K)} \cdot \dfrac{(2 N - 3)L^3  +  3(N - 1) L^2 + NL  + 6 N}{3(L+1)^2N} 
%	\\&=& 
	ML \left(
	1 -\right.\\ &\left.\dfrac{(N+6)L^3 + 3(N-1)L^2 +(2N-3) L - 6N}{3L(L+1)^2(N-1)}
	\right).
	\end{split}
	\end{equation}
\end{itemize}
Comparing the bounds as derived in Case 1 and Case 2,  we reach the desired results.
%The detailed of the above derivations for the proof have been presented in %Appendix A.
	\end{IEEEproof}
\end{corollary}
In Remark \ref{rinksa1}, we compare the correlation lower bounds derived in Corollary \ref{crol1} with that of bound reported in \cite{crlbzl}.
\begin{remark}\label{rinksa1}
	%	[Comparison Between the Results in \textit{Corollary 1} and \cite{crlbzl}]
	From \cite{crlbzl}, the lower-bound on $\theta$ is given by
	\begin{equation}\label{zrprnt}
	\theta^2\geq ML\left(1-\frac{2}{\sqrt{3N}}\right),
	\end{equation}
	where $N=K/M\geq 3$.	
	For $N=3$, it can easily be verified that the derived lower bound in \textit{Corollary \ref{crol1}} is tighter than the bound in (\ref{zrprnt}).   
	%reported in \cite{crlbzl}, which can be represented as 
	%\begin{equation}\label{zrprnt}
	%	\theta^2\geq ML\left(1-\frac{2}{\sqrt{3N}}\right).
	%\end{equation}
	For $N>3$, we have
	\begin{equation}\label{llchu}
	\frac{\left( \frac{2\sqrt{{3L^2N}-N^2}-3L }{3L(N-1)}\right)}{\frac{2}{\sqrt{3N}}}\leq \frac{\sqrt{N}-\sqrt{3}}{\sqrt{N}-\frac{1}{\sqrt{N}}}<1.
	\end{equation}
	From (\ref{Eqtt}), (\ref{zrprnt}), and (\ref{llchu}), it is clear that our derived lower bound on $\theta$ in \textit{Corollary \ref{crol1}} is tighter than the lower bound in \cite{crlbzl}.   
\end{remark}
%There are some other weight vector can be found in 
\smallskip 

Although we already have a tighter lower bound on $\theta$ with respect to the weight vector in (\ref{wv1}), another weight vector,
termed \textit{positive-cycle-of-a-sine-wave} weight vector in \cite{zlbnd_lvstn}, may yield a  tighter lower bound. Below, we present another corollary to derive the proposed bound of \textit{Theorem 1} with respect to the positive-cycle-of-a-sine-wave weight vector. 
\begin{corollary}\label{crol2}
	The positive-cycle-of-a-sine-wave weight vector is given by 
	\begin{equation}\nonumber
	\begin{split}
	w_j=\begin{cases}
	\tan \frac{\pi}{2t} \sin \frac{\pi j}{t}, & j\in \{0,1,\hdots,t-1\},\\
	0, & j\in\{t,t+1, 2L-2  \},
	\end{cases}
	\end{split}
	\end{equation}
	where $1< t\leq L$. Assume $L\geq N$. Then the lower bounds for $\theta$ are given as follows:
	\begin{itemize}
		%	\item For $N=2,3$, $$	\theta^2\geq f(x_1),$$ where 
		%$x_1=(x_0+3L+3)/4$, $x_0=\frac{3b_2+\sqrt{9b_2^2-32b_1b_3}}{(-4b_1)}$,
		%$b_0=0.22$, $b_1=\frac{1.2L^2}{N}-0.62+3.3L-3.8L^2$, %$b_2=3.9L^3-8.7L^2+3.2L$,
		%$b_3=9.3L^3-4.6L^2$,  $a_0=4\pi^2-26\pi+60$, %$a_1=5\pi^2-\frac{10L^2\pi^2}{N}-10L\pi^2-4L^2\pi+32L^2\pi^2-52L\pi$, %$a_2=10L^2\pi^2+8L^2\pi^3-16L^3\pi^2-13L\pi^2$, $a_3=4L^2\pi^3-8L^3\pi^3$, %and 
		%$a_4=40-156L+26\pi-4\pi^2-20L\pi^2+78L\pi$.
		%$f(x_1)=\frac{MN}{80(N-1)}\left(a_0x_1+\frac{a_1}{x_1}+\frac{a_2}{x_1^2}+\frac{a_3}{x_1^3}+a_4\right)$, $x_1$ is an approximation of the root of first-order %derivative of $f$, and $a_0,a_1,\hdots,a_4$ are the functions of $L$ and $N$. 
		 \item when $2\leq L\leq 4$,
		 \[
		 \theta^2\geq \frac{ML}{N-1}\left( \frac{N}{4}\left( 3+ \tan^2\frac{\pi}{2L} \right)\!\! - \!\!\frac{L^2}{2}\tan^2\frac{\pi}{2L}   \right),
		 \]
		\item when $L\geq 5$ and $N=2, 3, 4$,
		% \text{ or } \left\lceil \frac{2\pi^2L^2}{4L^2+\pi^2} \right\rceil$,
		$$\theta^2\geq ML\left(1-\frac{L^2(2\pi^2+4N-16)-N\pi^2}{16L^2(N-1)}\right),$$
		\item when  $L\geq 5$ and $N\geq 5$, $$\theta^2\geq ML\left(1-\frac{\pi\sqrt{N(2L^2-N)}-4L}{4(N-1)L}\right).$$
	\end{itemize}
	\begin{IEEEproof}
	%\section{Proof of Corollary 2}
	We have the following results from \cite{zlbnd_lvstn}:
	\begin{equation}\label{cr2wu2}
	\sum_{u=0}^{2L-2}w_u^2=\frac{t}{2}\tan^2\frac{\pi}{2t}.	
	\end{equation}
	For $2\leq t\leq L$, 
	\begin{equation}\label{pcs_1}
	\sum_{u,v=0}^{t-1}\tau_{u,v,L}w_uw_v=\frac{t}{4}\left(1-\tan^2\frac{\pi}{2t}\right).
	\end{equation}
	From (\ref{lbcccderiv}), (\ref{cr2wu2}), and (\ref{pcs_1}), we have
	\begin{equation}\label{neq_sw_theta}
	\theta^2\geq \frac{ML\left(1-\frac{t}{4L}\left(1+\frac{2L^2-N}{N}\tan^2\frac{\pi}{2t}\right)\right)}{1-\frac{1}{N}}.
	\end{equation}
	Assume $\varphi(x)=1-\frac{x}{4L}\left(1+\frac{2L^2-N}{N}\tan^2\frac{\pi}{2x}\right).$
	We shall find the maximum value of $\varphi(x)$ in $[2:L]$. 
 Note that for $x \geq 5$, the function $\tan^2\frac{\pi}{2x}$ can be approximated as $(\frac{\pi}{2x})^2$ since their difference is roughly $0.0987$ at $x=5$ and becomes smaller for as $x$ increases.
In order to find the maximum value of $\varphi(x)$ in $[2:L]$, we 
divide the derivation in the following two cases: $L\leq 4$ and $L\geq 5$.
	\begin{case}[$L\leq 4$]
In this case, we need to determine the maximum value of $\varphi(x)$ within the interval $[2:L]$. This can be easily determined since $L\leq 4$ and $x$ can only take on $L-1$ values $2$, $3$, and $L$. Table \ref{amrbunkita2} lists the values $\varphi(x)$ for $x=2, 3, 4$, where $N\leq L$.
\begin{table}[!t]
		\caption{Value of $\varphi(x)$ at $x=2,3,$ and $4$}\label{amrbunkita2}	
		\begin{tabular}{|c|c|c|c|}
			\hline
			$x$    & $2$             & $3$                           & $4$                                       \\ \hline
			$\varphi(x)$ & $1-\frac{L}{N}$ & $1-\frac{L}{2N}-\frac{1}{2L}$ & $1-\frac{429L}{1250N}-\frac{2071}{2500L}$ \\ \hline
		\end{tabular}
	\end{table}			
	 It 
	 can be easily verified that $\varphi(2)\leq \varphi(3)\leq \varphi(4)$, indicating that
	 the function $\varphi(x)$ at $[2:L]$ achieves it's maximum value at $x=L$. Therefore, we can express the maximum as follows:
\[\max_{x\in [2,L]} \varphi(x)=\varphi(L)=\frac{1}{4}\left( 3+ \tan^2\frac{\pi}{2L} \right) - \frac{L^2}{2N}\tan^2\frac{\pi}{2L} \]	
\end{case}	
	\begin{case}[$L\geq 5$]
		In this case, we consider the integer interval $[2: L] = [2:4] \cup [5: L]$, indicating
		\begin{equation}
		\max_{x\in [2:L]} \varphi(x)=\max\left\{\max_{x\in [2:4]} \varphi(x), \max_{x\in [5:L]} \varphi(x) \right\}. 
		\end{equation}
For the integer interval $[2:4]$, since $\varphi(4)\geq \varphi(3)\geq \varphi(2)$, we have
	\[\max_{x\in [2:4]} \varphi(x)=\varphi(4)=1-\frac{429L}{1250N}-\frac{2071}{2500L}.\]
		Next we find maximum of $\varphi(x)$ over the integer interval $[5:L]$. To this end, we will consider it over the interval $[5, L]$ instead. As discussed in the beginning, we have $\tan^2\frac{\pi}{2x}\approx\frac{\pi^2}{4x^2}$ for $x \in [5, L]$, thereby we  approximate $\varphi(x)$ as $$\varphi(x)=1-\frac{1}{4L}\left(x+\frac{\pi^2(2L^2-N)}{4xN}\right).$$
		The derivative function $\varphi'(x)$ has two zeros $x_0=\frac{\pi}{2}\sqrt{\frac{2L^2-N}{N}}$, and $-x_0$. 
		Note that $x_0$ lies in $[5, L)$ if $L \geq N \geq 5$ and $x_0\geq L$ otherwise.
		When $L\geq N \geq 5$, $\varphi'(x)>0$ for $x\in [5,x_0)$ and $\varphi'(x)<0$ for $x\in [x_0,L)$, this implies
		that $\varphi(x)$ is monotonically increasing over $[5,x_0]$ and monotonically decreasing over $[x_0, L) $.
		Hence
		$f$ attains maximum value at $x=x_0$, when $N\geq 5$.  
		When $N\leq  4$, the function $f$ is is monotonically increasing over $[5, L]$ and therefore attains maximum value at $x=L$. 
      Furthermore, 
		we can easily verify that $\varphi(L)\geq \varphi(4)$ when $N\leq 4$ and $\varphi(x_0)\geq \varphi(4)$ when $N\geq 5$.
		 That is to say, in Case 2 the function $\varphi(x)$ achieves its maximum value at $x=L$ when $N\leq 4$ and at $x=x_0$ when $N\geq 5$.
	\end{case}	
	Combining the two cases,  we have the following simplified lower bounds for  $2\leq t\leq L$:
	\begin{itemize}
		\item when $2\leq L\leq 4$, the lower bound in (\ref{neq_sw_theta}) is given by
		\begin{equation}\label{ankitabu}
		\theta^2\geq \frac{ML}{N-1}\left( \frac{N}{4}\left( 3+ \tan^2\frac{\pi}{2L} \right) - \frac{L^2}{2}\tan^2\frac{\pi}{2L}   \right),
		\end{equation}
		\item when $L\geq 5$ and $N\leq  4$, the maximum lower bound in (\ref{neq_sw_theta}) is approximately given by
		% \text{ or } \left\lceil \frac{2\pi^2L^2}{4L^2+\pi^2} \right\rceil$,
		\begin{equation}\label{komp11}\theta^2\geq ML\left(1-\frac{L^2(2\pi^2+4N-16)-N\pi^2}{16L^2(N-1)}\right),\end{equation}
		\item when $L\geq N\geq 5$, by properly choosing $t$ around $x_0$,  we obtain the maximum lower bound in (\ref{neq_sw_theta}) and it is approximately given by 
		%\text{ or } \left\lceil \frac{2\pi^2L^2}{4L^2+\pi^2} \right\rceil$,
		\begin{equation}\label{komp10}
		\theta^2\geq ML\left(1-\frac{\pi\sqrt{N(2L^2-N)}-4L}{4(N-1)L}\right).
		\end{equation}
	\end{itemize}
	
\ccn
\end{IEEEproof}

\end{corollary}
%%%%%%%%%%%%%%%%%%%%%%%%%%%%%%%%%%%%%%%
%%%%%%%%%%%%%%%%%%%%%%%%%%%%%%%%%%%%%%%
\begin{remark}\label{remoo1}
This remark compares the lower bounds derived in \textit{Corollary \ref{crol1}} and \textit{Corollary \ref{crol2}}. 
We start with the case of $L\leq 4$. In this case we have
\begin{itemize}
	\item for $N=2, 3$ and $N\leq L$, the lower bound in (\ref{EqBound-N2komp}) is tighter than  the lower bound in (\ref{ankitabu}),
	\item for $N=L=4$, the bound in (\ref{ankitabu}) is tighter than that of the lower bound in (\ref{Eqtt}).
\end{itemize}
Now we compare the bounds for $L\geq 5$. According to the bounds in \eqref{EqBound-N2komp} and \eqref{komp11}, 
it suffices to consider the sign of
\begin{equation}\nonumber\label{kamaka}
\begin{split}
	\frac{L-1}{2L-1}-\frac{L^2(2\pi^2+4N-16)-N\pi^2}{16L^2(N-1)}.
\end{split}
\end{equation}
A routine calculation indicates that
\begin{itemize}
\item for $N=2$, the lower bound in \eqref{EqBound-N2komp}  is tighter than that in \eqref{komp11},
\item for $N=3$, the lower bound in \eqref{EqBound-N2komp} is tighter than that in \eqref{komp11} for $5\leq L \leq 25$, and 
	for $L>25$, the lower bound in \eqref{komp11} is tighter.
\item for $N=4$, the the lower bound in (\ref{komp11}) is tighter than that in (\ref{Eqtt}) for $L\geq 5$,
\item for $L\geq N\geq 5$, the tighter bound is given by \eqref{komp10}, which is, according to \eqref{Eqtt} and \eqref{komp10}, determined by
\begin{equation}\nonumber
\begin{split}
&\frac{2\sqrt{{3L^2N}-N^2}-3L }{3L(N-1)} 
\cdot \frac{4(N-1)L}{\pi\sqrt{N(2L^2-N)}-4L}\\ =& \frac{\sqrt{{192L^2N}-64N^2}-12L }{\sqrt{18\pi^2 L^2N-9\pi^2N^2}-12L}>1.
\end{split}
\end{equation}

\end{itemize}
\end{remark}
\medskip  
In Appendix \ref{Appendix:ank}, we discuss correlation lower bound for the positive-cycle-of-a-sine-wave when $L+1\leq t\leq 2L-1$, for which the analysis is lengthy. We also compare the result with Remark \ref{remoo1} based on the asymptotic behaviour of the lower bounds. 
%\end{remark}
\medskip 

Finally we summarize the newly derived tighter lower bounds below. 
\begin{remark}\label{newrem}
For a $(K, M, L, \theta)$-QCSS as a collection of $N\geq 2$ different $(M,L)$ CCCs, where the sequence length $L\geq N$, 
the lower bounds on the maximum aperiodic correlation magnitude $\theta$ are improved as follows:
\begin{equation*}
\theta^2 \geq 
\begin{cases}
ML(1-\frac{L-1}{2L-1}), & 
N = 2 \text{ or } N = 3, \\&N \leq L\leq 25, \\
		{ML\left(1-\frac{L^2(2\pi^2+4N-16)-N\pi^2}{16L^2(N-1)}\right)},& N =3,L>25 \\&\text{ or }\!  N\! =\!4,L\!\geq \! 5,\\	
		ML(1-\frac{L-1.2}{2L-1}), & N=L=4, \\
		{ML\left(1-\frac{\pi\sqrt{N(2L^2-N)}-4L}{4(N-1)L}\right)},& L\geq N \geq 5,
\end{cases}
\end{equation*} where the bound for $N=L=4$ is transformed for a more direct comparison. Furthermore, for sufficiently large $L$,
the lower bounds may be roughly given as follows:
\begin{equation*}\label{key}
\theta^2 \geq  \begin{cases}
ML/2,&  N=2,\\
ML\big(1-\frac{1}{4(N-1)}\big),&  N=3, 4,\\
ML\big(1-\frac{\pi\sqrt{2N}-4}{4(N-1)}\big),&  N\geq 5.
\end{cases}
\end{equation*}
\end{remark}
\section{Construction of Asymptotically Optimal QCSSs Comprised of CCCs}
In this section, we shall first present a construction of CCCs using $q$-ary functions, and then we will show that the collection of those CCCs forms asymptotically optimal QCSSs with respect to 
the correlation lower bounds derived in the previous section.
\smallskip

	We first introduce the $q$-ary functions which will be used in our construction.
    For a subset $J=\{j_0, \dots, j_{n-1}\}\subset \Z_m$ with $n \leq m-1$, 
	consider an $m$-variable $q$-ary function $f:\Z_p^m\rightarrow \mathbb{Z}_q$  such that for each 
	$\mathbf{c}\in\Z_p^n$, the graph $G(f\arrowvert_{\mathbf{x}_J=\mathbf{c}})$ is a Hamiltonian path with edges having identical weight $q/p$. For the case of $m=n+1$, $f\arrowvert_{\mathbf{x}_J=\mathbf{c}}$ is a linear function of one variable, which forms a simplest Hamiltonian path. For $m>n+1$, the function $f\arrowvert_{\mathbf{x}_J=\mathbf{c}}$ can be algebraically expressed as 
	\begin{equation}\label{Eq_f}
	f\arrowvert_{\mathbf{x}_J=\mathbf{c}}=
	\frac{q}{p}\sum_{\alpha=0}^{m-n-2}x_{l_{\pi(\alpha)}}x_{l_{\pi(\alpha+1)}} +\sum_{\alpha=0}^{m-n-1}c_{l_\alpha}x_{l_\alpha}+c,
	\end{equation}
	where $\{l_0,\hdots,l_{m-n-1}\}=\Z_m\setminus \{j_0, \dots, j_{n-1}\}$, $\pi$ is a permutation on $\Z_{m-n}$, and $c_{l_0}$, $c_{l_1}$, $ \hdots $, $c_{l_{m-n-1}}$, and $c$ $\in \mathbb{Z}_q$.\ccn
	
	Let $k$ be an integer such that $1\leq k < p$. For an integer $t$ with $0\leq t <p^{n+1}$, denote its vector representation w.r.t base-$p$ 
	as $(\mathbf{t}, t_n) \in \Z_p^{n+1}$. 
	Let us define the following set of $q$-ary functions:
	\begin{equation}\label{defctk}
	\begin{split}
	C_t^k&=\left\{f_{d,t} = f+\frac{kq}{p}\left(\mathbf{d}\cdot\mathbf{x}_J+d_n x_{l_{\pi(0)}}\right)+\right.\\  &\left.\frac{q}{p}\left(\mathbf{t}\cdot\mathbf{x}_J+t_n x_{l_{\pi(m-n-1)}}\right): 0\leq d<{p^{n+1}}\right\},
	\end{split}
	\end{equation} where $(\mathbf{d}, d_n)=(d_0,d_1, \dots, d_n)\in \mathbb{Z}_p^{n+1}$ is the vector representation of the integer $d$.
	Then we can define a code as $
	\psi (C_t^k) = \{
	\psi(f_{d,t})\,|\, f_{d,t} \in C_t^k
	\}
	$ \ccn and thereby a set of codes as follows:
	\begin{equation}\label{Eq_CCC}
	\mathcal{C}_k=\left\{\psi (C_t^k)\,|\,0\leq t<p^{n+1}\right\}.
	\end{equation} 
	The following theorem characterizes the correlation proerties of the code sets $\mathcal{C}_k$.
	\begin{theorem}\label{thly} Let $f$ be a $q$-ary function as characterized in \eqref{Eq_f}.
		Then,  the code set $\mathcal{C}_k$ defined in \eqref{Eq_CCC}
		is a $(p^{n+1},p^m)$-CCC over $\mathcal{A}_q$ for any integer $k$ with $1\leq k <p$.
		\begin{IEEEproof}
	As the full proof is lengthy, here we only provide a sketch of the proof and the full proof can be found in Appendix \ref{appendix:B}.
		
According to the defintion of $\mathcal{C}_k$, we reprsent each set of $q$-ary functions $C_t^k$ given in (\ref{defctk}) as follows:
$C_t^k=\left\{f_{d,t}: 0\leq d<{p^{n+1}}\right\}$, where  
\begin{equation}
\begin{split}
f_{d,t}&=f+\frac{q}{p}(kd_n x_{l_{\pi(0)}}+t_n x_{l_{\pi(m-n-1)}}) +\frac{q}{p}(k\mathbf{d}+\mathbf{t})\cdot\mathbf{x}_J \\&=f_{d_n,t_n} +\frac{q}{p}(k\mathbf{d}+\mathbf{t})\cdot\mathbf{x}_J.
\end{split}
\end{equation} 
Let $\tau$ be an integer satisfying 
$0\leq |\tau|<p^m $. The ACCF between two codes  $\psi(C_t^k)$ and $\psi(C_{t'}^k)$ in $\mathcal{C}_k$ at the time shift $\tau$ can be expressed as 
\begin{equation}\label{skthmainmainres}
\begin{split}
&\Theta\left(\psi(C_{t}^k),\psi(C_{t'}^k)\right)(\tau)\\
=&\sum_{d=0}^{p^{n+1}-1} \sum_{\mathbf{c}_1,\mathbf{c}_2\in \Z_p^n} \Theta\left(\psi(f_{d,t}\arrowvert_{\mathbf{x}_J=\mathbf{c}_1}),\psi(f_{d,t'}\arrowvert_{\mathbf{x}_J=\mathbf{c}_2})\right)(\tau)\\
=&\mathcal{S}_1+\mathcal{S}_2,
\end{split}
\end{equation}
where 
\begin{equation}\nonumber\label{skths1}
\mathcal{S}_1=\sum_{d=0}^{p^{n+1}-1} \sum_{\mathbf{c}_1=\mathbf{c}_2}\Theta\left(\psi(f_{d,t}\arrowvert_{\mathbf{x}_J=\mathbf{c}_1}),\psi(f_{d,t'}\arrowvert_{\mathbf{x}_J=\mathbf{c}_2})\right)(\tau),
\end{equation}
and 
\begin{equation}\nonumber\label{skths2}
\mathcal{S}_2=        
\sum_{d=0}^{p^{n+1}-1} \sum_{\mathbf{c}_1\neq\mathbf{c}_2}\Theta\left(\psi(f_{d,t}\arrowvert_{\mathbf{x}_J=\mathbf{c}_1}),\psi(f_{d,t'}\arrowvert_{\mathbf{x}_J=\mathbf{c}_2})\right)(\tau).   
\end{equation}
A routine calculation shows that $\mathcal{S}_2=0$. 
The calculation of $\mathcal{S}_1$ is more complicated. Assume $\mathbf{c}_1=\mathbf{c}_2=\mathbf{c}\in\Z_p^n$. Then we have
\begin{equation}\label{skthnews1}
\begin{split}
\mathcal{S}_1
&=\sum_{d=0}^{p^{n+1}-1} \sum_{\mathbf{c}\in\Z_p^n}\Theta\left(\psi(f_{d,t}\arrowvert_{\mathbf{x}_J=\mathbf{c}}),\psi(f_{d,t'}\arrowvert_{\mathbf{x}_J=\mathbf{c}})\right)(\tau)
\\
&=\sum_{(\mathbf{d},d_n)\in\Z_{p}^{n+1}}\sum_{\mathbf{c}}\Theta\left(\psi(f_{d,t}\arrowvert_{\mathbf{x}_J=\mathbf{c}}),\psi(f_{d,t'}\arrowvert_{\mathbf{x}_J=\mathbf{c}})\right)(\tau)\\
\\&
=p^n\sum_{\mathbf{c}}\xi_p^{(\mathbf{t}-\mathbf{t}')\cdot\mathbf{c}}\mathcal{S}_3,
\end{split}
\end{equation}
where
\begin{equation}\nonumber\label{skths3}
\begin{split}
\mathcal{S}_3&=\sum_{d_n}\Theta\left(\psi\left(f_{d_n,t_n}\arrowvert_{\mathbf{x}_J=\mathbf{c}}\right),\psi\left(f_{d_n,t'_n}\arrowvert_{\mathbf{x}_J=\mathbf{c}}\right)\right)(\tau).
\end{split}
\end{equation}
In Appendix \ref{appendix:B}, we consider the calculation of $\mathcal{S}_3$ in three cases and obtain
\begin{equation}\nonumber\label{skthprofullo}
	\begin{split}
\mathcal{S}_3=\begin{cases}
p^{m-n+1},& \tau=0,~ t_n=t_n',\\
0,& \tau=0,~t_n\neq t_n',\\
0,& \tau\neq 0.
\end{cases}
	\end{split}
\end{equation}
The above result, combined with (\ref{skthmainmainres}), \eqref{skthnews1}  and $\mathcal{S}_2=0$, implies that
\begin{equation}\nonumber
\begin{split}
\Theta\left(\psi(C_{t}^k),\psi(C_{t'}^k)\right)(\tau)
&=\begin{cases}
p^{m+n+1}, & \tau=0, ~t=t',\\
0,& \textnormal{otherwise.}
\end{cases}\\
\end{split}
\end{equation}
Therefore, $\mathcal{C}_k$ forms a $(p^{n+1},p^m)$-CCC for any choice of $k$ in $\{1,2,\hdots,p-1\}$.				
%The proof is given in detailed in Appendix B.
		\end{IEEEproof}
	\end{theorem}

\begin{table*}[!t]
	\caption{Sets $C_t^k$ for the ($27,27$)-CCCs $\mathcal{C}_1$ and $\mathcal{C}_2$} \label{exthm1C}
	\centering
	\footnotesize
\resizebox{\textwidth}{!}{	
	\begin{tabular}{|l|l|l|}
		\hline
		\hspace{4cm}$\mathcal{C}_1$ &\hspace{4cm}	$\mathcal{C}_2$ \\ \hline
		$C_0^1= \{ f + 2(d_0x_0+d_1x_1+d_2x_2): 0\leq d<27\}$  & $C_0^2= \{ f + 4(d_0x_0+d_1x_1+d_2x_2): 0\leq d<27\}$ \\
		$C_1^1=\{ f +2(d_0x_0+d_1x_1+d_2x_2) + 2x_2: 0\leq d<27\}$ & $C_1^2= \{ f + 4(d_0x_0+d_1x_1+d_2x_2) + 2x_2: 0\leq d<27\}$ \\
		$C_2^1=\{ f +2(d_0x_0+d_1x_1+d_2x_2) + 4x_2: 0\leq d<27\}$ & $C_2^2= \{ f + 4(d_0x_0+d_1x_1+d_2x_2) + 4x_2: 0\leq d<27\}$ \\
		$C_3^1= \{ f + 2(d_0x_0+d_1x_1+d_2x_2) + 2x_1: 0\leq d<27\}$& $C_3^2= \{ f + 4(d_0x_0+d_1x_1+d_2x_2) + 2x_1: 0\leq d<27\}$ \\
		$C_4^1=\{ f + 2(d_0x_0+d_1x_1+d_2x_2)+ 2(x_1+x_2): 0\leq d<27\}$ & $C_4^2= \{ f + 4(d_0x_0+d_1x_1+d_2x_2) +2(x_1+x_2): 0\leq d<27\}$ \\
		$C_5^1=\{ f +2(d_0x_0+d_1x_1+d_2x_2) + 2(x_1+2x_2): 0\leq d<27\}$& $C_5^2= \{ f + 4(d_0x_0+d_1x_1+d_2x_2) +2(x_1+2x_2): 0\leq d<27\}$ \\
		$C_6^1= \{ f + 2(d_0x_0+d_1x_1+d_2x_2)+ 4x_1: 0\leq d<27\}$& $C_6^2= \{ f + 4(d_0x_0+d_1x_1+d_2x_2) +4x_1: 0\leq d<27\}$ \\
		$C_7^1=\{ f + 2(d_0x_0+d_1x_1+d_2x_2) + 2(2x_1+x_2): 0\leq d<27\}$ & $C_7^2= \{ f + 4(d_0x_0+d_1x_1+d_2x_2) + 2(2x_1+x_2): 0\leq d<27\}$ \\
		$C_8^1=\{ f +2(d_0x_0+d_1x_1+d_2x_2) + 2(2x_1+2x_2): 0\leq d<27\}$& $C_8^2= \{ f + 4(d_0x_0+d_1x_1+d_2x_2) + 2(2x_1+2x_2): 0\leq d<27\}$ \\ 
		$C_9^1=\{ f +2(d_0x_0+d_1x_1+d_2x_2) + 2x_0: 0\leq d<27\}$& $C_9^2= \{ f + 4(d_0x_0+d_1x_1+d_2x_2) + 2x_0: 0\leq d<27\}$ \\ 
		$C_{10}^1=\{ f +2(d_0x_0+d_1x_1+d_2x_2) + 2(x_0+x_2): 0\leq d<27\}$& $C_{10}^2= \{ f + 4(d_0x_0+d_1x_1+d_2x_2) + 2(x_0+x_2): 0\leq d<27\}$ \\
        $C_{11}^1=\{ f +2(d_0x_0+d_1x_1+d_2x_2) + 2(x_0+2x_2): 0\leq d<27\}$& $C_{11}^2= \{ f + 4(d_0x_0+d_1x_1+d_2x_2) + 2(x_0+2x_2): 0\leq d<27\}$ \\	
        $C_{12}^1=\{ f +2(d_0x_0+d_1x_1+d_2x_2) + 2(x_0+x_1): 0\leq d<27\}$& $C_{12}^2= \{ f + 4(d_0x_0+d_1x_1+d_2x_2) + 2(x_0+x_1): 0\leq d<27\}$ \\
        $C_{13}^1=\{ f +2(d_0x_0+d_1x_1+d_2x_2) + 2(x_0+x_1+x_2): 0\leq d<27\}$& $C_{13}^2= \{ f + 4(d_0x_0+d_1x_1+d_2x_2) + 2(x_0+x_1+x_2): 0\leq d<27\}$ \\
        $C_{14}^1=\{ f +2(d_0x_0+d_1x_1+d_2x_2) + 2(x_0+x_1+2x_2): 0\leq d<27\}$& $C_{14}^2= \{ f + 4(d_0x_0+d_1x_1+d_2x_2) + 2(x_0+x_1+2x_2): 0\leq d<27\}$ \\	
        $C_{15}^1=\{ f +2(d_0x_0+d_1x_1+d_2x_2) + 2(x_0+2x_1): 0\leq d<27\}$& $C_{15}^2= \{ f + 4(d_0x_0+d_1x_1+d_2x_2) + 2(x_0+2x_1): 0\leq d<27\}$ \\
        $C_{16}^1=\{ f +2(d_0x_0+d_1x_1+d_2x_2) + 2(x_0+2x_1+x_2): 0\leq d<27\}$& $C_{16}^2= \{ f + 4(d_0x_0+d_1x_1+d_2x_2) + 2(x_0+2x_1+x_2): 0\leq d<27\}$ \\
        $C_{17}^1=\{ f +2(d_0x_0+d_1x_1+d_2x_2) + 2(x_0+2x_1+2x_2): 0\leq d<27\}$& $C_{17}^2= \{ f + 4(d_0x_0+d_1x_1+d_2x_2) + 2(x_0+2x_1+2x_2): 0\leq d<27\}$ \\
        $C_{18}^1=\{ f +2(d_0x_0+d_1x_1+d_2x_2) + 4x_0: 0\leq d<27\}$& $C_{18}^2= \{ f + 4(d_0x_0+d_1x_1+d_2x_2) + 4x_0: 0\leq d<27\}$ \\
        $C_{19}^1=\{ f +2(d_0x_0+d_1x_1+d_2x_2) + 2(2x_0+x_2): 0\leq d<27\}$& $C_{19}^2= \{ f + 4(d_0x_0+d_1x_1+d_2x_2) + 2(2x_0+x_2): 0\leq d<27\}$ \\
        $C_{20}^1=\{ f +2(d_0x_0+d_1x_1+d_2x_2) + 2(2x_0+2x_2): 0\leq d<27\}$& $C_{20}^2= \{ f + 4(d_0x_0+d_1x_1+d_2x_2) + 2(2x_0+2x_2): 0\leq d<27\}$ \\
        $C_{21}^1=\{ f +2(d_0x_0+d_1x_1+d_2x_2) + 2(2x_0+x_1): 0\leq d<27\}$& $C_{21}^2= \{ f + 4(d_0x_0+d_1x_1+d_2x_2) + 2(2x_0+x_1): 0\leq d<27\}$ \\
        $C_{22}^1=\{ f +2(d_0x_0+d_1x_1+d_2x_2) + 2(2x_0+x_1+x_2): 0\leq d<27\}$& $C_{22}^2= \{ f + 4(d_0x_0+d_1x_1+d_2x_2) + 2(2x_0+x_1+x_2): 0\leq d<27\}$ \\
        $C_{23}^1=\{ f +2(d_0x_0+d_1x_1+d_2x_2) + 2(2x_0+x_1+2x_2): 0\leq d<27\}$& $C_{23}^2= \{ f + 4(d_0x_0+d_1x_1+d_2x_2) + 2(2x_0+x_1+2x_2): 0\leq d<27\}$ \\
        $C_{24}^1=\{ f +2(d_0x_0+d_1x_1+d_2x_2) + 2(2x_0+2x_1): 0\leq d<27\}$& $C_{24}^2= \{ f + 4(d_0x_0+d_1x_1+d_2x_2) + 2(2x_0+2x_1): 0\leq d<27\}$ \\
        $C_{25}^1=\{ f +2(d_0x_0+d_1x_1+d_2x_2) + 2(2x_0+2x_1+x_2): 0\leq d<27\}$& $C_{25}^2= \{ f + 4(d_0x_0+d_1x_1+d_2x_2) + 2(2x_0+2x_1+x_2): 0\leq d<27\}$\\ $C_{26}^1=\{ f +2(d_0x_0+d_1x_1+d_2x_2) + 2(2x_0+2x_1+2x_2): 0\leq d<27\}$& $C_{26}^2= \{ f + 4(d_0x_0+d_1x_1+d_2x_2) + 2(2x_0+2x_1+2x_2): 0\leq d<27\}$ \\
		\hline
	\end{tabular}}
\end{table*}  
\begin{figure}
	\centering
	\includegraphics[width=7cm]{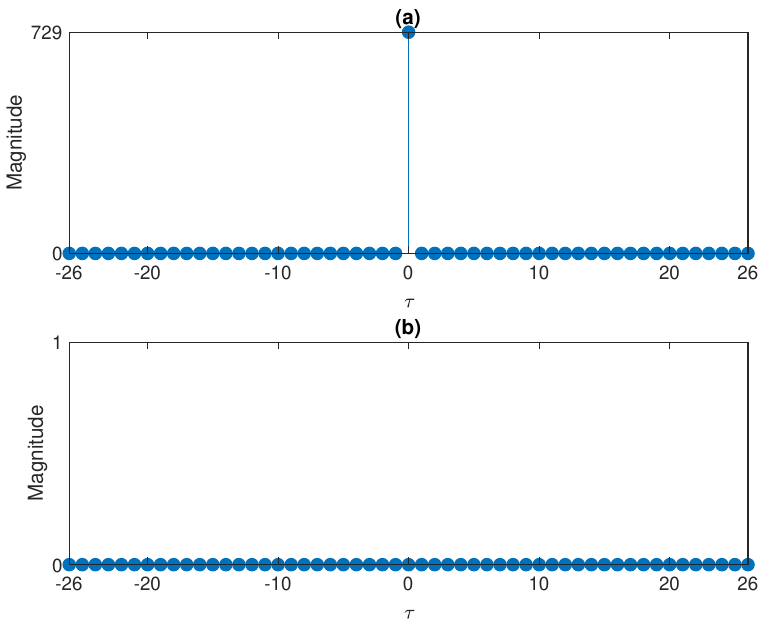}
	\caption{Correlation plot for $\mathcal{C}_k$}\label{thm1cp}
\end{figure}

	\smallskip
The following example illustrates the CCCs constructed in Theorem \ref{thly}. 
\begin{example}\label{exthm1097}	
For $m=3$, $p=3$, and $q=6$, let us consider the following function:
$$f(x_0,x_1,x_2)=x_0x_2+2x_2x_1+x_1x_0+x_0+2x_1+x_2+1.$$
%Note that $f\arrowvert_{x_0=0}=2x_1x_2+2x_1+x_2+1$, $f\arrowvert_{x_0=1}=2x_1x_2+3x_1+2x_2+2$, and $f\arrowvert_{x_0=2}=2x_1x_2+4x_1+3x_2+3$. 
%Then the graphs $G(f\arrowvert_{x_0=0}), G(f\arrowvert_{x_0=1})$, and $G(f\arrowvert_{x_0=2})$ are Hamiltonian paths identified by the quadratic term $2x_1x_2$. Hence the %function $f$ satisfies  the properties required for Theorem 2 with $n=1$ and $J=\{0\}$. 
Taking $J=\{0,1\}$, from (\ref{defctk}), we construct the following set of $6$-ary functions: 
\begin{equation}\tag{45}
\begin{split}
C_t^k&=\left\{f+2k(d_0x_0+d_1x_1+d_2x_2)\right. \\  &~~~~~~~ \left.+2(t_0x_0+t_1x_1+t_2x_2):d\in \mathbb{Z}_3^3\right\},
\end{split}
\end{equation} 
where  $1\leq k<3$, $(t_0,t_1,t_2) \in\mathbb{Z}_3^3$ corresponds to integers $t=0,1,\hdots,26$, and $(d_0,d_1,d_2)$ corresponds to integers $d=0,1,\hdots,26$.
Following (\ref{Eq_CCC}), we obtain $\mathcal{C}_k$, $k=1,2$, as below:
\begin{equation}\label{codethm1}\tag{46}
\begin{split}
%\mathcal{C}_k&=\left\{\left\{\psi(f+2k(d_0x_0+d_1x_1)+(t_0x_0+t_1x_1)): 0\leq d<9\right\}: 0\leq t < 9\right\}\\
\mathcal{C}_k&=\left\{\psi(C_t^k): 0\leq t < 27\right\}\\
&=\left\{\psi(C_0^k),\psi(C_1^k),\hdots, \psi(C_{26}^k)\right\}.
\end{split}
\end{equation} 
%where $(d_0,d_1)\in \Z_{3}^2$ is the vector representation of the integer $d$.
From (\ref{codethm1}), it is clear that both $\mathcal{C}_1$ and $\mathcal{C}_2$ contain $27$ codes of length $27$ over $\mathcal{A}_6$. In Table \ref{exthm1C}, we present the function sets for generating codes in $\mathcal{C}_1$ and $\mathcal{C}_2$.	
Besides, we list the codes $\psi(C_0^1)$, $\psi(C_1^1)$, and $\psi(C_2^1)$ from $\mathcal{C}_1$ in Table \ref{exthm1C1}, and $\psi(C_0^2)$, $\psi(C_1^2)$, and $\psi(C_2^2)$ from $\mathcal{C}_2$ in Table $\ref{exthm1C2}$.
%list the code sets  and $\mathcal{C}_2$, respectively.  
In addition, as shown in Figure \ref{thm1cp}, the AACF and ACCF of the codes in $\mathcal{C}_k$ are ideal. Hence $\mathcal{C}_k$ forms a $(27,27)$-CCC for $k=1,2$, which is consistent with Theorem \ref{thly}. 
\begin{table*}[]
	\caption{$\psi(C_0^1)$, $\psi(C_1^1)$ and $\psi(C_2^1)$ from  ($27,27$)-CCC $\mathcal{C}_1$ over the alphabet $\mathcal{A}_6$, where $\xi_6^i$ is given as $i$ for simplicity} \label{exthm1C1}
	\centering
	\small
	%\resizebox{\textwidth}{!}{

	%}
\end{table*}
\newpage
\begin{table*}[]
	\caption{$\psi(C^2_0)$, $\psi(C^2_1)$ and $\psi(C^2_2)$ from ($27,27$)-CCC $\mathcal{C}_2$ over the alphabet $\mathcal{A}_6$, where $\xi_6^i$ is given as $i$ for simplicity}\label{exthm1C2}
	\centering
	\small
	%\resizebox{\textwidth}{!}{
	%
 \\ \hline
	\end{tabular}
	%}
\end{table*}
\end{example}
	According to \textit{Theorem \ref{thly}}, the sets $\mathcal{C}_1,\mathcal{C}_2,\hdots,\mathcal{C}_{p-1}$ are $(p^{n+1},p^m)$-CCCs over $\mathcal{A}_q$. 
	In the sequel, we will show that the maximum aperiodic cross-correlation magnitude between two codes from any two distinct CCCs among $\mathcal{C}_1,\mathcal{C}_2,\hdots,\mathcal{C}_{p-1}$ is upper bounded by $p^m$. 
	To this end, we need the following proposition.
	\begin{proposition}
		Let $g$ and $h$ be two $q$-ary functions from $\Z_p^m$ to $\mathbb{Z}_q$. For any two different integers $1\leq k_1, k_2 <p$, define a set $\mathcal{S}$ as
		\begin{equation}\nonumber
		\mathcal{S}=\left\{(\mathbf{e}_1,\mathbf{e}_2):\mathbf{e}_1,\mathbf{e}_2\in\Z_p^{w}, k_1\mathbf{e}_1-k_2\mathbf{e}_2\equiv \mathbf{0}_{w}(\!\!\!\!\!\!\mod p)\right\}.
		\end{equation}
		Then for $J_1=\Z_w$ with $w< m$ and $\mathbf{x}_{J_1}=(x_0,x_1,\hdots,x_{w-1})$, we have
		\begin{equation}\nonumber
		\Big |\sum_{(\mathbf{e}_1,\mathbf{e}_2)\in\mathcal{S}}\Theta(\psi(g\arrowvert_{\mathbf{x}_{J_1}=\mathbf{e}_1}),\psi(h\arrowvert_{\mathbf{x}_{J_1}=\mathbf{e}_2}))(\tau)\Big|\leq p^{m-w}.
		\end{equation}
		\begin{IEEEproof}
				The full proof is lengthy, so here we only provide important steps of the proof. The details of the full proof can be found in Appendix \ref{Appendix:C}.

				Let us define a mapping $\Lambda:\mathcal{S}\rightarrow\mathbb{Z}$ as follows:
				$$\Lambda(\mathbf{e}_1,\mathbf{e}_2)= \sum_{t=0}^{w-1} e_{2,t}p^{w-1-t} - \sum_{t=0}^{w-1} e_{1,t}p^{w-1-t}.$$ 
				It can be shown that the above mapping is injective. Define two sets 
				\begin{equation}\nonumber
					\begin{split}
				\mathcal{S}'&=\{(\mathbf{e}_1,\mathbf{e}_2)\in\mathcal{S}:\Lambda(\mathbf{e}_1,\mathbf{e}_2)\geq 0\} \text{ and }\\ \mathcal{S}''&=\{(\mathbf{e}_1,\mathbf{e}_2)\in\mathcal{S}:\Lambda(\mathbf{e}_1,\mathbf{e}_2)\leq 0\}.
					\end{split}
				\end{equation}
				 Clearly they
				satisfy the following properties:
				$$\mathcal{S}=\mathcal{S}'\cup \mathcal{S}'',~\textnormal{and}~\mathcal{S}'\cap\mathcal{S''}=\{(\mathbf{0}_w,\mathbf{0}_w)\},$$
implying $|\mathcal{S}'|=|\mathcal{S}''|=\frac{p^w+1}{2}=E$ since $\left |\mathcal{S}\right |=p^w$. 

Assume that $(\mathbf{e}_1^i,\mathbf{e}_2^i)$ is
an element of $\mathcal{S}'$ and $\Lambda(\mathbf{e}_1^i,\mathbf{e}_2^i)=D_i$, where 	$\mathbf{e}_j^i=(e_{j,0}^i,e_{j,1}^i,\hdots, e_{j,n-1}^i)$, $i=1,2,\hdots,E$, and $j=1,2$. Since, $(\mathbf{0}_w,\mathbf{0}_w)\in \mathcal{S}'$ and $\Lambda$ is an injective mapping, without loss of generality, we can assume that $0=D_1<D_2<\cdots<D_E$. For $0\leq \tau\leq p^m-1$, following (\ref{cross_corr_restrict_c1_c2}), we have
\begin{equation}
	\begin{split}
\mathbf{B}_\tau(\mathbf{e}_1^i,\mathbf{e}_2^i)=&\{(\gamma,\delta):\delta=\gamma+\tau, 0\leq \gamma \leq p^m-\tau-1, \\&~~\gamma_\alpha=e_{1,\alpha}^i, \delta_\alpha=e_{2,\alpha}^i, 0 \leq \alpha<w\},
	\end{split}
\end{equation}
where
$(\gamma_0,\gamma_1,\hdots,\gamma_{m-1})$ and $(\delta_0,\delta_1,\hdots,\delta_{m-1})$ are the base-$p$ vector representations of the non-negative integers $\gamma$ and $\delta$, respectively. \ccn
Denote $ I_{D_i}= [p^{m-w}(D_i-1)+1:p^{m-w}(D_i+1)-1]$. Then the set $\mathbf{B}_\tau(\mathbf{e}^i_1,\mathbf{e}^i_2)$ is non-empty if $\tau$ is taken from $I_{D_i}$. 
In addition, it can be shown that for $1\leq i_1<i_2\leq E$,
%\begin{equation}\label{skthupd1}
%\begin{split}
%I_{D_{i_1}}\cap I_{D_{i_2}}=\begin{cases}
%[p^{m-w}D_{i_1}+1,p^{m-w}(D_{i_1}+1)-1],&~\textnormal{if}~D_{i_2}=D_{i_1}+1,\\
%\emptyset,&~ \textnormal{if}~D_{i_2}>D_{i_1}+1.
%\end{cases}
%\end{split}
%\end{equation}	
%Then
$I_{D_{i_1}}\cap I_{D_{i_2}}\neq \emptyset ~\textnormal{iff}~D_{i_2}=D_{i_1}+1.$
%where $\triangle_{D_{i_j}}=\mathbb{Z}\cap I_{D_{i_j}}$. 
%Assume that  $(\mathbf{e}^{i_1}_1,\mathbf{e}^{i_1}_2)$ and $(\mathbf{e}^{i_2}_1,\mathbf{e}^{i_2}_2)$ are two distinct elements 
%in $\mathcal{S}'$ with
%$D_{i_1}<D_{i_2}$.

\medskip

For a fixed value of $\tau$ in $[0:p^m-1]$, we need to consider the following three cases:

\textit{Case 1:}	$\tau\notin \cup_{i=1}^E I_{D_i}$. 
In this case,  since $\tau\geq 0$ and $\mathbf{B}_\tau(\mathbf{e}^i_1,\mathbf{e}^i_2)=\emptyset$ for all $i\in [1:E]$, we have
\begin{equation}\nonumber
\begin{split}
\sum_{(\mathbf{e}_1,\mathbf{e}_2)\in\mathcal{S}}\Theta(\psi(g\arrowvert_{\mathbf{x}_{J_1}=\mathbf{e}_1}),\psi(h\arrowvert_{\mathbf{x}_{J_1}=\mathbf{e}_2}))(\tau)
%&=\sum_{(\mathbf{e}_1,\mathbf{e}_2)\in\mathcal{S}'}\Theta(\psi(g\arrowvert_{\mathbf{x}_{J_1}=\mathbf{e}_1}),\psi(h\arrowvert_{\mathbf{x}_{J_1}=\mathbf{e}_2}))(\tau%)\\
%&=\sum_{i=1}^E \Theta(\psi(g\arrowvert_{\mathbf{x}_{J_1}=\mathbf{e}^i_1}),\psi(h\arrowvert_{\mathbf{x}_{J_1}=\mathbf{e}^i_2}))(\tau)\\
%&=\sum_{i=1}^E \sum_{(\gamma,\delta)\in \mathbf{B}_\tau(\mathbf{e}^i_1,\mathbf{e}^i_2) }\xi_q^{g_\gamma-h_\delta}\\
=0.
\end{split}
\end{equation} 
%Therefore, $	\left|\sum_{(\mathbf{e}_1,\mathbf{e}_2)\in\mathcal{S}}\Theta(\psi(g\arrowvert_{\mathbf{x}_{J_1}=\mathbf{e}_1}),\psi(h\arrowvert_{\mathbf{x}_{J_1}=\mathbf{e}_2}))(\tau)\right|=0$ when $\tau\notin \cup_{i=1}^E \triangle_{D_i}$.	

\textit{Case 2:}
$\tau\in I_{D_i}$ and $\tau\notin I_{D_j}$ for all $i\neq j$ in $[1:E]$. In this case, 
since $\tau$ does not belong to any $I_{D_j}$, we have
$\mathbf{B}_\tau(\mathbf{e}^j_1,\mathbf{e}^j_2)=\emptyset$ for all $j\in [1:E] \setminus \{i\}$. 
Partition the set $I_{D_j}$ into two as
$I_{D_i}^{(1)}= \{v \in I_{D_j}\,:\, v \leq  p^{m-w}D_{i}\}$ and 
$I_{D_i}^{(2)}= \{v \in I_{D_j}\,:\, v >  p^{m-w}D_{i}\}$. 
Now $\tau$ can be expressed as follows:
\begin{equation}\nonumber
\begin{split}
\tau=\begin{cases}
p^{m-w}(D_{i}-1)+\tau_1,& \textnormal{if}~ \tau\in I_{D_i}^{(1)},\\
p^{m-w}D_{i}+\tau_2, & \textnormal{if}~ \tau\in I_{D_i}^{(2)},
\end{cases}
\end{split}
\end{equation}
where $\tau_1=1,2,\hdots,p^{m-n}$ and $\tau_2=1,2,\hdots,p^{m-n}-1$. Then we have 
\begin{equation}\label{skthbc1c2con}
\begin{split}
|\mathbf{B}_\tau(\mathbf{e}^i_1,\mathbf{e}^i_2)|=\begin{cases}
\tau_1, & \textnormal{if}~ \tau\in I_{D_i}^{(1)},\\
p^{m-w}-\tau_2,& \textnormal{if}~ \tau\in I_{D_i}^{(2)}. 
\end{cases}
\end{split}
\end{equation}
We can show that 
\begin{equation}\label{skthcrthet}
\begin{split}
&\sum_{(\mathbf{e}_1,\mathbf{e}_2)\in\mathcal{S}}\Theta(\psi(g\arrowvert_{\mathbf{x}_{J_1}=\mathbf{e}_1}),\psi(h\arrowvert_{\mathbf{x}_{J_1}=\mathbf{e}_2}))(\tau)\\
%&=\sum_{(\mathbf{e}_1,\mathbf{e}_2)\in\mathcal{S}'}\Theta(\psi(g\arrowvert_{\mathbf{x}_{J_1}=\mathbf{e}_1}),\psi(h\arrowvert_{\mathbf{x}_{J_1}=\mathbf{e}_2}))(\tau%)\\
%&=\Theta(\psi(g\arrowvert_{\mathbf{x}_{J_1}=\mathbf{e}^i_1}),\psi(h\arrowvert_{\mathbf{x}_{J_1}=\mathbf{e}^i_2}))(\tau)\\&~~~~~~~~+\sum_{\substack{j=1\\j\neq %i}}^E \Theta(\psi(g\arrowvert_{\mathbf{x}_{J_1}=\mathbf{e}^j_1}),\psi(h\arrowvert_{\mathbf{x}_{J_1}=\mathbf{e}^j_2}))(\tau)\\
%&=\sum_{(\gamma,\delta)\in \mathbf{B}_\tau(\mathbf{e}^i_1,\mathbf{e}^i_2) }\xi_q^{g_\gamma-h_\delta}+\sum_{\substack{j=1\\ j\neq i}}^E \sum_{(\gamma,\delta)\in %\mathbf{B}_\tau(\mathbf{e}^j_1,\mathbf{e}^j_2) }\xi_q^{g_\gamma-h_\delta}\\
=&\sum_{(\gamma,\delta)\in \mathbf{B}_\tau(\mathbf{e}^i_1,\mathbf{e}^i_2) }\xi_q^{g_\gamma-h_\delta}.
\end{split}
\end{equation} 
From (\ref{skthbc1c2con}) and (\ref{skthcrthet}), we have
\begin{equation}\nonumber\label{skthnxtcrbd}
\begin{split}
&\left|\sum_{(\mathbf{e}_1,\mathbf{e}_2)\in\mathcal{S}}\Theta(\psi(g\arrowvert_{\mathbf{x}_{J_1}=\mathbf{e}_1}),
\psi(h\arrowvert_{\mathbf{x}_{J_1}=\mathbf{e}_2}))(\tau)\right|\\
%&=\left|\sum_{(\gamma,\delta)\in \mathbf{B}_\tau(\mathbf{e}^i_1,\mathbf{e}^i_2) }\xi_q^{g_\gamma-h_\delta}\right|
%\\ 
\leq &\begin{cases}
\tau_1, & \textnormal{if}~ \tau\in I_{D_i}^{(1)},\\
p^{m-w}-\tau_2,& \textnormal{if}~ \tau\in I_{D_i}^{(2)},
\end{cases}	
\end{split}
\end{equation}
implying
\begin{equation}\nonumber\label{Eq:case2}
|\sum_{(\mathbf{e}_1,\mathbf{e}_2)\in\mathcal{S}}\Theta(\psi(g\arrowvert_{\mathbf{x}_{J_1}=\mathbf{e}_1}),\psi(h\arrowvert_{\mathbf{x}_{J_1}=\mathbf{e}_2}))(\tau)|\leq p^{m-w}.
\end{equation}
\textit{Case 3:}
$\tau\in I_{D_i}\cap I_{D_{i+1}}$ for some $i\in\{1,2,\hdots,E\}$, where $D_{i+1}=D_{i}+1$. In this case we have 
\begin{equation}\nonumber\label{skthudf2}
\begin{split}
I_{D_i}\cap I_{D_{i+1}}
%&=(I_{D_i}\cap I_{D_{i+1}})\cap \mathbb{Z}\\
%=[p^{m-n}(D_{i_1}-1)+1, p^{m-n}(D_{i_1}+1)-1]\cap [p^{m-n}D_{i_1}+1, %p^{m-n}(D_{i_1}+2)-1]\\
&=[p^{m-w}D_{i}+1:p^{m-w}(D_{i}+1)-1].
\end{split}	
\end{equation}\ccn
It can be observed that
\begin{equation}\nonumber\label{skthcrthet1}
\begin{split}
&\sum_{(\mathbf{e}_1,\mathbf{e}_2)\in\mathcal{S}}\Theta(\psi(g\arrowvert_{\mathbf{x}_{J_1}=\mathbf{e}_1}),\psi(h\arrowvert_{\mathbf{x}_{J_1}=\mathbf{e}_2}))(\tau)\\
%&=\sum_{(\mathbf{e}_1,\mathbf{e}_2)\in\mathcal{S}'}\Theta(\psi(g\arrowvert_{\mathbf{x}_{J_1}=\mathbf{e}_1}),\psi(h\arrowvert_{\mathbf{x}_{J_1}=\mathbf{e}_2}))(\tau%)\\
%&=\Theta(\psi(g\arrowvert_{\mathbf{x}_{J_1}=
%	\mathbf{e}^{i}_1}),\psi(h\arrowvert_{\mathbf{x}_{J_1}=\mathbf{e}^{i}_2}))(\tau)+\Theta(\psi(g\arrowvert_{\mathbf{x}_{J_1}=
%	\mathbf{e}^{i+1}_1}),\psi(h\arrowvert_{\mathbf{x}_{J_1}=\mathbf{e}^{i+1}_2}))(\tau)+\\&\sum_{\substack{j=1\\j\neq i,i+1}}^E %\Theta(\psi(g\arrowvert_{\mathbf{x}_{J_1}=\mathbf{e}^j_1}),\psi(h\arrowvert_{\mathbf{x}_{J_1}=\mathbf{e}^j_2}))(\tau)\\
%&=\Theta(\psi(g\arrowvert_{\mathbf{x}_{J_1}=
%	\mathbf{e}^{i}_1}),\psi(h\arrowvert_{\mathbf{x}_{J_1}=\mathbf{e}^{i}_2}))(\tau)+\Theta(\psi(g\arrowvert_{\mathbf{x}_{J_1}=
%	\mathbf{e}^{i+1}_1}),\psi(h\arrowvert_{\mathbf{x}_{J_1}=\mathbf{e}^{i+1}_2}))(\tau)\\
=&\sum_{(\gamma,\delta)\in \mathbf{B}_\tau(\mathbf{e}^{i}_1,\mathbf{e}^{i}_2) }\xi_q^{g_\gamma-h_\delta}+\sum_{(\gamma,\delta)\in \mathbf{B}_\tau(\mathbf{e}^{i+1}_1,\mathbf{e}^{i+1}_2) }\xi_q^{g_\gamma-h_\delta}.
\end{split}
\end{equation}
%From (\ref{skthudf2}), $\tau\in [p^{m-w}D_{i}+1, p^{m-w}(D_{i}+1)-1]\cap\mathbb{Z}=[p^{m-w}(D_{i+1}-1)+1, p^{m-w}(D_{i+1})-1]\cap\mathbb{Z}$, 
Note that
$\tau$ can be expressed as $\tau=p^{m-w}D_{i}+\tau_3=p^{m-w}(D_{i+1}-1)+\tau_3$, where $\tau_3=1,2,\hdots,p^{m-w}-1$. 
Therefore, $\left|\mathbf{B}_\tau(\mathbf{e}^{i}_1,\mathbf{e}^{i}_2)\right|=p^{m-w}-\tau_3$ and  
$\left|\mathbf{B}_\tau(\mathbf{e}^{i+1}_1,\mathbf{e}^{i+1}_2)\right|=\tau_3$.
Similarly to Case 2, we have
\begin{equation}\nonumber
\begin{split}
&\left|\sum_{(\mathbf{e}_1,\mathbf{e}_2)\in\mathcal{S}}\Theta(\psi(g\arrowvert_{\mathbf{x}_{J_1}
	=\mathbf{e}_1}),\psi(h\arrowvert_{\mathbf{x}_{J_1}=\mathbf{e}_2}))(\tau)\right|\\
%&=\left|\sum_{(\gamma,\delta)\in %\mathbf{B}_\tau(\mathbf{e}^{i}_1,\mathbf{e}^{i}_2) }\xi_q^{g_\gamma-h_\delta}+\sum_{(\gamma,\delta)\in \mathbf{B}_\tau(\mathbf{e}^{i+1}_1,\mathbf{e}^{i+1}_2) %}\xi_q^{g_\gamma-h_\delta}\right|\\
%&\leq \left|\sum_{(\gamma,\delta)\in \mathbf{B}_\tau(\mathbf{e}^{i}_1,\mathbf{e}^{i}_2) }\xi_q^{g_\gamma-h_\delta}\right|+\left|\sum_{(\gamma,\delta)\in \mathbf{B}_\tau(\mathbf{e}^{i+1}_1,\mathbf{e}^{i+1}_2) }\xi_q^{g_\gamma-h_\delta}\right|\\
\leq &(p^{m-w}-\tau_3)+\tau_3=p^{m-w}.
\end{split}
\end{equation}
%Therefore, $\left|\sum_{(\mathbf{e}_1,\mathbf{e}_2)\in\mathcal{S}}\Theta(\psi(g\arrowvert_{\mathbf{x}_{J_1}=\mathbf{e}_1}),\psi(h\arrowvert_{\mathbf{x}_{J_1}=\mathbf{e}_2}))(\tau)\right|\leq p^{m-w}$.

\medskip

Combining Cases 1, 2, 3 gives
\begin{equation}
\begin{split}
&\left|\sum_{(\mathbf{e}_1,\mathbf{e}_2)\in\mathcal{S}}\Theta(\psi(g\arrowvert_{\mathbf{x}_{J_1}=\mathbf{e}_1}),\psi(h\arrowvert_{\mathbf{x}_{J_1}=\mathbf{e}_2}))(\tau)\right|\\\leq &p^{m-w}, \forall\, \tau\in [0:p^{m}-1].
\end{split}
\end{equation}
%$$$$ 
The statement for $\tau\in [-(p^{m-w}-1):0]$ \ccn can be similarly shown.
%Similarly, it can be shown that $$\left|\sum_{(\mathbf{e}_1,\mathbf{e}_2)\in\mathcal{S}}\Theta(\psi(g\arrowvert_{\mathbf{x}_{J_1}=\mathbf{e}_1}),\psi(h\arrowvert_{\mathbf{x}_{J_1}=\mathbf{e}_2}))(\tau)\right|\leq p^{m-w}, \forall\, \tau\in [-(p^{m-w}-1),0].$$
%The detailed of this proof is presented in Appendix C.
		\end{IEEEproof}
	\end{proposition}
		\begin{theorem}\label{thule}
		Consider $\mathcal{C}_1, \dots, \mathcal{C}_{p-1}$ given in Theorem \ref{thly}. Then for $J=\{0, \dots, n-1\}$ and $l_{\pi(0)}=n$,
		the union $\cup_{k=1}^{p-1}\mathcal{C}_k$ forms a $(p^{n+1}(p-1),p^{n+1},p^m,p^m)$-QCSS over $\mathcal{A}_q$. 
	\end{theorem}

\begin{IEEEproof}
	Let $f$ be the function as in \eqref{Eq_f}.
	Take $w=n+1$, $J_1=\{0, 1, \dots, n\}=J \cup \{n\}$ and $\mathbf{d}_1=(\mathbf{d}, d_n)$.
		Then we have   $\mathbf{d}\cdot\mathbf{x}_J+d_n x_n = \mathbf{d}_1\cdot\mathbf{x}_{J_1}$. 
Setting 
\begin{equation}
\begin{split}
g&=f+\frac{q}{p}\left(\mathbf{t}\cdot\mathbf{x}_J+t_n x_{l_{\pi(m-n-1)}}\right) , \\
 h&=f+\frac{q}{p}\left(\mathbf{t}'\cdot\mathbf{x}_J+t_n' x_{l_{\pi(m-n-1)}}\right).
\end{split}
\end{equation}
		One can express the
		ACCF sum between two codes \\$\psi(C_t^{k_1})\in \mathcal{C}_{k_1}$ and $\psi(C_{t'}^{k_2})\in \mathcal{C}_{k_2}$, where $k_1\neq k_2$, as 
		\begin{equation}\label{th2corr}
		\begin{split}
		&\Theta\left(\psi(C_{t}^{k_1}),\psi(C_{t'}^{k_2})\right)(\tau)\\=& 
		\sum_{\mathbf{d}_1}\Theta\left(\psi\left(g+\frac{k_1q}{p}\left(\mathbf{d}_1\cdot\mathbf{x}_{J_1}\right)\right),\right. \\ &\left.~~~~~~~~~~~~~~~~~~~~   \psi\left(h+\frac{k_2q}{p}\left(\mathbf{d}_1\cdot\mathbf{x}_{J_1}\right)\right)\right)(\tau)\\
		=&\sum_{\mathbf{d}_1}\sum_{\mathbf{e}_1,\mathbf{e_2}\in\Z_p^{n+1}}\xi_q^{\frac{k_1q}{p}\mathbf{d}_1\mathbf{e}_1+ \frac{k_2}{p}\mathbf{d}_1\mathbf{e}_2}\Theta(\psi(g\arrowvert_{\mathbf{x}_{J_1}=\mathbf{e}_1}),\\&~~~~~~~~~~~~~~~~~~~~~~~~~~~~~~~~~~~~\psi(h\arrowvert_{\mathbf{x}_{J_1}=\mathbf{e}_2}))(\tau)\\
		=&\sum_{\mathbf{d}_1}\sum_{\mathbf{e}_1,\mathbf{e_2}\in\Z_p^{n+1}}\xi_p^{\mathbf{d}_1\cdot(k_1\mathbf{e}_1-k_2\mathbf{e}_2)}\Theta(\psi(g\arrowvert_{\mathbf{x}_{J_1}=\mathbf{e}_1}),\\&~~~~~~~~~~~~~~~~~~~~~~~~~~~~~~~~~~~\psi(h\arrowvert_{\mathbf{x}_{J_1}=\mathbf{e}_2}))(\tau)
     \end{split}
	\end{equation}
	
	\begin{equation}\nonumber
	\begin{split}
	&=\sum_{\mathbf{e}_1,\mathbf{e_2}\in\Z_p^{n+1}}\Theta(\psi(g\arrowvert_{\mathbf{x}_{J_1}=\mathbf{e}_1}),\psi(h\arrowvert_{\mathbf{x}_{J_1}=\mathbf{e}_2}))(\tau)\times\\&~~~~~~~~~~~~~~~~~~~~~~~~~~~\sum_{\mathbf{d}_1}\xi_p^{\mathbf{d}_1\cdot(k_1\mathbf{e}_1-k_2\mathbf{e}_2)}.
		\end{split}
		\end{equation}
		
		According to (\ref{th2corr}), we consider the following cases:

		\textit{Case 1:} $k_1\mathbf{e}_1-k_2\mathbf{e}_2\not\equiv \mathbf{0}_{n+1}~ (\!\!\!\!\mod p)$. In this case,
		we have $$\sum_{\mathbf{d}_1}\xi_p^{\mathbf{d}_1\cdot(k_1\mathbf{e}_1-k_2\mathbf{e}_2)}=0,$$
		which implies
		$\Theta \left(\psi(C_{t}^{k_1}),\psi(C_{t'}^{k_2})\right)(\tau)=0.$

		\textit{Case 2:} $k_1\mathbf{e}_1-k_2\mathbf{e}_2\equiv \mathbf{0}_{n+1}~(\!\!\!\!\mod p)$. In this case we have $$\sum_{\mathbf{d}_1}\xi_p^{\mathbf{d}_1\cdot(k_1\mathbf{e}_1-k_2\mathbf{e}_2)}=p^{n+1}.$$
		Denote 
		\begin{equation}
		\begin{split}
		\mathcal{S}&=\{(\mathbf{e}_1,\mathbf{e}_2):\mathbf{e}_1,\mathbf{e}_2\in\Z^{n+1},\\&~~~~~~~~~~~k_1\mathbf{e}_1-k_2\mathbf{e}_2\equiv \mathbf{0}_{n+1}~(\!\!\!\!\mod p)\}.
		\end{split}
		\end{equation}
		Then the result in (\ref{th2corr}) reduces to the following:
		\begin{equation}\label{condc1c2}
		\begin{split}
	&	\Theta \left(\psi(C_{t}^{k_1}),\psi(C_{t'}^{k_2})\right)(\tau)\\
		%		&=\sum_{(\mathbf{e}_1,\mathbf{e}_2)\in\mathcal{S}} \Theta\left(\psi\left(g\arrowvert_{\mathbf{x}_{J_1}=\mathbf{e}_1}\right),\psi\left(h\arrowvert_{\mathbf{x}_{J_1}=\mathbf{e}_2}\right) \right)(\tau)\sum_{\mathbf{d}_1}\xi_p^{\mathbf{d}_1\cdot(k_1\mathbf{e}_1-k_2\mathbf{e}_2)}\\
		=& p^{n+1}\sum_{(\mathbf{e}_1,\mathbf{e}_2)\in\mathcal{S}} \Theta\left(\psi\left(g\arrowvert_{\mathbf{x}_{J_1}=\mathbf{e}_1}\right),\psi\left(h\arrowvert_{\mathbf{x}_{J_1}=\mathbf{e}_2}\right) \right)(\tau).
		\end{split}
		\end{equation}
		
		Applying \textit{Proposition 1} in (\ref{condc1c2}), we obtain
		\begin{equation}\nonumber\label{thm2fns}
		\begin{split}
	&	\Big|\Theta \left(\psi(C_{t}^{k_1}),\psi(C_{t'}^{k_2})\right)(\tau)\Big|\\
		=& p^{n+1}\Big|\sum_{(\mathbf{e}_1,\mathbf{e}_2)\in\mathcal{S}} \Theta\left(\psi\left(g\arrowvert_{\mathbf{x}_{J_1}=\mathbf{e}_1}\right),\psi\left(h\arrowvert_{\mathbf{x}_{J_1}=\mathbf{e}_2}\right) \right)(\tau)\Big| \\\leq & p^m.
		\end{split}
		\end{equation}

	Combining the above cases, it is clear that 
	\begin{equation}\label{thm2fns2}
	\Big|\Theta \left(\psi(C_{t}^{k_1}),\psi(C_{t'}^{k_2})\right)(\tau)\Big|\leq p^m ~ \textnormal{for}~ |\tau|<p^m,k_1\neq k_2.
	\end{equation}
	
	As per Theorem \ref{thly}, each $\mathcal{C}_k$  is a $(p^{n+1},p^m)$-CCC set for $k\in\{1,2,\hdots,p-1\}$. From (\ref{thm2fns2}), it is clear that the maximum magnitude of the ACCFs between the codes from two distinct sets of CCCs $\mathcal{C}_{k_1}$ and $\mathcal{C}_{k_2}$ is upper bounded by $p^m$. Therefore, the set of codes  $\cup_{k=1}^{p-1}\mathcal{C}_k$ forms a $(p^{n+1}(p-1),p^{n+1},p^m,p^m)$-QCSS over $\mathcal{A}_q$.
\end{IEEEproof}

\bigskip

\begin{remark}
	As a comparison, Table \ref{comtab} lists the existing constructions of aperiodic QCSSs and our proposed construction in this paper. 
	It is clear that in all existing constructions, the alphabets have sizes no less than the length of constituent sequences. Therefore, 
	for any integer $q$ smaller than $p^m$ with $m>1$, the known constructions cannot genetrate the QCSSs as reported in \textit{Theorem \ref{thule}}.
\end{remark}

The following corollary discusses the optimality of the proposed QCSSs w.r.t the newly derived lower bounds in this paper.
\begin{corollary}[Asymptotic Optimality of the Proposed Construction]\label{cor3}
	The proposed construction produces $(p^{n+1}(p-1),p^{n+1},p^m,p^m)$-QCSS over $\mathcal{A}_q$. We check optimality for $N> 4$, with respect to our newly derived tighter lower bound given in \textit{Remark \ref{newrem}}. The optimality factor $\rho$
	can be expressed as follows:
	\begin{equation}\label{opro1}
	\begin{split}
	\rho &=\frac{\theta}{\sqrt{ML\left(1-\frac{\pi\sqrt{N(2L^2-N)}-4L}{4(N-1)L}\right)}}\\
	&=\frac{p^{\frac{m-n-1}{2}}}{\sqrt{1-\frac{\pi\sqrt{\frac{(p-1)}{p^2}\left(2-\frac{p-1}{p^{2m}}\right)}-\frac{4}{p}}{1-\frac{2}{p}}}}.
	\end{split}
	\end{equation}
	In particular, for $m=n+1$, 
	it can be observed from (\ref{opro1}) that the optimality factor $\rho$ achieve the value $1$ for a sufficiently large value of $p$. 
\end{corollary}
	
\begin{figure}[t]
	\includegraphics[width=9cm]{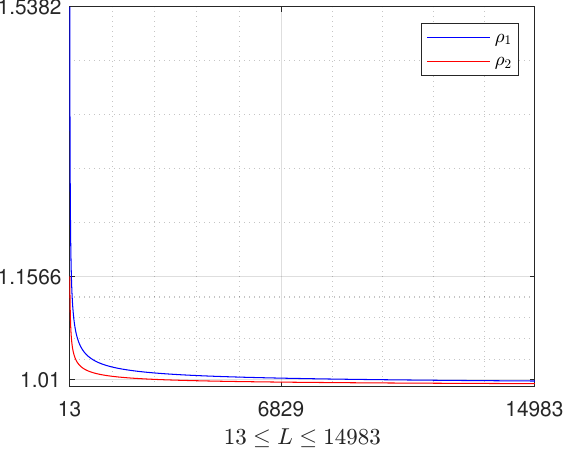}
	\caption{Comparison between the optimality factors $\rho_1$ and $\rho_2$ with respect to $((p-1)p,p,p,p)$-QCSS for $13\leq p<15000$}\label{boutkita1}
\end{figure}

Corollary \ref{cor3} shows that the proposed QCSSs are asympototically optimal w.r.t to the lowers bound in Corollaries \ref{crol1} and \ref{crol2}. 
As a matter of fact, when the prime $p$ and $m$ take small values, the resulting QCSSs are near optimal. 
With respect to the lowers bound in \cite{crlbzl} and those in Remark \ref{newrem}, we denote optimal factors 
\begin{equation*}
\begin{split}
\rho_1&=\frac{\theta}{\sqrt{ML\left(1-2\sqrt{\frac{M}{3K}}\right)}}~~  \text{ and }\\
 \rho_2 & = 
	\begin{cases}
	 \frac{\theta}{\sqrt{\frac{ML^2}{2L-1}}}, & N=2, \\
	 \frac{\theta}{{ML\left(1-\frac{L^2(2\pi^2+4N-16)-N\pi^2}{16L^2(N-1)}\right)}}, & N=4,\\
	 \frac{\theta}{\sqrt{ML\left(1-\frac{\pi\sqrt{N(2L^2-N)}-4L}{4(N-1)L}\right)}}, & N>4.
	\end{cases}
	\end{split}
\end{equation*} 
Table \ref{corrtab} lists the values of optimality factors $\rho_1$, $\rho_2$, respectively, for certain parameters $p$ and $m$. 
For $p=3$, the entries are denoted as ``--'' as the bound in \cite{crlbzl} is valid for $K\geq 3M$.
Table \ref{corrtab} clearly shows that the proposed QCSS tends to optimality faster with respect to the proposed bound. Furthermore, we also have compared $\rho_1$ and $\rho_2$ for $N>4$, with respect to the proposd $(p(p-1), p,p,p)$-QCSS, where $13\leq p< 15000 $ in Figure \ref{boutkita1}. In this figure, the horizontal axis represents sequence lengths in the form of prime numbers ranging from $13$ to $14983$, while the vertical axis represents the values of $\rho_1$ and $\rho_2$, which range from $1$ to $1.5382$ (the value of $\rho_1$ at $p=13$, where $\rho_2=1.1566$). It is evident that $\rho_2$ tends to converge to $1$ faster than $\rho_1$. 
\begin{table}[t]
	\caption{Optimality factors for the proposed QCSS with respect to the proposed lower bound and the lower bound in \cite{crlbzl} }\label{corrtab}
	\arraycolsep=1 pt
	\begin{tabular}{|l|l|l|l|l|l|l|c|c|}
		\hline
		$p$ & $m$ & $K$ & $M$ & $N$ & $L$ & $\theta$ & $\rho_1$ & $\rho_2$ \\ \hline
		\multirow{2}{*}{$3$}&$1$&$6$&$3$&$2$&$3$&$3$&$-$&$1.29$\\\cline{2-9}
		&$2$&$18$&$9$&$2$&$9$&$9$&$-$&$1.37$
		\\\hline
		\multirow{2}{*}{$5$}	& $1$    & $20$    & $5$    & $4$    & $5$    & $5$         &   $1.54$       &   $1.27$   \\ \cline{2-9}
		&$2$&$100$&$25$&$4$&$25$&$25$&$1.54$&$1.3$ 
		\\\hline
		\multirow{2}{*}{$7$}&$1$&$42$&$7$&$6$&$7$&$7$&$1.38$&$1.22$\\\cline{2-9}   
		& $2$&$294$&$49$&$6$&$49$&$49$&$1.38$&$1.23$ 
		\\\hline  
		\multirow{2}{*}{$11$} &$1$&$110$&$11$&$10$&$11$&$11$&$1.25$&$1.17$\\\cline{2-9}
		& $2$&$1210$&$121$&$10$&$121$&$121$&$1.25$&$1.18$
		\\\hline
	\end{tabular}
\end{table}

In the end we provide two examples for the proposed QCSSs.
\begin{example}
	\begin{figure}[t]
		\centering
		\includegraphics[width=9cm]{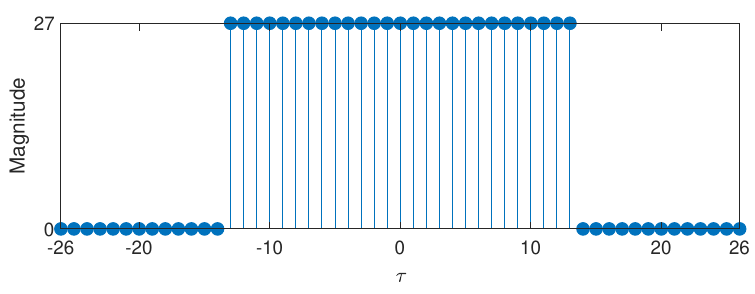}
		\caption{ Correlation plot between the codes $\psi(C_0^1)$ and $\psi(C_1^2)$}\label{thm3097}
	\end{figure}
Recall from Example \ref{exthm1097} that $m=3$, $p=3$, $q=6$, $J=\{0,1\}$, $n=2$, and the function appears as follows:
	$$f(x_0,x_1,x_2)=x_0x_2+2x_2x_1+x_1x_0+x_0+2x_1+x_2+1.$$
In (\ref{codethm1}), we have derived $(27,27)$-CCCs $\mathcal{C}_1$ and $\mathcal{C}_2$ from Theorem \ref{thly}. We present three codes from each of $\mathcal{C}_1$ and $\mathcal{C}_2$ in Table \ref{exthm1C1} and Table \ref{exthm1C2}, respectively, in $\mathbb{Z}_6$-valued form. 

In this example, we shall consider the maximum  ACCF magnitude for codes drawn from $\mathcal{C}_1$ and $\mathcal{C}_2$. As each of $\mathcal{C}_1$ and $\mathcal{C}_2$ contains $27$ codes, there are $729$ possible ACCFs. For clarity, we choose two codes $\psi(C_0^1)\in \mathcal{C}_1$ and $\psi(C_1^2)\in \mathcal{C}_2$ and plot their ACCF magnitudes in Figure \ref{thm3097}. It can be verified that the maximum magnitude among the remaining $728$ ACCFs is also given by $27$, thus verifying the correlation properties as stated in 
Theorem \ref{thule}. Therefore $\mathcal{C}_1\cup\mathcal{C}_2$ forms a $(54,27,27,27)$-QCSS over the alphabet $\mathcal{A}_6$. As $N=p-1=2$, from Remark \ref{newrem}, the optimality factor is 
$\rho=\frac{\theta}{\sqrt{\frac{ML^2}{2L-1}}}=1.40$. Therefore, the code set $(54,27,27,27)$-QCSS forms near-optimal QCSS over $\mathcal{A}_6$. 
\end{example}
Below we present another example to derive a near-optimal $QCSS$ from Theorem \ref{thule}.
\begin{example}
	\begin{table*}[!ht]\small 
		\caption{($18,9,9,9$)-QCSS over the alphabet $\mathcal{A}_3$, where $\xi_3^i$ is given as $i$ for simplicity} \label{Tab3}
		\resizebox{\textwidth}{!}{                                     \\ \hline
		\end{tabular}}
	\end{table*}
	Let $f:\Z_3^2\rightarrow\mathbb{Z}_3$ be a ternary function given by
	$$f(x_0,x_1)=x_0x_1+x_0^2+x_1.$$
	%\begin{figure}[h]
	%	\centering
	%	\includegraphics{empl3.pdf}
	%	\caption{Graph corresponding to $x_0x_1+x_0^2+x_1$ }\label{palm}
	%	\label{exmple2}
	%\end{figure}
	%From Figure \ref{palm}, we observe that $G(f)$ contains a cycle around the vertex $x_0$. Therefore for each $c\in \Z_3 $, $G(f\arrowvert_{x_0=c})$ contains only one vertex $x_1$, and hence it is a path. 
	Taking $J=\{0\}$, from (\ref{defctk}), we construct the following sets of ternary functions for $k=1, 2$:
	$$
	\begin{array}{l}
	C_t^k=\left\{f+k(d_0x_0+d_1x_1)+(t_0x_0+t_1x_1):d_0,d_1\in\Z_3 \right\}, 
	\end{array}
	$$
	where $(t_0, t_1) \in Z_3^2$ corresponds to integers $t=0, 1, \dots ,8$.
	Table \ref{Tab3} lists the sequences associated with the functions in $C_t^k$.

	Numerical results show that the sets 
	$\mathcal{C}_1=\left\{\psi(C_t^1):0\leq t< 9\right\}$, $\mathcal{C}_2=\left\{\psi(C_t^2): 0\leq t< 9\right\}$ are $(9,9)$-CCCs.
	It is also confirmed that the maximum magnitude of ACCF between any two codes from $\mathcal{C}_1$ and $\mathcal{C}_2$, respectively, is upper bounded by $9$,
	indicating that $\mathcal{C}_1\cup\mathcal{C}_2$ forms an $(18,9,9,9)$-QCSS over the alphabet $\mathcal{A}_3$. As an illustration, Figure \ref{paln} (a) and Figure \ref{paln} (b) represent 
	the absolute value of AACF and ACCF, respectively, for the CCCs $\mathcal{C}_1$ and 
	$\mathcal{C}_2$. In Figure \ref{paln} (c), we present the absolute value of ACCF between 
	$\psi(C^1_0)$ and $\psi(C^2_0)$. 
	These numerical results are consistent with \textit{Theorem \ref{thly}} and  \textit{Theorem \ref{thule}}.
	
	In addition, since $N=2$, the optimality factor appears as:
	$
	\rho=\frac{\theta}{\sqrt{\frac{ML^2}{2L-1}}}=1.37.
	$
	This indicates that the derived  $(18,9,9,9)$-QCSS $\mathcal{C}_1\cup\mathcal{C}_2$ is near-optimal. 
	
	\begin{figure}
		\centering
		\includegraphics[width=7cm]{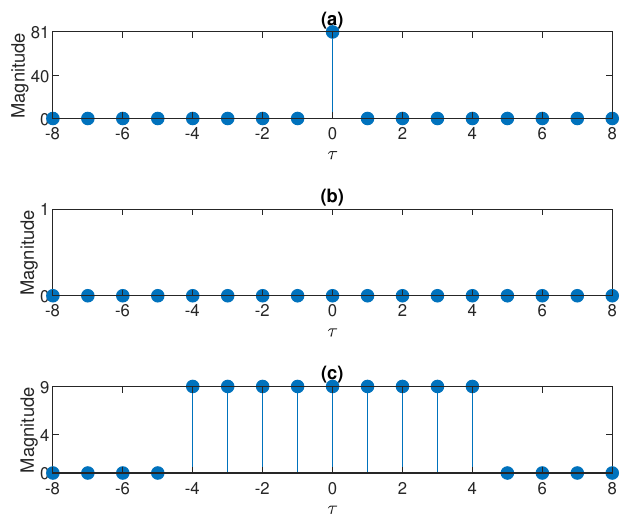}
		\caption{Correlation plot}\label{paln}
	\end{figure}

\end{example}
%%%%%%%%%%%%%%%%%%%%%%%%%%%%%%%%%%%%%%
\section{Conclusion}
In this paper, we have first studied the lower bound on the maximum magnitude of aperiodic auto- and cross-correlation functions for those QCSSs that appear as the collection of multiple CCCs. By selecting appropriate weight vectors into the bounding function, we have shown that the derived bound is tighter than the bound reported in \cite{crlbzl}.
Then, we have studied $q$-ary functions through a graphical point-of-view to produce aperiodic QCSSs over a small alphabet. The proposed construction generates aperiodic QCSSs over the alphabet $\mathcal{A}_q$, where $q$ is divisible by $p$. 
Unlike the existing aperiodic QCSSs, the proposed construction can maintain a small alphabet size with increasing set size and sequence lengths. It is also to be noted that the obtained QCSSs appears in the form of the collection of multiple sets of CCCs which may gurantee multipath interference free communication in MC-CDMA system as the multipath interference is closely related to the AACFs of the codes assigned to the users. 
%With respect to the newly derived lower bound, the proposed {\color{blue}QCSSs} tend to optimality as the seuqence length increases. 
As the sequence length increases, the proposed QCSSs tend to be asymptotically optimal with respect to the derived lower bound.  
%%%%%%%%%%%%%%%%%%%%%%%%%%%%%%%%%%%%%%
\appendices
\section{Proof of Corollary 1}\label{Appendix:A}
\textit{Case 1 ($0<t\leq L$):} 
		For $0\leq u\leq t-1$, we obtain
		\begin{equation}\nonumber
		\begin{split}
\sum_{v=0}^{t-1}\tau_{u,v,L}&=\sum_{v=0}^{u}(u-v)+\sum_{v=u+1}^{t-1}(v-u)
		\\
		&=\frac{u(u+1)}{2}+\frac{(t-u-1)(t-u)}{2}.
		\end{split}
		\end{equation}
		Therefore,
		\begin{equation}\nonumber
		\begin{split}
		\sum_{u,v=0}^{t-1}\tau_{u,v,L}&=\sum_{u=0}^{t-1}\frac{u(u+1)}{2}+\sum_{u=0}^{t-1}\frac{(t-u-1)(t-u)}{2}
		\\&
		=\frac{t(t^2-1)}{3}.
		\end{split}
		\end{equation}
		Substituting $\sum_{u,v=0}^{t-1}\tau_{u,v,L}=\frac{t(t^2-1)}{3}$ in (\ref{lbcccderiv}), we have 
		\begin{equation}\label{bndm<L}
		\begin{split}
		\theta^2&\geq \frac{ML\left(1-\frac{ML}{Kt}-\frac{(t^2-1)}{3Lt}\right)}{1-\frac{M}{K}}\\&= \frac{ML}{1-\frac{M}{K}} \left(1 - \frac{1}{3L}\left(t + \frac{3ML^2-K}{K t}\right)\right).
		\end{split}
		\end{equation}
		
		Define a function $f(x) = x + \frac{3ML^2-K}{K x}$. We are interested in the minimum value of $f(x)$ over the interval $[1, L]$. Observe that $f(x)$ is continuous on $[1, L]$. 
		Consider its derivative $f'(x) = 1 - \frac{3ML^2-K}{K x^2}$. It has two zeros $x_0 = \sqrt{\frac{3ML^2}{K}-1}$ and $-x_0$, and is monotonically increasing over the interval $[1, L]$. We now consider the minimum value of $f(x)$ over the interval $[1, L]$.
		\begin{enumerate}
			\item When $N=K/M = 2$, one has $x_0> L$, which means that $f'(x)<0$ over $[1, L]$. In this case $f(x)$ is 
			monotonically decreasing over $[1, L]$ and achieves the minimum value at $x=L$. Then we have 
			\begin{equation}\nonumber\label{key}
			f(x)\geq f(L)= 2L  - \frac{1}{L}  \text{ for } x \in [1, L],
			\end{equation}
			which implies
			\begin{equation}\nonumber\label{Eq_bound_N=2}
			\begin{split}
			\theta^2 & \geq  \frac{ML}{1-M/K}\left(\frac{2}{3}-\frac{M}{K}+\frac{1}{3L^2}\right)\\& = ML\left( \frac{1}{3} + \frac{2}{3L^2} \right).
			\end{split}
			\end{equation}
			\item When $N=K/M= 3$, one has $x_0=\sqrt{\frac{3ML^2}{K}-1} = \sqrt{L^2-1}\leq L$, which is very close to $L$. In this case, $f'(x)<0$ over the interval $[1,L]$, and hence monotonically decreasing. Therefore, $f(x)$ achieve minimum value at $x=L$, and we have the following lower bound:
			\begin{equation}\nonumber\label{EqBound-N3}
			\theta^2 \geq ML\left(\frac{1}{2}+\frac{1}{2L^2}\right).
			\end{equation} 
			\item When $N=K/M >3$, one has $
			x_0=\sqrt{\frac{3ML^2}{K}-1}< L$. That is to say, $f'(x)<0$ over $[1, x_0]$ and $f'(x)>0$ over $[x_0, L]$.
			This implies that the function $f(x)$ achieves the minimum value at $x=x_0$. 
			Hence we have 
			\begin{equation}\nonumber\label{key1}
			f(x)\geq f(x_0)= 2\sqrt{\frac{3ML^2}{K}-1} \text{ for } x \in [1, L]. 
			\end{equation}
			In this case, by properly choosing $x$ around $\sqrt{\frac{3L^2}{N}-1}$, we have the following simplified lower bound:
			\begin{equation}\nonumber\label{bndmaxm}
			\theta^2\geq ML\left(1- \frac{2\sqrt{N(3L^2-N)}-3L}{3L(N-1)}\right).
			\end{equation}
		\end{enumerate}			
\textit{Case 2 ($L<t\leq 2L-1$):}			
	For $0\leq u\leq t-L-1$, we obtain
	\begin{equation}\nonumber
	\begin{split}
	\sum_{v=0}^{t-1}\tau_{u,v,L}&=\sum_{v=0}^{u-1}(u-v)+\sum_{v=u}^{u+L-1}(v-u)\\&+\sum_{v=u+L}^{t-1}(2L-1-v+u)\\
	&=u(t+1-2L)-L^2+2Lt-\frac{t^2+t}{2}.
	\end{split}
	\end{equation}
	Hence, \begin{equation}\nonumber
	\sum_{u=0}^{t-L-1}\sum_{v=0}^{t-1}\tau_{u,v,L}=\frac{(t-L)(t+1)(L-1)}{2}.
	\end{equation}				
	For $t-L\leq u\leq L-1$, we obtain
	\begin{equation}\nonumber
	\begin{split}
	\sum_{v=0}^{t-1}\tau_{u,v,L}&=\sum_{v=0}^{u}(u-v)+\sum_{v=u+1}^{t-1}(v-u)
	\\&
	=u^2-(t-1)u+\frac{t^2-t}{2}.
	\end{split}
	\end{equation}	
	Hence, 
	\begin{equation}\nonumber
	\sum_{u=t-L}^{L-1}\sum_{v=0}^{t-1}\tau_{u,v,L}=\frac{3Lt^2-t^3-3L^2t+t+2L^3-2L}{3}.
	\end{equation}	
	Similarly, we obtain
	\begin{equation}\nonumber
	\begin{split}
	\sum_{u=L}^{t-1}\sum_{v=0}^{t-1}\tau_{u,v,L}=\frac{(t+1)(t-L)(L-1)}{2}.
	\end{split}
	\end{equation}
	Therefore,
	\begin{equation}\label{tau_L+1_2L-1}
	\begin{split}
	\sum_{u,v=0}^{t-1}\tau_{u,v,L}&=(t+1)(t-L)(L-1)\\&+\frac{3Lt^2-t^3-3L^2t+t+2L^3-2L}{3}.
	\end{split}
	\end{equation}	
	From (\ref{lbcccderiv}) and	(\ref{tau_L+1_2L-1}), we have
	\begin{equation}\label{Eq4}
	\theta^2 \geq \frac{M}{3(1-M/K)}\left(
	t + \dfrac{a}{t} - \frac{b}{t^2} - 3(L-1)
	\right),
	\end{equation}
	where \begin{equation}\nonumber
	\begin{split}
	a&= (6L^2-6L+2)-\frac{3ML^2}{K},  \\b&=L(L-1)(2L-1).
	\end{split}
	\end{equation}
	Similarly we define a function $$f(x) = x + \frac{a}{x} - \frac{b}{x^2},$$ and will discuss the property of $f(x)$ on the interval $[L+1, 2L-1]$.
	
	We shall consider both the 1st-order and 2nd-order derivatives of $f(x)$, which are given by 
	$$
	f'(x) = 1 - \frac{a}{x^2}+\frac{2b}{x^3} \text{ and } f''(x) = \frac{2a}{x^3} -\frac{6b}{x^4}=\frac{2a}{x^4}\left(x-\frac{3b}{a}\right).
	$$
	The properties of $f'(x)$ and $f''(x)$ on the interval $[L+1, 2L-1]$ will be used to determine the maximum value of $f(x)$ in $[L+1, 2L-1]$.
	
	\medskip
	
	We start with the property of $f''(x)$ on $[L+1, 2L+1]$. Note that $f''(x)$ has zero 
	$$\begin{array}{rcl}
	x_0 &=& \dfrac{3b}{a}\\ &=& \dfrac{3L(L-1)(2L-1)}{(6L^2-6L+2)-\frac{3ML^2}{K}} \\
	&=&\dfrac{6L^3-9L^2+3L}{(6-\frac{3}{N})L^2 -6L + 2}
	\\&=& L \left(1+ \dfrac{(\frac{3}{N}L^2-3L+1)}{(6-\frac{3}{N})L^2-6L+2}\right).
	\end{array}
	$$ 
	We need to consider whether $x_0$ lies in between $L+1$ and $2L-1$. Hence we divide the discussion into three subcases:  $N=2$, $3 \leq N \leq L/3$ and $N > L/3$.
	\begin{enumerate}
		\item  In the case of $N=2$, we have 
		$$
		\frac{1}{L} \leq \frac{(\frac{3}{N}L^2-3L+1)}{(6-\frac{3}{N})L^2-6L+2} < 1.
		$$
		In this case,  the root
		$x_0=\frac{3b}{a}$ lies in the interval $[L+1, 2L-1]$. This implies
		$f''(x)<0$ for $x\in [L+1, x_0)$ and $f''(x)>0$ for $x\in (x_0, 2L-1]$.  Consequently,
		the first order derivative $f'(x)$ is monotonically decreasing over $[L+1, x_0]$ and  monotonically increasing over $[x_0, 2L-1]$. Furthermore, we have
		\begin{equation}\label{Eq5a}
		\begin{split}
		f'(2L-1) =& \dfrac{1}{(2L-1)^3}\left( (2L-1)^3 \!-\! a(2L\!-\!1) \!+\! 2b\right) \\
		=&\dfrac{\frac{3}{N}L^2-1}{(2L-1)^2}\\ =& \dfrac{\frac{3}{2}L^2-1}{(2L-1)^2} > 0
		\end{split}
%		\\&>& 0
		\end{equation}
		and 
		\begin{equation}\label{Eq5b}
		\begin{split}
	&	f'(L+1)\\
		=& \dfrac{ (L+1)^3 - a(L+1) + 2b}{(L+1)^3}
		\\
		=& \frac{(L+1)^3 - ((6L^2-6L+2)-\frac{3}{N}L^2)(L+1)}{(L+1)^3}\\&~~~~+ \frac{2L(L-1)(2L-1)}{(L+1)^3} \\
		=& \dfrac{-(1-\frac{3}{N})L^3 -L^2 (3-\frac{3}{N}) + 9 L - 1}{(L+1)^3}
		\\=& \dfrac{L^3 -3L^2 + 18 L - 2}{2(L+1)^3} > 0.
		\end{split}
		\end{equation}
		Since for $x\in [L+1, 2L-1]$,  the derivative function $f'(x) $ satisfies 
		$
		f'(x) \geq f'\left(\frac{3b}{a}\right) 
		=1 - \frac{a\cdot a^2}{9b^2} + \frac{2a^3}{27b^3} 
		= 1 - \frac{a^3}{27b^2}
		>0
		$ for $L\geq 2$. Hence
		the function $f(x)$ is monotonically increasing on $[L+1, 2L-1]$ and it achieves the maximum value at $x=2L-1$.
		Since 
		\begin{equation}
		\begin{split}
		f(2L-1) &= (2L-1) + \frac{a}{2L-1} - \frac{b}{(2L-1)^2}\\& = \frac{(9L^2-9L+3)-\frac{3}{N}L^2}{2L-1}.
		\end{split}
		\end{equation}
		It follows that (\ref{lbcccderiv}) becomes
		\begin{equation}\nonumber\label{Eq_BoundN2}
		\theta^2 \geq \frac{M}{(1-M/K)} \cdot \frac{(1-\frac{1}{N})L^2}{(2L-1)} =\frac{ML^2}{2L-1}.
		\end{equation}
		\item When $3 \leq N \leq L/3$, similarly one has  $$
		\frac{1}{L} < \frac{(\frac{3}{N}L^2-3L+1)}{(6-\frac{3}{N})L^2-6L+2} < 1,$$
		implying that $f'(x)$ is monotonically decreasing over $[L+1, x_0]$ and  monotonically increasing over $[x_0, 2L-1]$. 
		From the calculations in \eqref{Eq5a} and \eqref{Eq5b}, one has
		$$\begin{array}{rcl}
		f'(2L-1) = \dfrac{\frac{3}{N}L^2-1}{(2L-1)^2}> 0
		\end{array}
		$$
		and 
		\begin{equation}\nonumber
		\begin{split}
		f'(L+1)&=\dfrac{-(1-\frac{3}{N})L^3 -L^2 (3-\frac{3}{N}) + 9 L - 1}{(L+1)^3}\\& <0.
		\end{split}
		\end{equation}
		
		This implies the minimum value $f'(x_0)<0$.
		Then the function $f'(x)$ has a zero $x_1$ in the interval $[x_0, 2L-1]$. That is to say, 
		$f'(x)<0 \text{ for } x\in [L+1, x_1] \text{ and } f'(x)\geq 0 \text{ for } x\in [x_1, 2L-1].$ Hence the function $f(x)$ is monotonically decreasing on $[L+1, x_1]$ and is monotonically increasing on $[x_1, 2L-1]$. Consequently, the maximum value of $f(x)$ is attained either at $x=L+1$ or $x=2L-1$. 
		
		Note that 
		
		\begin{equation}\nonumber
		\begin{split}
		&f(L+1)- f(2L-1) 
		\\=& 
		(L+1) + \frac{a}{L+1} - \frac{b}{(L+1)^2}\\& - \left((2L-1) + \frac{a}{2L-1} - \frac{b}{(2L-1)^2} \right)
%		\\&=& \dfrac{(L-2)}{(L+1)^2(2L-1)^2} \left(
%		(L+1)(2L-1) a - 3b L-  (L+1)^2(2L-1)^2  \right)
%		\\& = & \dfrac{(L-2)}{(L+1)^2(2L-1)} \left(
%		(L+1) a - 3L^2(L-1)-  (L+1)^2(2L-1)  \right)
%		\\& = & \dfrac{(L-2)}{(L+1)^2(2L-1)} \left(
%		(L+1) a - 3L^2(L-1)-  (L+1)^2(2L-1)  \right)
		\\=&  \dfrac{(L-2)}{(L+1)^2(2L-1)} 
		\left(
		\frac{N-3}{N} L^3 - \frac{3}{N}L^2 - 4L +3
		\right).
		\end{split}
		\end{equation}
		
		This implies that for $L\geq 3N$, $f(L+1)<f(2L-1)$ for $N=3$, and $f(L+1) > f(2L-1)$ for $N>3$.
		Therefore, for $N=3$ one has 
		\begin{equation}\nonumber\label{Eq_BoundN3abc}
		\theta^2 \geq \frac{M}{(1-M/K)} \cdot \frac{(1-\frac{1}{N})L^2}{(2L-1)} =\frac{ML^2}{2L-1},
		\end{equation}
		and for $N>3$, we have
		\begin{equation}\label{Eq_BoundN3}
		\begin{split}
		\theta^2 &\geq \dfrac{M}{(1-M/K)} \times\\& \dfrac{(2 N - 3)L^3  +  3(N - 1) L^2 + NL  + 6 N}{3(L+1)^2N} 
		\\&= 
		ML \left(
		1 -\right. \\ &\left. \dfrac{(N+6)L^3 + 3(N-1)L^2 +(2N-3) L - 6N}{3L(L+1)^2(N-1)}
		\right).
		\end{split}
		\end{equation}
		\item When $N > L/3\geq 3$, one has $x_0=3b/a < L+1$ for $L\geq 9$. In this case $f''(L+1)>0$ and $f''(2L-1) >0$, implying that $f'(x)$ is monotonically increasing over $[L+1, 2L-1]$. Similarly, one has 
		$
		f'(L+1) < 0 \text{ and } f'(2L-1) >0. 
		$
		Hence $f(x)$ achieve the maximum value either at $L+1$ or $2L-1$. Similar to the previous discussion for the case, we know $f(L+1) > f(2L-1)$ when $N>L/3 \geq 3$. Then the lower bound is as given in \eqref{Eq_BoundN3}.
	\end{enumerate}
	In summary, for $L+1\leq t\leq 2L-1$,  when $N=2, 3$, the bound is given by
	\begin{equation}\nonumber\label{EqCase2i}
	\theta^2 \geq  \frac{ML^2}{2L-1},
	\end{equation}
	and  when $N >3$, the bound is given by 
	\begin{equation}\label{eqCase2ii}
	\begin{split}
	\theta^2 &\geq ML \left(
	1 -\right.\\&\left. \dfrac{(N+6)L^3 + 3(N-1)L^2 +(2N-3) L - 6N}{3L(L+1)^2(N-1)}
	\right).
	\end{split}
	\end{equation}
Comparing the lower bounds in \textit{Case 1} and \textit{Case 2}, we have the following result:
	\begin{itemize}
		\item when $N=2, 3$, the maximum value of the lower bounds is achieved from (\ref{Eq4}) at $t=2L-1$, namely,
		\begin{equation}\nonumber\label{EqBound-N2}
		\theta^2 \geq \frac{ML^2}{2L-1},
		\end{equation}
%		\item when $N=3$,
%		\begin{itemize}
%			\item for $L\leq 8$, the maximum value of the lower bound in (\ref{bndm<L}) is achieved at $t=L$, namely, 
%		\begin{equation}\label{EqBound-N3}
%		\theta^2 \geq ML\left(\frac{1}{2}+\frac{3}{2L^2}\right)
%		\end{equation}
%		\item for $L\geq 9$, the maximum value of the lower bounds is achieved from (\ref{Eq4}) at $t=2L-1$, namely,
%		\begin{equation}\label{snig*}
%		\theta^2 \geq \frac{ML^2}{2L-1}
%		\end{equation}
%		\end{itemize}
		\item when $N>3$,		
		the maximum value of the lower bounds is achieved from (\ref{bndm<L}) at
		%$t=\left\lceil \sqrt{\frac{3L^2}{N}-1}\right\rceil$
		$\left\lceil\sqrt{\frac{3L^2}{N}-1}\right\rfloor$. The lower bound is approximately given by
		\begin{equation}\nonumber
		\theta^2   \geq ML \left( 1 - \frac{2\sqrt{{3L^2N}-N^2}-3L }{3L(N-1)}\right).
		\end{equation}
		%	since in \eqref{Eq_BoundN3} one has
		%	$$
		%	\dfrac{(N+6)L^3 + 3(N-1)L^2 +(2N-3) L - 6N}{(L+1)^2} > (2\sqrt{{3N}}-3)L.
		%	$$ 
\end{itemize}	
\section{Proof of Theorem 2} \label{appendix:B}
According to the defintion of $\mathcal{C}_k$, we reprsent each set of $q$-ary functions $C_t^k$ given in (\ref{defctk}) as follows:
$C_t^k=\left\{f_{d,t}: 0\leq d<p^{n+1}\right\}$, where $f_{d,t}=f_{d_n,t_n}+\frac{q}{p}(k\mathbf{d}+\mathbf{t})\cdot\mathbf{x}_J,f_{d_n,t_n}=f+\frac{q}{p}(kd_n x_{l_{\pi(0)}}+t_n x_{l_{\pi(m-n-1)}})$, and
$(d_0,d_1,\hdots,d_n)$ is the vector representation of $d$ with respect to base-$p$.
Let $\tau$ be an integer satisfying 
$0\leq |\tau|<p^m $. The ACCF between two codes  $\psi(C_t^k)$ and $\psi(C_{t'}^k)$ in $\mathcal{C}_k$ at the time shift $\tau$ can be expressed as 
\begin{equation}\label{mainmainres}
    \begin{split}
        &\Theta\left(\psi(C_{t}^k),\psi(C_{t'}^k)\right)(\tau)\\=&\sum_{d=0}^{p^{n+1}-1}\Theta\left(\psi(f_{d,t}),\psi(f_{d,t'})\right)(\tau)\\
        =&\sum_{d=0}^{p^{n+1}-1}\sum_{\mathbf{c}_1,\mathbf{c}_2\in \Z_p^n} \Theta\left(\psi(f_{d,t}\arrowvert_{\mathbf{x}_J=\mathbf{c}_1}),\psi(f_{d,t'}\arrowvert_{\mathbf{x}_J=\mathbf{c}_2})\right)(\tau)\\
        =&\sum_{d=0}^{p^{n+1}-1}\sum_{\mathbf{c}_1=\mathbf{c}_2}\Theta\left(\psi(f_{d,t}\arrowvert_{\mathbf{x}_J=\mathbf{c}_1}),\psi(f_{d,t'}\arrowvert_{\mathbf{x}_J=\mathbf{c}_2})\right)(\tau)+\\&
        \sum_{d=0}^{p^{n+1}-1}\sum_{\mathbf{c}_1\neq\mathbf{c}_2}\Theta\left(\psi(f_{d,t}\arrowvert_{\mathbf{x}_J=\mathbf{c}_1}),\psi(f_{d,t'}\arrowvert_{\mathbf{x}_J=\mathbf{c}_2})\right)(\tau)\\
        =&\mathcal{S}_1+\mathcal{S}_2,
    \end{split}
\end{equation}
where 
\begin{equation}\nonumber\label{s1}
\mathcal{S}_1=\!\!\!\sum_{d=0}^{p^{n+1}-1}\sum_{\mathbf{c}_1=\mathbf{c}_2}\Theta\left(\psi(f_{d,t}\arrowvert_{\mathbf{x}_J=\mathbf{c}_1}),\psi(f_{d,t'}\arrowvert_{\mathbf{x}_J=\mathbf{c}_2})\right)(\tau),
\end{equation}
and 
\begin{equation}\label{s2}
\mathcal{S}_2= \!\!\!       
\sum_{d=0}^{p^{n+1}-1}\sum_{\mathbf{c}_1\neq\mathbf{c}_2}\Theta\left(\psi(f_{d,t}\arrowvert_{\mathbf{x}_J=\mathbf{c}_1}),\psi(f_{d,t'}\arrowvert_{\mathbf{x}_J=\mathbf{c}_2})\right)(\tau).   
\end{equation}
Now, 
\begin{equation}\label{Frest1}
    \psi(f_{d,t}\arrowvert_{\mathbf{x}_J=\mathbf{c}_1})=
    \xi_p^{k(\mathbf{d}\cdot\mathbf{c}_1)}\xi_p^{\mathbf{t}\cdot\mathbf{c}_1}\psi\left(f_{d_n,t_n}\arrowvert_{\mathbf{x}_J=\mathbf{c}_1}\right),
\end{equation}
and
\begin{equation}\label{Frest2} 
    \psi(f_{d,t'}\arrowvert_{\mathbf{x}_J=\mathbf{c}_2})= \xi_p^{k(\mathbf{d}\cdot\mathbf{c}_2)}\xi_p^{\mathbf{t}'\cdot\mathbf{c}_2}\psi\left(f_{d_n,t_n'}\arrowvert_{\mathbf{x}_J=\mathbf{c}_2}\right),
\end{equation}
where $\mathbf{t}'=(t_0',t_1',\hdots,t_{n-1}')$, and $(t_0',t_1',\hdots,t_{n-1}',t_n')$ is the vector representation of $t'$ with respect to base-$p$.
Let us first start with $\mathcal{S}_2$ in (\ref{s2}),
\begin{equation}\nonumber
\begin{split}
\mathcal{S}_2&=\sum_{d=0}^{p^{n+1}-1}\sum_{\mathbf{c}_1\neq\mathbf{c}_2}\Theta\left(\psi(f_{d,t}\arrowvert_{\mathbf{x}_J=\mathbf{c}_1}),\psi(f_{d,t'}\arrowvert_{\mathbf{x}_J=\mathbf{c}_2})\right)(\tau)\\       %\sum_{\mathbf{d}d_n}\sum_{\mathbf{c}_1\neq\mathbf{c}_2}\Theta\left(\psi_\lamb%da(F_1\arrowvert_{\mathbf{x}_J=\mathbf{c}_1}),\psi_q(F_2\arrowvert_{\ma%thbf{x}_J=\mathbf{c}_2})\right)(\tau)\\
&=\sum_{(\mathbf{d},d_n)\in      \mathbb{Z}_p^{n+1}}\sum_{\mathbf{c}_1\neq\mathbf{c}_2}\xi_p^{k(\mathbf{d}\cdot(\mathbf{c}_1-\mathbf{c}_2))}\xi_p^{(\mathbf{t}\cdot\mathbf{c}_1-\mathbf{t}'\cdot\mathbf{c}_2)}\\&~~~~~~~~ \times
\Theta\left(\psi\left(f_{d_n,t_n}\arrowvert_{\mathbf{x}_J=\mathbf{c}_1}\right),\psi\left(f_{d_n,t'_n}\arrowvert_{\mathbf{x}_J=\mathbf{c}_2}\right)\right)(\tau) \\
&=\sum_{d_n}\sum_{\mathbf{c}_1\neq\mathbf{c}_2}\xi_p^{(\mathbf{t}\cdot\mathbf{c}_1-\mathbf{t}'\cdot\mathbf{c}_2)}
\Theta\left(\psi\left(f_{d_n,t_n}\arrowvert_{\mathbf{x}_J=\mathbf{c}_1}\right),\right.\\&~~~~~~~~~~ \left.\psi\left(f_{d_n,t'_n}\arrowvert_{\mathbf{x}_J=\mathbf{c}_2}\right)\right)(\tau)\sum_{\mathbf{d}} \xi_p^{k(\mathbf{d}\cdot(\mathbf{c}_1-\mathbf{c}_2))}.
\end{split}
\end{equation}
Since, $1\leq k\leq p-1$ and $\mathbf{c}_1\neq \mathbf{c}_2$,
$\sum_{\mathbf{d}} \xi_p^{k(\mathbf{d}\cdot(\mathbf{c}_1-\mathbf{c}_2))}=0$. Therefore, $\mathcal{S}_2=0$. Now let us move to $\mathcal{S}_1$.
Let us assume $\mathbf{c}_1=\mathbf{c}_2=\mathbf{c}\in\Z_p^n$. Then
\begin{equation}\label{news1}
\begin{split}
\mathcal{S}_1&=\sum_{d=0}^{p^{n+1}-1}\sum_{\mathbf{c}_1=\mathbf{c}_2}\Theta\left(\psi(f_{d,t}\arrowvert_{\mathbf{x}_J=\mathbf{c}_1}),\psi(f_{d,t'}\arrowvert_{\mathbf{x}_J=\mathbf{c}_2})\right)(\tau)\\
&=\sum_{(\mathbf{d},d_n)\in\Z_{p^{n+1}}}\sum_{\mathbf{c}}\Theta\left(\psi(f_{d,t}\arrowvert_{\mathbf{x}_J=\mathbf{c}_1}),\psi(f_{d,t'}\arrowvert_{\mathbf{x}_J=\mathbf{c}_2})\right)(\tau).
\end{split}
\end{equation}
From (\ref{news1}), (\ref{Frest1}), and (\ref{Frest2}), we have
\begin{equation}\label{mains1}
    \begin{split}
        \mathcal{S}_1 &= \sum_{(\mathbf{d},d_n) \in \mathbb{Z}_p^{n+1}}\sum_{\mathbf{c}}\xi_p^{k(\mathbf{d}\cdot(\mathbf{c}-\mathbf{c}))}\xi_p^{(\mathbf{t}-\mathbf{t}')\cdot\mathbf{c}}
\\&~~~~~~~~ \times\Theta\left(\psi\left(f_{d_n,t_n}\arrowvert_{\mathbf{x}_J=\mathbf{c}}\right),\psi\left(f_{d_n,t'_n}\arrowvert_{\mathbf{x}_J=\mathbf{c}}\right)\right)(\tau)\ccn\\
&=p^n\sum_{d_n}\sum_{\mathbf{c}}\xi_p^{(\mathbf{t}-\mathbf{t}')\cdot\mathbf{c}}
\Theta\left(\psi\left(f_{d_n,t_n}\arrowvert_{\mathbf{x}_J=\mathbf{c}}\right),\right.\\&~~~~~~~~~~~~~~~~~~~~~~~~~~~~ \left.\psi\left(f_{d_n,t'_n}\arrowvert_{\mathbf{x}_J=\mathbf{c}}\right)\right)(\tau)\\
&=p^n\sum_{\mathbf{c}}\xi_p^{(\mathbf{t}-\mathbf{t}')\cdot\mathbf{c}}\mathcal{S}_3,
    \end{split}
\end{equation}
where
\begin{equation}\nonumber\label{s3}
\begin{split}
    \mathcal{S}_3&=\sum_{d_n}\Theta\left(\psi\left(f_{d_n,t_n}\arrowvert_{\mathbf{x}_J=\mathbf{c}}\right),\psi\left(f_{d_n,t'_n}\arrowvert_{\mathbf{x}_J=\mathbf{c}}\right)\right)(\tau).
    \end{split}
\end{equation}
% Let us assume $\gamma$ and $\delta$ are two non-negative integers such that $0\leq \gamma,\delta<p^m$.
 %and 
%$\delta=\gamma+\tau$.
Let us recall $\gamma$ and $\delta$ and their base-$p$ vector representations $(\gamma_0,\gamma_1,\hdots,\gamma_{m-1})$ and $(\delta_0,\delta_1,\hdots,\delta_{m-1})$, respectively, and $\mathbf{A}_\tau(\mathbf{c})=\{(\gamma,\delta):0\leq \gamma\leq p^m-\tau-1,\delta=\gamma+\tau,\gamma_{j_\alpha}=c_\alpha, \delta_{j_\alpha}=c_\alpha, \alpha=0,1,\hdots,n-1\}$ as defined in Section II-B.
%NEED CHANGE: Let us assume $(\gamma_0,\gamma_1,\hdots,\gamma_{m-1})$ and $(\delta_0,\delta_1,\hdots,\delta_{m-1})$ are vector representations 
%of $\gamma$ and $\delta$ with respect to base $p$.
Based on the definition of complex-valued sequence corresponding to restricted $q$-ary function given in Section II-A, the $\gamma$th component 
of $\psi\left(f_{d_n,t_n}\arrowvert_{\mathbf{x}_J=\mathbf{c}}\right)$ 
is given by $\xi_q^{(f_{d_n,t_n})_\gamma}$ if $\gamma_{j_\alpha}=c_\alpha$ for $\alpha=0,1,\hdots,n-1$, else the value is zero, where $(f_{d_n,t_n})_\gamma=f_{d_n,t_n}(\gamma_0,\gamma_1,\hdots,\gamma_{m-1})$. Similarly, the $\gamma$th component of 
$\psi\left(f_{d_n,t'_n}\arrowvert_{\mathbf{x}_J=\mathbf{c}}\right)$ is given by $\xi_q^{(f_{d_n,t'_n})_\gamma}$ if $\gamma_{j_\alpha}=c_\alpha$ for $\alpha=0,1,\hdots,n-1$, else the value is zero, where $(f_{d_n,t'_n})_i=f_{d_n,t'_n}(\gamma_0,\gamma_1,\hdots,\gamma_{m-1})$. 
%Let us define the following set:
%\begin{equation}\nonumber
%    \mathbf{A}_\tau(\mathbf{c})=\{(\gamma,\delta):0\leq \gamma\leq p^m-\tau-1,\delta=\gamma+\tau,\gamma_{j_\alpha}=c_\alpha, \delta_{j_\alpha}=c_\alpha, \alpha=0,1,\hdots,n-1\}.
%\end{equation}}
Then $\mathcal{S}_3$ can be expressed as 
\begin{equation}\label{s3der}
    \begin{split}
      \mathcal{S}_3&=\sum_{d_n}\Theta\left(\psi\left(f_{d_n,t_n}\arrowvert_{\mathbf{x}_J=\mathbf{c}}\right),\psi\left(f_{d_n,t'_n}\arrowvert_{\mathbf{x}_J=\mathbf{c}}\right)\right)(\tau)  \\
      &=\sum_{d_n} \sum_{(\gamma,\delta)\in \mathbf{A}_\tau(\mathbf{c})}\xi_q^{(f_{d_n,t_n})_\gamma-(f_{d_n,t'_n})_\delta}.
    \end{split}
\end{equation}
\textit{Case 1} ($\tau\neq 0,~ \gamma_{l_{\pi(0)}}\neq \delta_{l_{\pi(0)}}$):
Since, $0<k<p$, $\sum_{d_n} \xi_p^{kd_n(\gamma_{l_{\pi(0)}}-\delta_{l_{\pi(0)}})}=0$. From (\ref{s3der}), we have 
\begin{equation}\nonumber
    \begin{split}
      \mathcal{S}_3&
      =\sum_{d_n} \sum_{(\gamma,\delta)\in \mathbf{A}_\tau(\mathbf{c})}\xi_q^{(f_\gamma-f_\delta)+\frac{\gamma k d_n}{p}(\gamma_{l_{\pi(0)}}-\delta_{l_{\pi(0)}})}\\&~~~~~~~~~~~~~~~~~~~~~~~~\times\xi_q^{\frac{q}{p}(t_n\gamma_{l_{\pi(m-n-1)}}-t'_n \delta_{l_{\pi(m-n-1)}})}\\
      &=\sum_{d_n} \sum_{(\gamma,\delta)\in \mathbf{A}_\tau(\mathbf{c})}\xi_q^{f_\gamma-f_\delta}\xi_p^{kd_n(\gamma_{l_{\pi(0)}}-\delta_{l_{\pi(0)}})}\\&~~~~~~~~~~~~~~~\times\xi_p^{(t_n\gamma_{l_{\pi(m-n-1)}}-t'_n \delta_{l_{\pi(m-n-1)}}}\\
      &=\sum_{(\gamma,\delta)\in \mathbf{A}_\tau(\mathbf{c})}\xi_q^{f_\gamma-f_\delta} \xi_p^{(t_n\gamma_{l_{\pi(m-n-1)}}-t'_n \delta_{l_{\pi(m-n-1)}}} \\&~~~~~~~~~~~~~~~~~\times\sum_{d_n} \xi_p^{kd_n(\gamma_{l_{\pi(0)}}-\delta_{l_{\pi(0)}})}\\
     & =0,
    \end{split}
\end{equation}
where
$ f_\gamma=f(\gamma_0,\gamma_1,\hdots,\gamma_{m-1})$ 
and
$f_\delta=f(\delta_0,\delta_1,\hdots,\delta_{m-1}).$

\textit{Case 2} ($\tau\neq 0,~\gamma_{l_{\pi(0)}}= \delta_{l_{\pi(0)}}$):
We assume $u$ is the smallest positive integer for which $\gamma_{l_{\pi(u)}}\neq \delta_{l_{\pi(u)}}$. Let us also assume $\gamma^v$ to be an integer whose base-$p$
vector representation is given by
$(\gamma_0,\gamma_1,\hdots,\kappa p+\gamma_{l_{\pi(u-1)}}-v,\hdots,\gamma_{m-1})$, where $v\in\{1,2,\hdots,p-1\}$, and $\kappa=0$ when $\gamma_{l_{\pi(u)}}-v\geq 0$ and $\kappa=1 $ when $\gamma_{l_{\pi(u)}}-v<0$. Similarly, we assume $\delta^v$ to be an integer whose base-$p$
vector representation is given by
$(\delta_0,\delta_1,\hdots,\kappa p+\delta_{l_{\pi(u-1)}}-v,\hdots,\delta_{m-1})$. It is clear that, $\gamma^v$ and $\delta^v$ differs from $\gamma$ and $\delta$, respectively, only at $l_{\pi(u-1)}$th position. It is also to be noted that, it can easily be drawn an invertible between any two pairs in $\{(\gamma,\delta), (\gamma^0,\delta^0),\hdots,(\gamma^{p-1},\delta^{p-1})\}$. Therefore, each of the $p$ pairs contributes to $\mathcal{S}_3$. 
Now
\begin{equation}\label{hdif1}
\begin{split}
&(f_{d_n,t_n})_{\gamma^v}-(f_{d_n,t'_n})_{\delta^v}-((f_{d_n,t_n})_\gamma-(f_{d_n,t'_n})_\delta)\\
=&f_{\gamma^v}-f_{\delta^v}-(f_\gamma-f_\delta)
\end{split}
\end{equation}
Since $\gamma^v$ and $\gamma$ differs only at the position $l_{\pi(u-1)}$, and $\gamma_{j_\alpha}=c_\alpha$ for $\alpha=0,1,\hdots,n-1$,
\begin{equation}\label{hdif2}
    \begin{split}
      f_{\gamma^v}-f_\gamma&=q \kappa (\gamma_{l_{\pi(u-2)}}+\gamma_{l_{\pi(u)}})+\kappa p       g_{l_{\pi(u-1)}}\\&~~~~ -v\left(\frac{q}{p}\gamma_{l_{\pi(u-2)}}+\frac{q}{p}\gamma_{l_{\pi(u)}}+g_{l_{\pi(u-1)}}\right).
    \end{split}
\end{equation}
Similarly, 
\begin{equation}\label{hdif3}
    \begin{split}
      f_{\delta^v}-f_\delta=&q \kappa (\delta_{l_{\pi(u-2)}}+\delta_{l_{\pi(u)}})+\kappa p       g_{l_{\pi(u-1)}}\\&~~~ -v\left(\frac{q}{p}\delta_{l_{\pi(u-2)}}+\frac{q}{p}\delta_{l_{\pi(u)}}+g_{l_{\pi(u-1)}}\right).
    \end{split}
\end{equation}
Since $\gamma_{l_{\pi(\alpha)}}=\delta_{l_{\pi(\alpha)}}$ for $\alpha=0,1,\hdots,u-1$, from (\ref{hdif1}), (\ref{hdif2}), and (\ref{hdif3}), we have
\begin{equation}\label{hdif4}
    \begin{split}
       &(f_{d_n,t_n})_{\gamma^v}-(f_{d_n,t'_n})_{\delta^v}-((f_{d_n,t_n})_\gamma-(f_{d_n,t'_n})_\delta)\\=&f_{\gamma^v}-f_{\delta^v}-(f_\gamma-f_\delta) \\
       =&q \kappa (\gamma_{l_{\pi(u)}}-\delta_{l_{\pi(u)}})+\frac{vq}{p}(\delta_{l_{\pi(u)}}-\gamma_{l_{\pi(u)}}).
    \end{split}
\end{equation} 
Since $\xi_q^{q \kappa (\gamma_{l_{\pi(u)}}-\delta_{l_{\pi(u)}})}=1$, from (\ref{hdif4}), we have
\begin{equation}\nonumber
    \begin{split}
        &\sum_{v=1}^{p-1}\xi_q^{ (f_{d_n,t_n})_{\gamma^u}-(f_{d_n,t'_n})_{\delta^u}-((f_{d_n,t_n})_\gamma-(f_{d_n,t'_n})_\delta)}\\=&\sum_{v=1}^{p-1} \xi_q^{q \kappa (\gamma_{l_{\pi(u)}}-\delta_{l_{\pi(u)}})} \xi_p^{v (\delta_{l_{\pi(u)}}-\gamma_{l_{\pi(u)}})}\\
        =&\sum_{v=1}^{p-1}\xi_p^{v (\delta_{l_{\pi(u)}}-\gamma_{l_{\pi(u)}})}\\
        =&-1.
    \end{split}
\end{equation}
Therefore, 
\begin{equation}\nonumber
   \xi_q^{(f_{d_n,t_n})_\gamma-(f_{d_n,t'_n})_\delta}+\sum_{v=1}^{p-1}\xi_q^{ (f_{d_n,t_n})_{\gamma^v}-(f_{d_n,t'_n})_{\delta^v}}=0.
\end{equation}

\textit{Case 3} ($\tau=0$): 
Since $\tau=0$, $\gamma=\delta$, from (\ref{s3der}), we have 
\begin{equation}\nonumber
    \begin{split}
        \mathcal{S}_3&=\sum_{d_n}\Theta\left(\psi\left(f_{d_n,t_n}\arrowvert_{\mathbf{x}_J=\mathbf{c}}\right),\psi\left(f_{d_n,t'_n}\arrowvert_{\mathbf{x}_J=\mathbf{c}}\right)\right)(\tau)   \\
      &=\sum_{d_n} \sum_{(\gamma,\gamma)\in \mathbf{A}_\tau(\mathbf{c})}\xi_q^{(f_{d_n,t_n})_\gamma-(f_{d_n,t'_n})_\delta}\\
      &=\sum_{(\gamma,\gamma)\in \mathbf{A}_\tau(\mathbf{c})}\xi_q^{f_\gamma-f_\delta} \xi_p^{t_n\gamma_{l_{\pi(m-n-1)}}-t'_n \delta_{l_{\pi(m-n-1)}}} \\&~~~~~~~~~~~~~~~~~\times\sum_{d_n} \xi_p^{k d_n(\gamma_{l_{\pi(0)}}-\delta_{l_{\pi(0)}})}\\
      &=p\sum_{(\gamma,\gamma)\in \mathbf{A}_\tau(\mathbf{c})} \xi_p^{t_n\gamma_{l_{\pi(m-n-1)}}-t'_n \delta_{l_{\pi(m-n-1)}}}\\
      &=p\sum_{(\gamma,\gamma)\in \mathbf{A}_\tau(\mathbf{c})} \xi_p^{(t_n-t_n')\gamma_{l_{\pi(m-n-1)}}}\\
      &=p p^{m-n-1}\sum_{\gamma_{l_{\pi(m-n-1)}}=0}^{p-1} \xi_p^{(t_n-t_n')\gamma_{l_{\pi(m-n-1)}}}
      \\
      &=\begin{cases}
      p^{m-n+1},& t_n=t_n',\\
      0,&t_n\neq t_n'.
      \end{cases}
      \end{split}
      \end{equation}

Combining all the above cases in (\ref{mains1}), we have
\begin{equation}\label{lastmain}
    \begin{split}
        \mathcal{S}_1&=\begin{cases}
        p^{m-n+1}p^n\sum_{\mathbf{c}}\xi_p^{(\mathbf{t}-\mathbf{t}')\cdot\mathbf{c}},& \tau=0, t_n=t_n',\\
        0,& \textnormal{otherwise},
        \end{cases}\\
        &=\begin{cases}
        p^{m+n+1},& \tau=0, \mathbf{t}=\mathbf{t}',t_n=t_n',\\
        0, &\textnormal{otherwise.}
        \end{cases}
    \end{split}
\end{equation}
From (\ref{mainmainres}) and (\ref{lastmain}), we have 
\begin{equation}\nonumber
    \begin{split}
        &\Theta\left(\psi(C_{t}^k),\psi(C_{t'}^k)\right)(\tau)\\
        =&\begin{cases}
        p^{m+n+1}, & \tau=0, \mathbf{t}=\mathbf{t}',t_n=t_n',\\
        0,& \textnormal{otherwise},
        \end{cases}\\
        =&\begin{cases}
            p^{m+n+1}, & \tau=0, t=t',\\
            0,& 0<|\tau|<p^m,t= t',\\
            0,&0\leq |\tau|<p^m,t\neq t'.
        \end{cases}
    \end{split}
\end{equation}
Therefore, $\mathcal{C}_k$ forms $(p^{n+1},p^m)$-CCC for any choice of $k$ in $\{1,2,\hdots,p-1\}$.
\section{Proof of Proposition 1}\label{Appendix:C}
For any element $(\mathbf{e}_1,\mathbf{e}_2)$ in $\mathcal{S}$,
we have 
\begin{equation}\nonumber
\begin{split}
&k_1\mathbf{e}_1-k_2\mathbf{e}_2\equiv \mathbf{0}_w(\!\!\!\!\!\!\mod p)\\
&\Rightarrow \mathbf{e}_2\equiv \left(\frac{k_1}{k_2}\right)\mathbf{e}_1(\!\!\!\!\!\!\mod p),
\end{split}
\end{equation}
where $\frac{1}{k_2}$ represents the multiplicative inverse of $k_2$ with respect to modulo $p$ operation. 
Therefore, $|\mathcal{S}|=p^w$. Let us define a mapping $\Lambda:\mathcal{S}\rightarrow\mathbb{Z}$ as follows:
%$$\Lambda(\mathbf{e}_1,\mathbf{e}_2)=(\mathbf{e}_2-\mathbf{e}_1)\cdot (p^{w-1},p^{w-2},\hdots,1).$$
\begin{equation}\label{aninkta}
\begin{split}
 \Lambda(\mathbf{e}_1,\mathbf{e}_2)&=(\mathbf{e}_2-\mathbf{e}_1)\cdot (p^{w-1},p^{w-2},\hdots,1)\\& = \sum\limits_{t=0}^{w-1}e_{2,t}p^{w-1-t}-\sum\limits_{t=0}^{w-1}e_{1,t}p^{w-1-t}.
\end{split}
\end{equation}\ccn
For any two $(\mathbf{e}_1,\mathbf{e}_2)$ and $(\mathbf{e}'_1,\mathbf{e}'_2)$ in $\mathcal{S}$, we have
$k_1\mathbf{e}_1-k_2\mathbf{e}_2\equiv \mathbf{0}_w(\!\!\!\!\mod p)$ and $k_1\mathbf{e}_1'-k_2\mathbf{e}_2'\equiv \mathbf{0}_w(\!\!\!\!\!\!\mod p)$, where $\mathbf{e}_1=(e_{1,0},e_{1,1},\hdots,e_{1,w-1})$,
$\mathbf{e}_2=(e_{2,0},e_{2,1},\hdots,e_{2,w-1})$, 
$\mathbf{e}_1'=(e_{1,0}',e_{1,1}',\hdots,e_{1,w-1}')$, and 
$\mathbf{e}_2'=(e_{2,0}',e_{2,1}',\hdots,e_{2,w-1}')$.
Therefore,
\begin{equation}\label{dv}
\begin{split}
&\mathbf{e}_2-\mathbf{e}_1\equiv \left(\frac{k_1-k_2}{k_2}\right)\mathbf{e}_1(\!\!\!\!\!\!\mod p),\\
&\textnormal{and}\\
& \mathbf{e}'_2-\mathbf{e}'_1\equiv \left(\frac{k_1-k_2}{k_2}\right)\mathbf{e}'_1 (\!\!\!\!\!\!\mod p ). 
\end{split}
\end{equation}
From (\ref{dv}), it can be observed that $\Lambda$ is an injective mapping. Since, $0<k_1\neq k_2<p$, and $k_1\mathbf{e}_1-k_2\mathbf{e}_2\equiv \mathbf{0}_w(\!\!\!\!\mod p)$, $$\Lambda(\mathbf{e}_1,\mathbf{e}_2)=0~\textnormal{iff}~\mathbf{e}_1=\mathbf{e}_2=\mathbf{0}_w.$$
Now let us define two vectors $\bar{\mathbf{e}}_1$ and $\bar{\mathbf{e}}_2$ whose components are defined as follows:
\begin{equation}\label{coidc1c2}
\begin{split}
\bar{e}_{i,\alpha}=\begin{cases}
p-e_{i,\alpha},& \textnormal{if}~~ e_{i,\alpha}\neq 0,\\
0,& \textnormal{otherwise},
\end{cases} 
\end{split}
\end{equation}
where $i=1,2$, and $\alpha=0,1,\hdots,w-1$.
%Since, $k_1\mathbf{c}_1-k_2\mathbf{c}_2\equiv \mathbf{0}_n(\mod %p)$ and $0<k_1\neq k_2<p$, then $c_{2,\alpha}= %0,\textnormal{if}~~c_{1,\alpha}= 0$.
From (\ref{aninkta}) and (\ref{coidc1c2}), it 
is clear that $\Lambda(\mathbf{e}_1,\mathbf{e}_2)=-\Lambda(\bar{\mathbf{e}}_1,\bar{\mathbf{e}}_2)$.
From the mapping, define two sets $\mathcal{S}'=\{(\mathbf{e}_1,\mathbf{e}_2)\in\mathcal{S}:\Lambda(\mathbf{e}_1,\mathbf{e}_2)\geq 0\}$ and $\mathcal{S}''=\{(\mathbf{e}_1,\mathbf{e}_2)\in\mathcal{S}:\Lambda(\mathbf{e}_1,\mathbf{e}_2)\leq 0\}$.
Then the set $\mathcal{S}$ can be expressed as $\mathcal{S}=\mathcal{S}'\cup\mathcal{S''} $, where $\mathcal{S}'\cap\mathcal{S''}=\{\mathbf{0}_w\}$.
From (\ref{coidc1c2}), we have $|\mathcal{S}'|=|\mathcal{S}''|=\frac{p^w+1}{2}=E$.
We assume that $(\mathbf{e}_1^i,\mathbf{e}_2^i)$ is
%$,(\mathbf{c}_1^2,\mathbf{c}_2^2),\hdots,(\mathbf{c}_1^E,\mathbf{c%}_2^E)$ are $E$ 
an element of $\mathcal{S}'$ and $\Lambda(\mathbf{e}_1^i,\mathbf{e}_2^i)=D_i$, where 	$\mathbf{e}_j^i=(e_{j,0}^i,e_{j,1}^i,\hdots, e_{j,n-1}^i)$, $i=1,2,\hdots,E$, and $j=1,2$. Since, $(\mathbf{0}_w,\mathbf{0}_w)\in \mathcal{S}'$ and $\Lambda$ is an injective mapping, without loss of generality, let us assume that $0=D_1<D_2<\cdots<D_E$.  
%Let us consider $0\leq \tau \leq p^m-1$ and define the 

%%%%%%%%%%%%%%Need to modify%%%%%%%%%%%%%%%
For $0\leq \tau\leq p^m-1$, following (\ref{cross_corr_restrict_c1_c2}), we have
$\mathbf{B}_\tau(\mathbf{e}_1^i,\mathbf{e}_2^i)=\{(\gamma,\delta):\delta=\gamma+\tau, 0\leq \gamma \leq p^m-\tau-1, \gamma_\alpha=e_{1,\alpha}^i, \delta_\alpha=e_{2,\alpha}^i,\alpha=0,1,\hdots,w-1\}$, where
$(\gamma_0,\gamma_1,\hdots,\gamma_{m-1})$ and $(\delta_0,\delta_1,\hdots,\delta_{m-1})$ are the base-$p$ vector representations of the non-negative integers $\gamma$ and $\delta$, respectively. \ccn Now, 
\begin{equation}\nonumber
\begin{split}
\tau&=\delta-\gamma
\\&=\sum_{\alpha=0}^{m-1}(\delta_\alpha-\gamma_{\alpha})p^{m-\alpha-1}\\
&=\sum_{\alpha=0}^{w-1}(e^i_{2,\alpha}-e^i_{1,\alpha})p^{m-\alpha-1}+\sum_{\alpha=w}^{m-1}(\delta_\alpha-\gamma_\alpha)p^{m-\alpha-1}\\
&=p^{m-w}(\mathbf{e}^i_2-\mathbf{e}^i_1)\cdot (p^{w-1},p^{w-2},\hdots,1)\\&~~~~~~~~~~~+\sum_{\alpha=w}^{m-1}(\delta_\alpha-\gamma_\alpha)p^{m-\alpha-1}\\
&=p^{m-w}D_i+\sum_{\alpha=w}^{m-1}(\delta_\alpha-\gamma_\alpha)p^{m-\alpha-1}.
\end{split}
\end{equation} 
The set $\mathbf{B}_\tau(\mathbf{e}^i_1,\mathbf{e}^i_2)$ is non-empty if
$p^{m-w}D_i+\sum_{\alpha=w}^{m-1}((p-1)-0)p^{m-\alpha-1}\geq \tau\geq p^{m-w}D_i+\sum_{\alpha=w}^{m-1}(0-(p-1))p^{m-\alpha-1}$, or $\tau\in [p^{m-w}(D_i-1)+1: p^{m-w}(D_i+1)-1]=I_{D_i}$, say. 
%Now let us define a set $\triangle_{D_i}=\mathbb{Z}\cap I_{D_{i}}$. 
Hence, 
\begin{equation}\label{trio1}
\mathbf{B}_\tau(\mathbf{e}^i_1,\mathbf{e}^i_2)\neq \emptyset~\textnormal{iff}~\tau\in I_{D_i}.
\end{equation}
Let us assume that  $(\mathbf{e}^{i_1}_1,\mathbf{e}^{i_1}_2)$ and $(\mathbf{e}^{i_2}_1,\mathbf{e}^{i_2}_2)$ are two distinct elements 
in $\mathcal{S}'$ with
$D_{i_1}<D_{i_2}$.
Now
\begin{equation}\label{upd1}
\begin{split}
&I_{D_{i_1}}\cap I_{D_{i_2}}\\=&\begin{cases}
[p^{m-w}D_{i_1}+1,p^{m-w}(D_{i_1}+1)-1],~\textnormal{if}~D_{i_2}\!\!=\!D_{i_1}\!+\!1,\\
\emptyset,~~~~~~~~~~~~~~~~~~~~~~~~~~~~~~~~~~~~~~~~~~ \textnormal{if}~D_{i_2}\!>\!D_{i_1}\!+\!1.
\end{cases}
\end{split}
\end{equation}
From (\ref{upd1}), it is clear that 
$$I_{D_{i_1}}\cap I_{D_{i_2}}\neq \emptyset ~\textnormal{iff}~D_{i_2}=D_{i_1}+1.$$
Therefore, for a fixed value of $\tau$ in $[0,p^m-1]$, we need to consider the following three cases:
\textit{Case 1:}
In this case, we consider $\tau\notin \cup_{i=1}^E I_{D_i}$. From (\ref{trio1}), we have
$\mathbf{B}_\tau(\mathbf{e}^i_1,\mathbf{e}^i_2)=\emptyset\ \forall \ i \in [1:E]$. Since, $\tau\geq 0$ and $\mathbf{B}_\tau(\mathbf{e}^i_1,\mathbf{e}^i_2)=\emptyset\ \forall \ i\in [1:E]$,
\begin{equation}\nonumber
\begin{split}
&\sum_{(\mathbf{e}_1,\mathbf{e}_2)\in\mathcal{S}}\Theta(\psi(g\arrowvert_{\mathbf{x}_{J_1}=\mathbf{e}_1}),\psi(h\arrowvert_{\mathbf{x}_{J_1}=\mathbf{e}_2}))(\tau)\\
=&\sum_{(\mathbf{e}_1,\mathbf{e}_2)\in\mathcal{S}'}\Theta(\psi(g\arrowvert_{\mathbf{x}_{J_1}=\mathbf{e}_1}),\psi(h\arrowvert_{\mathbf{x}_{J_1}=\mathbf{e}_2}))(\tau)\\
=&\sum_{i=1}^E \Theta(\psi(g\arrowvert_{\mathbf{x}_{J_1}=\mathbf{e}^i_1}),\psi(h\arrowvert_{\mathbf{x}_{J_1}=\mathbf{e}^i_2}))(\tau)\\
=&\sum_{i=1}^E \sum_{(\gamma,\delta)\in \mathbf{B}_\tau(\mathbf{e}^i_1,\mathbf{e}^i_2) }\xi_q^{g_\gamma-h_\delta}\\
=&0.
\end{split}
\end{equation} 
Therefore, $	\left|\sum_{(\mathbf{e}_1,\mathbf{e}_2)\in\mathcal{S}}\Theta(\psi(g\arrowvert_{\mathbf{x}_{J_1}=\mathbf{e}_1}),\psi(h\arrowvert_{\mathbf{x}_{J_1}=\mathbf{e}_2}))(\tau)\right|=0$ when $\tau\notin \cup_{i=1}^E I_{D_i}$.

\textit{Case 2:}
In this case, we consider $\tau\in I_{D_i}$ and $\tau\notin I_{D_j}\ \forall \ i\neq j\in [1:E]$. 
Since $\tau\notin\displaystyle \cup_{\substack{j=1\\ j\neq i}}^E I_{D_j}$,
$\mathbf{B}_\tau(\mathbf{e}^j_1,\mathbf{e}^j_2)=\emptyset$ for all $j\in\{1,2,\hdots,E\}\setminus \{i\}$. Now $I_{D_i}$ can be expressed as $I_{D_i}=\left([p^{m-w}(D_{i}-1)+1: p^{m-w}D_{i}]\right)\cup \left((p^{m-w}D_{i}: p^{m-w}(D_{i}+1)-1]\right).$
Now $\tau$ can be expressed as follows:
\begin{equation}\nonumber
\begin{split}
&\tau\\=&\begin{cases}
p^{m-w}(D_{i}-1)+\tau_1, \\~~~~~~~~\textnormal{if}~\tau\in [p^{m-w}(D_{i}-1)+1: p^{m-w}D_{i}],\\
p^{m-w}D_{i}+\tau_2,  \\~~~~~~~~\textnormal{if}~ \tau\in (p^{m-w}D_{i}: p^{m-w}(D_{i}+1)-1],
\end{cases}
\end{split}
\end{equation}
where $\tau_1\in [1:p^{m-n}]$ and $\tau_2\in [1:p^{m-n}-1]$. Also, 
\begin{equation}\label{bc1c2con}
\begin{split}
&|\mathbf{B}_\tau(\mathbf{e}^i_1,\mathbf{e}^i_2)|\\=&\begin{cases}
\tau_1, & \textnormal{if}~ \tau\in [p^{m-w}(D_{i}-1)+1 : p^{m-w}D_{i}],\\
p^{m-w}-\tau_2,& \textnormal{if}~ \tau\in (p^{m-w}D_{i}: p^{m-w}(D_{i}+1)-1]. 
\end{cases}
\end{split}
\end{equation}
Now,
\begin{equation}\label{crthet}
\begin{split}
&\sum_{(\mathbf{e}_1,\mathbf{e}_2)\in\mathcal{S}}\Theta(\psi(g\arrowvert_{\mathbf{x}_{J_1}=\mathbf{e}_1}),\psi(h\arrowvert_{\mathbf{x}_{J_1}=\mathbf{e}_2}))(\tau)\\
=&\sum_{(\mathbf{e}_1,\mathbf{e}_2)\in\mathcal{S}'}\Theta(\psi(g\arrowvert_{\mathbf{x}_{J_1}=\mathbf{e}_1}),\psi(h\arrowvert_{\mathbf{x}_{J_1}=\mathbf{e}_2}))(\tau)\\
=&\Theta(\psi(g\arrowvert_{\mathbf{x}_{J_1}=\mathbf{e}^i_1}),\psi(h\arrowvert_{\mathbf{x}_{J_1}=\mathbf{e}^i_2}))(\tau)\\&+\sum_{\substack{j=1\\j\neq i}}^E \Theta(\psi(g\arrowvert_{\mathbf{x}_{J_1}=\mathbf{e}^j_1}),\psi(h\arrowvert_{\mathbf{x}_{J_1}=\mathbf{e}^j_2}))(\tau)\\
=&\sum_{(\gamma,\delta)\in \mathbf{B}_\tau(\mathbf{e}^i_1,\mathbf{e}^i_2) }\xi_q^{g_\gamma-h_\delta}+\sum_{\substack{j=1\\ j\neq i}}^E \sum_{(\gamma,\delta)\in \mathbf{B}_\tau(\mathbf{e}^j_1,\mathbf{e}^j_2) }\xi_q^{g_\gamma-h_\delta}\\
=&\sum_{(\gamma,\delta)\in \mathbf{B}_\tau(\mathbf{e}^i_1,\mathbf{e}^i_2) }\xi_q^{g_\gamma-h_\delta}.
\end{split}
\end{equation} 
From (\ref{bc1c2con}) and (\ref{crthet}), we have
\begin{equation}\label{nxtcrbd}
\begin{split}
&\left|\sum_{(\mathbf{e}_1,\mathbf{e}_2)\in\mathcal{S}}\Theta(\psi(g\arrowvert_{\mathbf{x}_{J_1}=\mathbf{e}_1}),\psi(h\arrowvert_{\mathbf{x}_{J_1}=\mathbf{e}_2}))(\tau)\right|\\=&\left|\sum_{(\gamma,\delta)\in \mathbf{B}_\tau(\mathbf{e}^i_1,\mathbf{e}^i_2) }\xi_q^{g_\gamma-h_\delta}\right|\\ \leq & \begin{cases}
\tau_1, & \textnormal{if}~ \tau\in [p^{m-w}(D_{i}-1)+1 : p^{m-w}D_{i}],\\
p^{m-w}-\tau_2,& \textnormal{if}~ \tau\in (p^{m-w}D_{i} : p^{m-w}(D_{i}+1)-1].
\end{cases}	
\end{split}
\end{equation}
Since $\tau_1\in [1: p^{m-w}]$ and $\tau_2\in [1: p^{m-w}-1]$, from (\ref{nxtcrbd}), we have
$$|\sum_{(\mathbf{e}_1,\mathbf{e}_2)\in\mathcal{S}}\Theta(\psi(g\arrowvert_{\mathbf{x}_{J_1}=\mathbf{e}_1}),\psi(h\arrowvert_{\mathbf{x}_{J_1}=\mathbf{e}_2}))(\tau)|\leq p^{m-w}.
$$

\textit{Case 3:}
In this case, we consider $\tau\in I_{D_i}\cap I_{D_{i+1}}$ for some $i\in [1:E]$, where $D_{i+1}=D_{i}+1$. From (\ref{upd1}), we have
\begin{equation}\label{udf2}
\begin{split}
%\triangle_{D_i}\cap \triangle_{D_{i+1}}
(I_{D_i}\cap I_{D_{i+1}})
%=[p^{m-n}(D_{i_1}-1)+1, p^{m-n}(D_{i_1}+1)-1]\cap [p^{m-n}D_{i_1}+1, %p^{m-n}(D_{i_1}+2)-1]\\
=[p^{m-w}D_{i}+1: p^{m-w}(D_{i}+1)-1].
\end{split}	
\end{equation}
Also, $\mathbf{B}_\tau(\mathbf{e}^i_1,\mathbf{e}^i_2)=\emptyset\ \forall \ i\in [1:E]\setminus\{i,i+1\}$. Therefore, 
\begin{equation}\label{crthet1}
\begin{split}
&\sum_{(\mathbf{e}_1,\mathbf{e}_2)\in\mathcal{S}}\Theta(\psi(g\arrowvert_{\mathbf{x}_{J_1}=\mathbf{e}_1}),\psi(h\arrowvert_{\mathbf{x}_{J_1}=\mathbf{e}_2}))(\tau)\\
=&\sum_{(\mathbf{e}_1,\mathbf{e}_2)\in\mathcal{S}'}\Theta(\psi(g\arrowvert_{\mathbf{x}_{J_1}=\mathbf{e}_1}),\psi(h\arrowvert_{\mathbf{x}_{J_1}=\mathbf{e}_2}))(\tau)\\
=&\Theta(\psi(g\arrowvert_{\mathbf{x}_{J_1}=
	\mathbf{e}^{i}_1}),\psi(h\arrowvert_{\mathbf{x}_{J_1}=\mathbf{e}^{i}_2}))(\tau)\\&+\Theta(\psi(g\arrowvert_{\mathbf{x}_{J_1}=
	\mathbf{e}^{i+1}_1}),\psi(h\arrowvert_{\mathbf{x}_{J_1}=\mathbf{e}^{i+1}_2}))(\tau)\\&+\sum_{\substack{j=1\\j\neq i,i+1}}^E \Theta(\psi(g\arrowvert_{\mathbf{x}_{J_1}=\mathbf{e}^j_1}),\psi(h\arrowvert_{\mathbf{x}_{J_1}=\mathbf{e}^j_2}))(\tau)\\
=&\Theta(\psi(g\arrowvert_{\mathbf{x}_{J_1}=
	\mathbf{e}^{i}_1}),\psi(h\arrowvert_{\mathbf{x}_{J_1}=\mathbf{e}^{i}_2}))(\tau)\\&+\Theta(\psi(g\arrowvert_{\mathbf{x}_{J_1}=
	\mathbf{e}^{i+1}_1}),\psi(h\arrowvert_{\mathbf{x}_{J_1}=\mathbf{e}^{i+1}_2}))(\tau)\\
=&\sum_{(\gamma,\delta)\in \mathbf{B}_\tau(\mathbf{e}^{i}_1,\mathbf{e}^{i}_2) }\xi_q^{g_\gamma-h_\delta}+\sum_{(\gamma,\delta)\in \mathbf{B}_\tau(\mathbf{e}^{i+1}_1,\mathbf{e}^{i+1}_2) }\xi_q^{g_\gamma-h_\delta}.
\end{split}
\end{equation}
From (\ref{udf2}), $\tau\in [p^{m-w}D_{i}+1, p^{m-w}(D_{i}+1)-1]=[p^{m-w}(D_{i+1}-1)+1, p^{m-w}(D_{i+1})-1]$, then
$\tau$ can be expressed as $\tau=p^{m-w}D_{i}+\tau_3=p^{m-w}(D_{i+1}-1)+\tau_3$, where $\tau_3\in [1:p^{m-w}-1]$. Therefore, $\left|\mathbf{B}_\tau(\mathbf{e}^{i}_1,\mathbf{e}^{i}_2)\right|=p^{m-w}-\tau_3$ and 
$\left|\mathbf{B}_\tau(\mathbf{e}^{i+1}_1,\mathbf{e}^{i+1}_2)\right|=\tau_3$. From (\ref{crthet1}), we have
\begin{equation}\nonumber
\begin{split}
&\left|\sum_{(\mathbf{e}_1,\mathbf{e}_2)\in\mathcal{S}}\Theta(\psi(g\arrowvert_{\mathbf{x}_{J_1}=\mathbf{e}_1}),\psi(h\arrowvert_{\mathbf{x}_{J_1}=\mathbf{e}_2}))(\tau)\right|\\=&\left|\sum_{(\gamma,\delta)\in \mathbf{B}_\tau(\mathbf{e}^{i}_1,\mathbf{e}^{i}_2) }\xi_q^{g_\gamma-h_\delta}+\sum_{(\gamma,\delta)\in \mathbf{B}_\tau(\mathbf{e}^{i+1}_1,\mathbf{e}^{i+1}_2) }\xi_q^{g_\gamma-h_\delta}\right|\\
\leq & \left|\sum_{(\gamma,\delta)\in \mathbf{B}_\tau(\mathbf{e}^{i}_1,\mathbf{e}^{i}_2) }\xi_q^{g_\gamma-h_\delta}\right|+\left|\sum_{(\gamma,\delta)\in \mathbf{B}_\tau(\mathbf{e}^{i+1}_1,\mathbf{e}^{i+1}_2) }\xi_q^{g_\gamma-h_\delta}\right|\\
\leq & p^{m-w}-\tau_3+\tau_3=p^{m-w}.
\end{split}
\end{equation}
Therefore, $$\left|\sum_{(\mathbf{e}_1,\mathbf{e}_2)\in\mathcal{S}}\Theta(\psi(g\arrowvert_{\mathbf{x}_{J_1}=\mathbf{e}_1}),\psi(h\arrowvert_{\mathbf{x}_{J_1}=\mathbf{e}_2}))(\tau)\right|\leq p^{m-w}$$.

Combining all the above cases, it is clear that 
\begin{equation}\nonumber
\begin{split}
&\left|\sum_{(\mathbf{e}_1,\mathbf{e}_2)\in\mathcal{S}}\Theta(\psi(g\arrowvert_{\mathbf{x}_{J_1}=\mathbf{e}_1}),\psi(h\arrowvert_{\mathbf{x}_{J_1}=\mathbf{e}_2}))(\tau)\right|\\ \leq & p^{m-w} ~\forall~ \tau\in [0:p^{m}-1].
\end{split}
\end{equation}
Similarly, it can be shown that
\begin{equation}
	\begin{split}
&\left|\sum_{(\mathbf{e}_1,\mathbf{e}_2)\in\mathcal{S}}\Theta(\psi(g\arrowvert_{\mathbf{x}_{J_1}=\mathbf{e}_1}),\psi(h\arrowvert_{\mathbf{x}_{J_1}=\mathbf{e}_2}))(\tau)\right|\\\leq & p^{m-w} ~\forall ~ \tau\in [-(p^{m-w}-1):0].
	\end{split}
\end{equation}
\section{Correlation Lower Bound with Respect to the Positive-Cycle-of-a-Sine-Wave Weight Vector when $t\in [L+1:2L-1]$ }\label{Appendix:ank}
\begin{figure}
		\includegraphics[width=8cm]{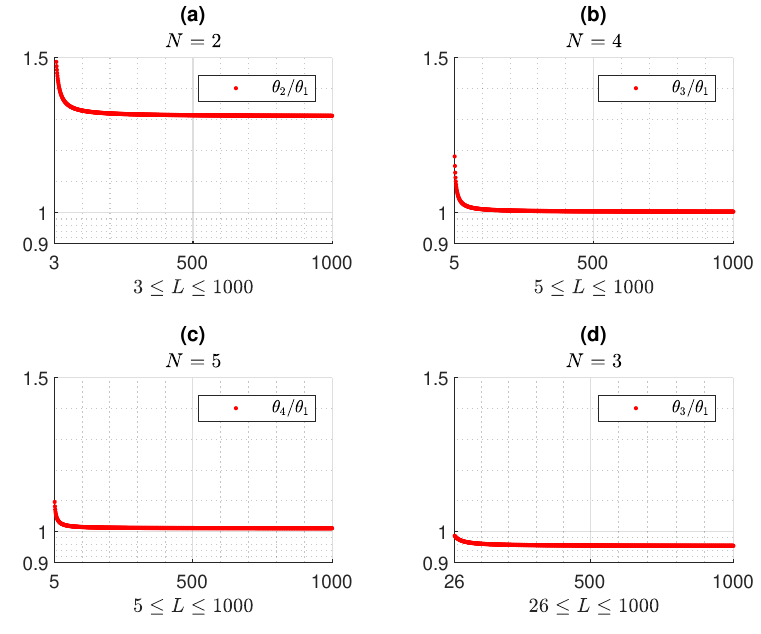}
		\caption{Comparison of the lower bound $\theta_1$ with the lower bounds $\theta_2$, $\theta_3$, and $\theta_4$ \ccn}\label{ekla_fig1}
	\end{figure}

	For $L+1\leq t\leq 2L-1$, we have the following results from \cite{zlbnd_lvstn}:
	\begin{equation}\label{pcs2}
	\begin{split}
	&\sum_{u,v=0}^{t-1}\tau_{u,v,L}w_uw_v\\ =&-\frac{3t-4L+2}{4}-\frac{t}{4}\tan^2\frac{\pi}{2t}+\frac{t-L-1}{2}\cos\frac{L\pi}{t}\\&+\left(\frac{2t-2L+1}{4}\tan\frac{\pi}{2t}
	+\frac{3}{4\tan\frac{\pi}{2t}}\right)\sin\frac{L\pi}{t}.
	\end{split}	
	\end{equation}	
	%\begin{figure}
	%	\includegraphics[width=7cm]{cos_matlab.pdf}
	%	\caption{Plot for $\cos\frac{L\pi}{t}$ and the approximation function $1.3\left(\frac{\pi}{2}-\frac{L\pi}{t}\right)-0.4\left(\frac{\pi}{2}-\frac{L\pi}{t}\right)^2$}\label{approx_cos}
	%\end{figure}
	To simplify the derivation, in (\ref{cr2wu2}) and (\ref{pcs2}), we consider the following approximations:
	\begin{equation}\nonumber\label{acr2wu2}
	\sum_{u=0}^{2L-2}w_u^2=\frac{t}{2}\tan^2\frac{\pi}{2t}\approx \frac{\pi^2}{8t}, 	
	\end{equation}	
	where $t$ is sufficiently large, 
	\begin{equation}\nonumber\label{apcs2}
	\begin{array}{c}
	\sin\frac{L\pi}{t}\approx \frac{1.3L\pi}{t}-\frac{0.4L^2\pi^2}{t^2},
	\end{array}	
	\end{equation}
	\begin{equation}\nonumber\label{approx_csin}
	\begin{array}{c}
	\cos\frac{L\pi}{t}\approx 1.3\left(\frac{\pi}{2}-\frac{L\pi}{t}\right)-0.4\left(\frac{\pi}{2}-\frac{L\pi}{t}\right)^2.
	\end{array}
	\end{equation}
	%	\begin{figure}[t!]
	%		\centering
	%		\includegraphics[width=7cm]{sin_mat.pdf}
	%		\caption{Plot for $\sin\frac{L\pi}{t}$ and the approximation function  $\frac{1.3L\pi}{t}-\frac{0.4L^2\pi^2}{t^2}$ }\label{approx_sin}
	%	\end{figure}
	%For a clear view on the above approximations, 
	%	in Figure \ref{approx_sin} (a), (b), and (c), we plot $\sin\frac{L\pi}{t}$ and it's approximation as presented in (\ref{apcs2})
	%	for $L=10$, $L=50$, and $L=100$, respectively.
	%	Furthermore in Figure \ref{approx_cos} (a), (b), and (c), we have %plotted  $\cos\frac{L\pi}{t}$ and it's approximation as presented in (\ref{approx_csin})
	%	for $L=10$, $L=50$, and $L=100$, respectively.
	With the above approximations, it follows from (\ref{lbcccderiv}), (\ref{cr2wu2}), and (\ref{pcs2}) that
	\begin{equation}\label{w2c2bnd}
	\begin{split}
	\theta^2\geq \frac{MN}{80(N-1)}\left(a_0t+\frac{a_1}{t}+\frac{a_2}{t^2}+\frac{a_3}{t^3}+a_4\right),
	\end{split}
	\end{equation}
	where 
	$a_0=4\pi^2-26\pi+60$, $a_1=5\pi^2-\frac{10L^2\pi^2}{N}-10L\pi^2-4L^2\pi+32L^2\pi^2-52L\pi$, $a_2=10L^2\pi^2+8L^2\pi^3-16L^3\pi^2-13L\pi^2$, $a_3=4L^2\pi^3-8L^3\pi^3$, and 
	$a_4=40-156L+26\pi-4\pi^2-20L\pi^2+78L\pi$. 
	It is challenging to get the maximum value of this lower bound.
	As an alternation, we provide a numerical comparisons between the above lower bound with the other lower bounds finalized in Remark \ref{remoo1}.	
In Figure \ref{ekla_fig1} (a), (b), (c), and (d), the lower bound in (\ref{w2c2bnd}) is denoted as $\theta_1$, 
	which is compared with the lower bounds $\theta_2$ in \eqref{EqBound-N2komp} for $N=2$ and $3\leq L\leq 1000$, $\theta_3$ in \eqref{komp11} for $N=3$, $1000\geq L \geq 26$, and for $N=4$, $5\leq L\leq 1000$,
	and the lower bound $\theta_4$ in \eqref{komp10} for $N=5$ and $5\leq L \leq 1000$.\ccn
%\section{Table \ref{exthm1C1} and Table \ref{exthm1C2}}\label{Appendix:D}

\bibliographystyle{IEEEtran}
\bibliography{QCSS_v2}
\end{document}